\documentclass[a4paper,12pt]{article}
\input cohmacros.sty

\def\at#1{[*** \att #1 ***]}  
\def\at#1{} 
\usepackage{citeref}

\advance\textheight4cm        
\advance\topmargin-2.5cm
\advance\textwidth2.6cm
\advance\evensidemargin-1.6cm
\advance\oddsidemargin-1.6cm

\begin{document}

\begin{center}
{\LARGE \bf Quantum tomography} \\

{\LARGE \bf explains quantum mechanics} \\

\vspace{1cm}

\centerline{\sl {\large \bf Arnold Neumaier}}

\vspace{0.5cm}

\centerline{\sl Fakult\"at f\"ur Mathematik, Universit\"at Wien}
\centerline{\sl Oskar-Morgenstern-Platz 1, A-1090 Wien, Austria}
\centerline{\sl email: Arnold.Neumaier@univie.ac.at}
\centerline{\sl \url{https://arnold-neumaier.at}}

\end{center}

arXiv:2110.05294
\hfill May 20, 2024

\bfi{Abstract.}
Starting from a new basic principle inspired by quantum tomography 
rather than from Born's rule, this paper gives an elementary and 
self-contained deductive approach to quantum mechanics and quantum 
measurement. A suggestive notion for what constitutes a quantum 
detector, and for the behavior of its responses leads to a logically 
impeccable definition of measurement. Applications to measurement 
schemes for optical states, position measurements, and particle tracks 
demonstrate that this definition is applicable to complex realistic 
experiments without any idealization. 

The various forms of quantum tomography for quantum states, quantum 
detectors, quantum processes, and quantum instruments are discussed. 
The traditional dynamical and spectral properties of quantum mechanics
are derived from a continuum limit of quantum processes. In particular, 
the Lindblad equation for the density operator of a mixing quantum 
system, and the Schr\"odinger equation for the state vector of a pure, 
nonmixing quantum system are shown to be consequences of this new 
approach. Normalized density operators are shown to play the role of 
quantum phase space variables, in complete analogy to the classical 
phase space variables position and momentum. A slight idealization of 
the measurement process leads to the notion of quantum fields, whose 
smeared quantum expectations emerge as reproducible properties of 
regions of space accessible to measurements. 

The new approach is closer to the actual practice than the traditional 
foundations. It is more general, and therefore more powerful. It is 
simpler and less technical than the traditional approach, and the 
standard tools of quantum mechanics are not difficult to derive. 
This makes the new approach suitable for introductory courses on 
quantum mechanics. 

A variety of quotes from the literature illuminate the formal exposition
with historical and philosophical aspects. 

\bigskip
{\bf Acknowledgment.}
This paper benefitted from discussions with Rahel Kn\"opfel.


\vfill

MSC Classification (2020): 
primary: 81P15, secondary: 81P16, 81P50, 81-01

\newpage
\tableofcontents 

\vspace{2cm}
\section{Introduction}\label{s.intro}

\nopagebreak
\hfill\parbox[t]{10.8cm}{\footnotesize

{\em Nous tenons la m\'ecanique des quanta pour une th\'eorie
compl\`ete, dont les hypoth\`eses fondamentales physiques et
math\'ematiques ne sont plus susceptibles de modification.\\
(We regard quantum mechanics as a complete theory for which the 
fundamental physical and mathematical hypotheses are no longer 
susceptible of modification.)
}

\hfill Born and Heisenberg, 1927 \cite[p.178]{BornHeisenberg1928}
}

\bigskip

In a few years, Born's statistical interpretation of quantum mechanics,
which completed the quantum revolution, will celebrate its hundredth 
birthday. Nothing shattered the view of Born and Heisenberg that 
quantum mechanics is a complete modeling framework. On the contrary, 
with a few unsettled exceptions in cosmology (the nonperturbative 
quantum treatment of gravitation being the most conspicuous part), the 
quantum mechanical setting proved to be overwhelmingly successful, 
encompassing now almost all domains of physics and chemistry.

But ...

\subsection{The meaning of quantum concepts}\label{ss.concepts}

\nopagebreak
\hfill\parbox[t]{10.8cm}{\footnotesize

{\em 
Die einzige bisherige annehmbare Interpretation der 
Schr\"odinger-Glei\-chung ist die von Born gegebene statistische 
Interpretation. Diese liefert jedoch keine Realbeschreibung f\"ur das 
Einzelsystem sondern nur statistische Aussagen \"uber 
System-Gesamtheiten.\\
Nach meiner Meinung ist es im Prinzip nicht befriedigend, eine 
der\-artige theoretische Einstellung der Physik zugrunde zu legen, zumal
auf die objektive Beschreibbarkeit der individuellen Makro-Systeme 
(Beschreibung des ''Realzustandes'') nicht verzichtet werden kann ohne 
dass das physikalische Weltbild gewissermassen sich in einen Nebel 
aufl\"ost. Schliesslich ist die Auffassung wohl unvermeidbar, dass die 
Physik nach einer Realbeschreibung des Einzel-Systems streben muss. 
Die Natur als Ganzes kann eben nur als individuelles (einmalig 
existierendes) System gedacht werden und nicht als eine 
''System-Gesamtheit''.
\\
(The only so far acceptable interpretation of the Schr\"odinger 
equation is the statistical interpretation given by Born. But this 
does not provide a realistic description for the single system but 
only statistical statements about ensembles of systems.\\
In my opinion it is unsatisfying in principle to base physics on such 
a theoretical conception, especially since one cannot dispense with the 
objective describability of individual macro systems (description of 
the 'real state'), without which the physical world view dissolves, so 
to say, into a nebula. Ultimately the conception appears unavoidable 
that physics must strive for a realistic description of the individual 
system. For Nature as a whole can only be thought of as an individual 
(uniquely existing) system, and not as an ''ensemble of systems''.)}

\hfill Albert Einstein, 1953 \cite[p.40]{Ein53}
}

\bigskip

\nopagebreak
\hfill\parbox[t]{10.8cm}{\footnotesize

{\em The wide-spread attitude that the claim for an objective 
description of physical reality must be given up, is rejected on the 
ground that the so-called external world is built up exclusively of 
elements of the single minds, and is characterized as what is common to 
all, recognized by every healthy and sane person. Hence the demand for 
a non-subjective description is inevitable, of course without prejudice 
whether it be deterministic or otherwise.}

\hfill Erwin Schr\"odinger, 1958 \cite[Summary]{Schroedinger1958}
}

\bigskip

\nopagebreak
\hfill\parbox[t]{10.8cm}{\footnotesize

{\em Personally I still have this prejudice against indeterminacy in 
basic physics.}

\hfill Paul Dirac, 1972 \cite[p.7]{Dir73}
}

\bigskip

\nopagebreak
\hfill\parbox[t]{10.8cm}{\footnotesize

{\em We always have had (secret, secret, close the doors!) we always
have had a great deal of difficulty in understanding the world view
that quantum mechanics represents. At least I do, because I'm
an old enough man that I haven't got to the point that this stuff
is obvious to me. Okay, I still get nervous with it... you know
how it always is, every new idea, it takes a generation or two until
it becomes obvious that there is no real problem. It has not yet
become obvious to me that there's no real problem. I cannot
define the real problem, therefore I suspect there's no real
problem, but I'm not sure there's no real problem.}

\hfill Richard Feynman, 1982 \cite[p.471]{Fey1982}
}

\bigskip

\nopagebreak
\hfill\parbox[t]{10.8cm}{\footnotesize

{\em Uncertainty over terms such as 'apparatus' is still rife in 
serious discussions of quantum mechanics, over 60 years after its 
conception.}

\hfill John Bell, 1990 \cite[p.33]{Bell.against}
}

\bigskip

\nopagebreak
\hfill\parbox[t]{10.8cm}{\footnotesize

{\em In spite of the seven decades since the discovery of quantum 
mechanics, and in spite of the variety of approaches developed with the 
aim of clarifying its content and improving the original formulation, 
quantum mechanics maintains a remarkable level of obscurity. [...]
Quantum mechanics will cease to look puzzling only when we will be able 
to derive the formalism of the theory from a set of simple physical 
assertions ("postulates," "principles") about the world. Therefore, we 
should not try to append a reasonable interpretation to the quantum 
mechanics formalism, but rather to derive the formalism from a set of 
experimentally motivated postulates.}

\hfill Carlo Rovelli, 1996 \cite[p.1638f]{Rov.rQM}
}

\bigskip

\nopagebreak
\hfill\parbox[t]{10.8cm}{\footnotesize

{\em Quantum theory as a weather-sturdy structure has been with us for 
75 years now. Yet, there is a sense in which the struggle for its 
construction remains. I say this because one can check that not a year 
has gone by in the last 30 when there was not a meeting or conference 
devoted to some aspect of the quantum foundations.}

\hfill Christopher Fuchs, 2003 \cite[p.987]{Fuc2003}
}

\bigskip

\nopagebreak
\hfill\parbox[t]{10.8cm}{\footnotesize

{\em Today, despite the great successes of quantum mechanics, arguments 
continue about its meaning, and its future. [...] It is a bad sign that 
those physicists today who are most comfortable with quantum mechanics 
do not agree with one another about what it all means.}

\hfill Steven Weinberg, 2017 \cite[p.1f]{Wei2017}
}

\bigskip

\nopagebreak
\hfill\parbox[t]{10.8cm}{\footnotesize

{\em We still lack any consensus about what one is actually talking 
about as one uses quantum mechanics. There is a gap between the 
abstract terms in which the theory is couched and the phenomena the 
theory enables each of us to account for so well. [...] The absence of 
conceptual clarity for almost a century suggests that the problem might 
lie in some implicit misconceptions.}

\hfill David Mermin, 2018 \cite[p.1f]{Mer}
}

\bigskip

\nopagebreak
\hfill\parbox[t]{10.8cm}{\footnotesize

{\em I consider it to be an intellectual scandal that, nearly one 
hundred years after the discovery of matrix mechanics by Heisenberg, 
Born, Jordan and Dirac, many or most professional physicists -- 
experimentalists and theorists alike -- admit to be confused about the 
deeper meaning of Quantum Mechanics (QM), or are trying to evade taking 
a clear standpoint by resorting to agnosticism or to overly abstract 
formulations of QM that often only add to the confusion.}

\hfill J\"urg Fr\"ohlich, 2020 \cite[p.1]{Fro}
}

\bigskip

\nopagebreak
\hfill\parbox[t]{10.8cm}{\footnotesize

{\em The question what the intermediate results of a calculation 
actually tell us about the physical processes that are going on, 
remained much more mysterious. Opinions diverged considerably, up to 
today, one hundred years later. [...] many more questions were asked, 
many of them very well posed, but the answers sound too ridiculous to 
be true. [...] I am not the only scientist who feels taken aback by
the imaginative ideas that were launched. [...] Almost a full century 
has passed since the equation was written down, and we still do not 
know what or whom to believe, while other scientists get irritated by 
all this display of impotence. Why is it that we still do not agree?}

\hfill Gerard 't Hooft, 2021 \cite[p.1f]{tHoo2021}
}

\bigskip

... many physicists, including an impressive array of Nobel prize 
winners such as Albert Einstein, Erwin Schr\"odinger, Paul Dirac, 
Richard Feynman, Steven Weinberg, and Gerard 't Hooft, have not been 
satisfied with the traditional foundations of quantum physics. 

Something seems missing that would give the basic quantum mechanical 
concepts a meaning to which everyone can agree.

Thus a fresh perspective seems worthwhile.

\subsection{A fresh perspective}\label{ss.perspective}

\nopagebreak
\hfill\parbox[t]{10.8cm}{\footnotesize

{\em We have developed a pure geometry, which is intended to be
descriptive of the relation-structure of the world. The
relation-structure presents itself in our experience as a physical world
consisting of space, time and things. The transition from the
geometrical description to the physical description can only be made by
identifying the tensors which measure physical quantities with tensors
occurring in the pure geometry; and we must proceed by inquiring first
what experimental properties the physical tensor possesses, and then
seeking a geometrical tensor which possesses these properties by virtue
of mathematical identities. \\
If we can do this completely, we shall have constructed out of the
primitive relation-structure a world of entities which behave in the
same way and obey the same laws as the quantities recognised in physical
experiments. Physical theory can scarcely go further than this.}

\hfill Arthur Eddington 1930 \cite[p.222]{Edd}
}

\bigskip

\nopagebreak
\hfill\parbox[t]{10.8cm}{\footnotesize

{\em Quantum mechanics as discussed in textbooks and in foundational
discussions has become somewhat removed from scientific practice, 
especially where the analysis of measurement is concerned.}

\hfill David Wallace, 2005 \cite[Abstract]{Wal12} 
}

\bigskip

\nopagebreak
\hfill\parbox[t]{10.8cm}{\footnotesize

{\em There is a fresh perspective to be taken on the axioms of quantum
mechanics that could yield a more satisfactory foundation for the
theory.}

\hfill Gilles Brassard, 2005 \cite[p.2]{Bras}
}

\bigskip

If one looks at the current practice of quantum mechanics -- especially 
in quantum information theory where the foundations of quantum mechanics
are probed most directly -- {\em one finds a single conspicuous place 
where the approach of the founding fathers has shown limitations}.
This is the place where one might look for a possible remedy. 

The Dirac--von Neumann quantum measurement theory from the early days 
of quantum mechanics  -- stated in terms of Born's rule for pure states 
collapsing to eigenstates when measured -- is the starting point of 
almost every textbook on quantum mechanics.\footnote{
Among them: Basdevant 2016; Cohen-Tannoudji, Diu and Laloe 1977; 
Dirac 1930, 1967; Gasiorowicz 2003; Greiner 2008;
Griffiths and Schroeter 2018; Landau and Lifshitz 1958, 1977; 
Liboff 2003; McIntyre 2012; Messiah 1961; Peebles 1992; 
Rae and Napolitano 2015; Sakurai 2010; Shankar 2016; Weinberg 2013.
} 
But it is well-known that, in this setting, a quantitatively correct 
description of realistic experiments cannot be obtained. For example, 
joint measurements of position and momentum with limited accuracy, 
ubiquitous in engineering practice, cannot be described in terms of the 
projective measurements encoded in the standard foundations. 
Born's rule in its pre-1970 forms does not even have an idealized 
terminology for them.
Nevertheless, in foundational studies, this projective idealization 
usually counts as the indisputable truth about everything measured on 
the most fundamental level.

In the spirit of Eddington's statement made long ago in the context of 
general relativity, it is shown here that ideas from quantum tomography 
provide an intuitive, self-contained approach to quantum measurement.

The purpose of this paper is to give a proper conceptual foundation of 
quantum physics with the same characteristic features as classical 
physics -- {\em except that the density operator takes the place of the 
phase space coordinates}. Since a convincing alternative to the 
traditional interpretation must be shown to treat correctly quite a 
number of different issues, the paper is quite long, effectively 
containing an extensive survey of a large collections of themes basic 
to quantum physics.

Like any physical theory, quantum mechanics gets its meaning from
the way it can be applied in practice to provide explanations and 
testable predictions for appropriate experiments.

This part describes how quantum physics relates to experiment, and
how the standard formalism of quantum mechanics arises very naturally 
from experimentally well motivated assumptions. It also derives the 
scope and limitations of the traditional textbook assumptions for
quantum mechanics -- pure states, Born's rule for projective 
measurements, and the Schr\"odinger equation. It is shown that these 
are in fact idealizations of their more realistic modern
generalizations -- density operators, Born's rule for quantum measures 
(POVMs), and the Lindblad equation. This gives a perspective on the 
foundations and the measurement problem of quantum mechanics that is 
quite different from the well-trodden path followed by most quantum 
mechanics textbooks.

This paper gives, for the first time, a formally precise definition of 
the concept of quantum measurement that

\pt
addresses the quantum mechanical framework from
an operational point of view, providing a clear and concrete meaning 
for the otherwise abstract and unintuitive (if not counterintuitive
or even magical sounding) concepts of quantum mechanics;

\pt 
is applicable without idealization to complex, realistic 
experiments;

\pt 
allows one to derive the standard quantum mechanical machinery from 
a single, well-motivated postulate;

\pt 
leads to objective (i.e., observer-independent, operational, and 
reproducible) quantum state assignments to all sufficiently stationary 
quantum systems; and

\pt 
gives fairly precise information about the measurement problem of 
quantum mechanics.

The paper shows that the amount of objectivity in quantum physics is no 
less than that in classical physics. Since everything follows from the
accepted technique of quantum tomography, our new approach may have the 
potential to lead in due time to a consensus on the foundations of 
quantum mechanics.

\subsection{A bird's eye view}\label{s.bird}

\nopagebreak
\hfill\parbox[t]{10.8cm}{\footnotesize

{\em (i) Find a set of simple assertions about the world, with clear 
physical meaning, that we know are experimentally true (postulates); 
(ii) analyze these postulates [...] 
(iii) derive the full formalism of quantum mechanics from these
postulates. I expect that if this program could be completed, we would 
at long last begin to agree that we have "understood" quantum 
mechanics.}

\hfill Carlo Rovelli, 1996 \cite[p.1640]{Rov.rQM}
}

\bigskip

\nopagebreak
\hfill\parbox[t]{10.8cm}{\footnotesize

{\em From what deep physical principles might we derive this exquisite 
mathematical structure? Those principles should be crisp; they should 
be compelling. They should stir the soul. When I was in junior high 
school, I sat down with Martin Gardner's book 'Relativity for the 
Millions' and came away with an understanding of the subject that
sustains me today; the concepts were strange, but they were clear enough
that I could get a grasp on them knowing little more mathematics than 
simple arithmetic. One should expect no less for a proper foundation to 
quantum theory.}

\hfill Christopher Fuchs, 2003 \cite[p.989]{Fuc2003}
}

\bigskip

We end the introduction with an extensive overview over the most 
important developments in the present paper.

Section \ref{s.measP} introduces the (Hermitian and positive 
semidefinite) density operator $\rho$ as the formal counterpart of the 
state of an arbitrary quantum source. This notion generalizes the 
polarization properties of light: In the case of the polarization of a 
source of light, the density operator represents a qubit. It is given 
by a $2\times 2$ matrix whose trace is the intensity of the light beam. 
If expressed as a linear combination of Pauli 
matrices, the coefficients define the Stokes vector. Its properties 
(encoded in the mathematical properties of the density operator) were 
first described by George Stokes (best known from the Navier--Stokes 
equations for fluid mechanics) who gave -- in 1852, well before the
birth of Maxwell's electrodynamics and long before quantum theory -- 
a complete description of the polarization phenomenon. For a stationary 
source, the density operator is independent of time.

A \bfi{quantum measurement device} is characterized by a collection of 
finitely many \bfi{detection elements} labelled by labels $k$ that 
respond statistically to the quantum source according to the following 
\bfi{detector response principle} \bfi{(DRP)}: 
{\em A detection element $k$ responds to an incident stationary source 
with density operator $\rho$ with a nonnegative mean rate $p_k$ 
depending linearly on $\rho$. The mean rates sum to the intensity of 
the source. Each $p_k$ is positive for at least one density operator 
$\rho$.}

The DRP, abstracted from the polarization properties of light,
relates theory to measurement. By its formulation, it allows one 
to discuss quantum measurements without the need for quantum mechanical 
models for the measurement process itself. The latter would involve the 
detailed dynamics of the microscopic degrees of freedom of the 
measurement device -- something clearly out of scope of a conceptual 
foundation on which to erect the edifice of quantum physics.

The main consequence of the DRP is the \bfi{detector response theorem}
(Theorem \ref{t.pPOVM}). It asserts that, for every measurement device, 
there are unique operators $P_k$ which determine the rates of response 
to every source with density operator $\rho$ according to the formula
\[
p_k=\<P_k\>:=\tr\rho P_k.
\]
This is a special case of the \bfi{quantum value}
\[
\<A\>:=\tr\rho A
\] 
of an arbitrary operator $A$.

The $P_k$ form a discrete quantum measure; i.e., they are Hermitian, 
positive semidefinite, and sum up to the identity operator $1$. This is 
the natural quantum generalization of a discrete probability measure, a 
collection of nonnegative numbers that sum up to 1.
(In more abstract terms, a discrete quantum measure is a simple 
instance of a so-called POVM, but the latter notion is not needed for 
understanding the main message of the paper.)

A quantum measurement device is characterized formally by means of 
a discrete quantum measure. To go from detection events to measured 
numbers, one needs to provide in addition a scale that assigns to each 
detection element $k$ a real or complex number (or vector) $a_k$. 
We call the combination of a measurement device with a scale a 
\bfi{quantum detector}. We say that the detector \bfi{measures} the 
\bfi{quantity}
\[
A:=\sum_{k\in K} a_kP_k,
\]
Equipped with different scales, the same quantum measurement device 
measures different quantities. 

If the density operator is normalized to intensity one the response 
rates $p_k$ form a discrete probability measure. In this case we refer 
to the response rates as \bfi{response probabilities} and to the quantum
value of $A$ as the \bfi{quantum expectation} of $A$. The statistical 
responses of a quantum detector define the \bfi{statistical expectation}
\[
\E(f(a_k)):=\sum_{k\in K} p_kf(a_k)
\]
of any function $f(a_k)$ of the scale values. It is easy to deduce 
from the main result the following version of \bfi{Born's rule} 
\bfi{(BR)}:
{\em The statistical expectation of the measurement results equals the
quantum expectation of the measured quantity.}
Moreover, the quantum expectations of the quantum measure constitute the
probability measure characterizing the response. 

This version of Born's rule applies without idealization to results of 
arbitrary quantum measurements. As always in statistics, the 
statistical expectation is operationally approximated by finite 
sample means of $f(a)$, where $a$ ranges over a sequence of actually 
measured values. However, the exact statistical expectation is an 
abstraction of this; it works with a nonoperational probabilistic limit 
of infinitely many measured values, so that the replacement of relative 
sample frequencies by probabilities is justified. 

\bigskip

The conventional version of Born's rule -- the traditional starting 
point relating quantum theory to measurement in terms of eigenvalues, 
found in all textbooks on quantum mechanics -- is obtained in Sections 
\ref{ss.proj} and \ref{ss.Born} by specializing our general result to
the case of projective measurements. The spectral notions do not 
appear as postulated input as in traditional expositions, but as 
consequences of the derivation for the special case where the components
of $A$ are commuting self-adjoint operators. Hence they have a 
joint spectral resolution with real eigenvalues $a_k$, and the $P_k$ 
are the projection operators to the eigenspaces of $A$. In this special 
case, we recover the traditional setting with all its ramifications 
together with its domain of validity.
This sheds a new light on the understanding of Born's rule, and 
eliminates the most problematic\footnote{
cf. Subsection \ref{ss.appBorn}. 
All problematic features known to me are collected in my recent book
(\sca{Neumaier} \cite[Section 14.3]{Neu.CQP}), following the preprint
\sca{Neumaier} \cite[Section 3.3]{Neu.Ifound}.
} 
features of its uncritical use.

The DRP leads naturally to all basic concepts and properties of modern 
quantum mechanics. 

\bigskip

Based on the detector response theorem, the paper gives in Section 
\ref{s.qTomo} an operational meaning to quantum states, quantum 
detectors, quantum processes, and quantum instruments, using the 
corresponding versions of \bfi{quantum tomography}. These techniques 
serve as foundations for far reaching derived principles. For quantum 
systems with a low-dimensional density matrix, they are also practically
relevant for the characterization of sources, detectors, and filters. 

A \bfi{quantum process}, also called a linear quantum filter, is 
formally described by a completely positive map. The operator sum 
expansion of completely positive maps (obtained in the somewhat 
technical Theorem  \ref{t.CP}) forms the basis for the derivation in 
Section \ref{s.qDyn} of the dynamical laws of quantum mechanics -- the
\bfi{quantum Liouville equation} for density operators, the 
conservative time-dependent \bfi{Schr\"odinger equation} for pure 
states in a nonmixing medium and the dissipative \bfi{Lindblad equation}
for states in mixing media -- by a continuum limit of a sequence of 
quantum filters. This derivation also reveals the conditions under 
which these laws are valid. 

It is shown that quantum physics has a natural phase space structure 
where normalized density operators play the role of quantum phase space 
variables. The resulting quantum phase space carries a natural Poisson 
structure. Like the dynamical equations of conservative classical 
mechanics, the quantum Liouville equation has the form of Hamiltonian 
dynamics in a Poisson manifold; only the manifold is different. 

Section \ref{s.systemId} discusses consequences for spectroscopy and 
high precision quantum measurements. The oscillations of quantum values 
of states satisfying the Schr\"odinger equation lead to the 
\bfi{Rydberg--Ritz combination principle} underlying spectroscopy, 
which marked the onset of modern quantum mechanics. The determination 
of the gyromagnetic ratio of the electron to 12 decimal digits by 
means of measurements of a single electron in a Penning trap is 
discussed, together with the imagined ensembles traditionally used for 
their analysis.

While high precision quantum measurements of system parameters are 
possible with in principle unrestricted accuracy, the measurement of 
quantities dependent on the quantum state of a system are limited,
not only by the law of large numbers but also by Heisenberg's 
uncertainty relation and measurement imperfections, 
discussed in Section \ref{s.unc}.

Many examples of realistic measurements are shown in Section \ref{s.ex}
to be measurements according to the DRP, but have no interpretation in 
terms of eigenvalues. For example, joint measurements of position and 
momentum with limited accuracy, essential for recording particle tracks 
in modern particle colliders, cannot be described in terms of 
projective measurements; Born's rule in its pre-1970 forms 
(i.e., before POVMs were introduced to quantum mechanics) does not even 
have an idealized terminology for them. Thus the scope of the DRP is 
far broader than that of the traditional approach based on highly 
idealized projective measurements. The new setting also accounts for 
the fact that, in many realistic experiments, the final measurement 
results are computed from raw observations, rather than being directly 
observed.

\bigskip

Observations with highly localized detectors naturally lead in Section 
\ref{s.qfields} to the notion of quantum fields whose quantum values 
encode the local properties of the universe. Among all quantum systems, 
\bfi{macroscopic systems} are characterized as those whose observable 
features can be correctly described by local equilibrium thermodynamics,
as predicted by nonequilibrium statistical mechanics. 

By shifting the attention from the microscopic structure to the 
experimentally accessible macroscopic equipment (sources, detectors, 
filters, and instruments), we get rid of all potentially subjective 
elements of quantum theory. This has implications for the foundations 
of quantum physics. The present findings lead to a new perspective on 
the \bfi{quantum measurement problem}, discussed in Section 
\ref{s.found}.

When a source is stationary, response rates, probabilities, and hence 
quantum values, can be measured in principle with arbitrary accuracy, 
in a reproducible way. Thus they are operationally quantifiable, 
independent of an observer. This makes them \bfi{objective properties} 
in the same sense as in classical mechanics, positions and momenta are 
objective properties. Thus quantum values are seen to be objective, 
reproducible \bfi{elements of reality} in the sense of 
\sca{Einstein, Podolski \& Rosen} \cite{EinPR}. The assignment of 
states to stationary sources is as objective as any assignment of 
classical properties to macroscopic objects. 

As in classical mechanics, probabilities appear only in the context of 
statistical measurements, never in the theoretical investigations 
(unless these are about manipulating measurement results). All 
probabilities are objective \bfi{frequentist probabilities} in the 
sense employed everywhere in experimental physics -- classical and 
quantum. Like all measurements, probability measurements are of limited 
accuracy only, approximately measurable as observed relative 
frequencies. 

This connects to the \bfi{thermal interpretation} of quantum
physics, discussed in detail in my 2019 book 'Coherent Quantum 
Physics' (\sca{Neumaier} \cite{Neu.CQP}). Indeed, the paper derives the 
main features of the thermal interpretation, including everything that
had only been postulated in my book, from the single and simple 
assumption (DRP), the detector response principle.

\bigskip

To summarize, this paper gives a new, elementary, and self-contained
deductive approach to quantum mechanics. A suggestive notion for what 
constitutes a quantum detector, and for the behavior of its responses 
leads to a definition of measurement from which the modern apparatus 
of quantum mechanics can be derived in full generality. 
The statistical interpretation of quantum mechanics is not assumed, but 
the version of it that emerges is discussed in detail. The standard  
dynamical, and spectral rules of introductory quantum mechanics are 
derived with little effort. At the same time, we find the conditions 
under which these standard rules are valid. A thorough and precise 
discussion is given of various quantitative aspects of uncertainty in 
quantum measurements. Normalized density operators play the role of 
quantum phase space variables, in complete analogy to the classical 
phase space variables position and momentum.

The new picture is simpler and more general than the traditional 
foundations, and closer to actual practice. This makes it suitable for 
introductory courses on quantum mechanics. Complex matrices are 
motivated from the start as a simplification of the mathematical 
description. Both conceptually and in terms of motivation, introducing 
the statistical interpretation of quantum mechanics through quantum 
measures is simpler than to introduce it in terms of eigenvalues. 
To derive the most general form of Born's rule from quantum measures,
one just needs simple linear algebra, whereas even to write down Born's 
rule in the traditional eigenvalue form, unfamiliar stuff about wave 
functions, probability amplitudes and spectral representations must be 
swallowed by the beginner -- not to speak of the difficult notion of 
self-adjointness and associated proper boundary conditions, which is 
traditionally simply suppressed in introductory treatments.

Thus there is no longer an incentive for basing quantum physics on 
measurements in terms of eigenvalues -- a special, highly idealized case
-- in place of the real thing.

\newpage
\section{The quantum measurement process}\label{s.measP}

\nopagebreak
\hfill\parbox[t]{10.8cm}{\footnotesize

{\em Unfortunately, the quantum measurement description of von Neumann 
is not valid for measurement schemes that employ real detectors, which 
are normally affected by several imperfections.}

\hfill Zavatta and Bellini, 2012 \cite[p.350]{ZavB}
}

\bigskip

In 1925, Heisenberg's paper initiating modern quantum physics appeared.
It took almost 50 years before a theoretical description of general 
quantum measurements was introduced in 1970 by \sca{Davies \& Lewis} 
\cite{DavL}. 4 years later, a very readable account was given by 
\sca{Ali \& Emch} \cite{AliE.meas} in terms of 
\bfi{positive operator valued measures} (\bfi{POVM}s). 
Since the latter paper appeared, another 50 years passed.

These general measurement schemes are based on the concept of a
\bfi{discrete quantum measure}\footnote{
This notion has no connection with quantum measures defined as measures
on the lattice of orthogonal projections of a Hilbert space as
discussed, e.g., in \sca{Hamhalter} \cite{Ham}. These measures have no
connection to most realistic quantum measurements.
} 
(also called a \bfi{discrete resolution of the identity}), a family of
finitely many Hermitian positive semidefinite operators $P_k$ on a
Hilbert space summing to 1,
\[
\sum_k P_k=1.
\]
This straightforward generalization of the concept of a discrete
probability measure, given by a family of finitely many nonnegative
numbers $p_k$ (probabilities) summing to 1, is a simplified version of
a discrete POVM; for the precise relation of these concepts see
Subsection \ref{ss.tomoMeas}. Quantum measures (in the form of POVMs) 
were soon found useful for concrete applications to the calibration of 
quantum systems (\sca{Helstrom} \cite{Hel}). For a fairly concise, 
POVM-based exposition of the foundations of quantum mechanics see, e.g.,
\sca{Englert} \cite{Eng}. A short, more substantive history can be 
found in \sca{Brandt} \cite{Bra}. Books discussing measurements in
terms of quantum measures include
\sca{Busch} et al. \cite{BusLM,BusGL,BusL2,BusLPY},
\sca{de Muynck} \cite{deMuy2002},
\sca{Holevo} \cite{Hol1982,Hol2001,Hol2012},
\sca{Nielsen \& Chuang} \cite{NieC} and \sca{Peres} \cite{Peres}.

Quantum measures are indispensable in quantum information theory.
Indeed, the well-known textbook by \sca{Nielsen \& Chuang} \cite{NieC}
introduces them (as POVMs) even before defining the traditional
projective measurements. Unlike Born's rule for projective measurements,
general quantum measures are able to account for things like losses, 
imperfect measurements, limited detection accuracy, dark detector 
counts, and the simultaneous measurement of position and momentum. 
Quantum measures are also needed to describe quite ordinary experiments 
without making the traditional textbook idealizations. For example, the
outcome of the original Stern-Gerlach experiment (see Subsection 
\ref{ss.appBorn}) cannot be described in terms of a projective 
measurement, but needs more general quantum measures.

In the remainder of this section, we give motivation and precise formal
definitions for states (positive linear functionals) and quantum
detectors (modeled by a quantum measure and a scale). We give a precise 
specification of what is measured by such a detector, covering the most 
general case. This leads to a precise operational meaning of states and 
detectors in terms of the detector response principle, in agreement 
with experimental practice. No idealization is involved.

\subsection{Quantum sources and states}\label{ss.states}

\nopagebreak
\hfill\parbox[t]{10.8cm}{\footnotesize

{\em
If you visit a real laboratory, you will never find there Hermitian
operators. All you can see are emitters (lasers, ion guns, synchrotrons
and the like) and detectors. The experimenter controls the emission
process and observes detection events. [...]
Quantum mechanics tells us that whatever comes from the emitter is
represented by a state $\rho$ (a positive operator, usually normalized
to 1). [...]
Traditional concepts such as ''measuring Hermitian operators'', that
were borrowed or adapted from classical physics, are not appropriate
in the quantum world. In the latter, as explained above, we have
emitters and detectors.}

\hfill Asher Peres, 2003 \cite[p.1545f]{Per2003}
}

\bigskip

We motivate the formal setting of this paper by considering the
polarization of light. Partially polarized light constitutes the
simplest macroscopic manifestation of quantum effects in Nature.
The Maxwell equations in vacuum have no partially polarized solutions;
hence classical electromagnetic fields only describe fully polarized
light.\footnote{
A semiclassical description of unpolarized light can be given in terms
of stochastic Maxwell equations for fluctuating fields; see
\sca{Mandel \& Wolf} \cite[Chapter 6]{ManW}). The dynamics of the
stochastic Maxwell equations is fully equivalent to the quantum dynamics
of a single photon in the infinite-dimensional Hilbert space of
transversal modes of the classical free electromagnetic field.
} 
The properties of unpolarized and partially polarized quasimonochromatic
beams of light were completely described in 1852 -- even before 
Maxwell's electromagnetic explanation of light -- by \sca{Stokes}
\cite{Sto}. Read with modern understanding, his account shows that such
a beam behaves exactly like a modern \bfi{quantum bit} or \bfi{qubit},
the minimal, 2-state quantum system. We might say that
classical ray optics in the form known in 1852 by Stokes is the
quantum physics of a single qubit passing through a medium,
complete with all bells and whistles. The modern subtleties of
quantum systems do not yet show up in Hilbert spaces of dimension two.
This is discussed in some more detail in \sca{Neumaier}
\cite{Neu.qubit} and Section 8.6 of \sca{Neumaier} \cite{Neu.CQP}, from
where we take the following few paragraphs.

A ray (quasimonochromatic, well-collimated beam) of partially polarized
light of fixed frequency is characterized (in the paraxial 
approximation) by a \bfi{state}, described, according to
\sca{Mandel \& Wolf} \cite[Section 6.2]{ManW}, by a real
\bfi{Stokes vector}
\[
S=(S_0,S_1,S_2,S_3)^T={S_0\choose \Sb}
\]
with
\lbeq{e.Sbound}
S_0\ge |\Sb| = \sqrt{S_1^2+S_2^2+S_3^2}.
\eeq
The Stokes vector is a macroscopic, observable vector quantity.
\at{see Wikipedia for expressions in terms of sums and differences 
of field expectations.} 
Equivalently, the state can be described by a \bfi{coherence matrix}, a
complex positive semidefinite
$2\times 2$ matrix $\rho$. These are related by
\[
\rho = \half(S_0\sigma_0 + \Sb \cdot\Bsigma)
=\half\pmatrix{S_0+S_3 & S_1-iS_2\cr S_1+iS_2 & S_0-S_3},
\]
where $\sigma_0$ is the $2\times 2$ identity matrix, and $\Bsigma$ is
the vector of \bfi{Pauli matrices} $\sigma_1,\sigma_2,\sigma_3$. Clearly
\[
S_k=\<\sigma_k\> \for k=0,\ldots,3,
\]
where
\[
\<X\>:=\tr\rho X
\]
denotes the \bfi{quantum value} of the matrix $X\in\Cz^{2\times 2}$.
Extending $\<X\>$ componentwise to vectors with matrix entries, the
Stokes vector $S$ is the quantum value of the \bfi{Stokes operator}
vector
\[
\Sigma:=\pmatrix{\sigma_0 \cr \Bsigma}.
\]
In particular,
\[
I:=\tr\rho=\<1\>=S_0
\]
is the \bfi{intensity} of the beam; it is nonnegative by
\gzit{e.Sbound}. Zero intensity ($I=0$) corresponds to the absence of a
beam; by \gzit{e.Sbound}, this is the case iff $S=0$, and hence 
$\rho=0$.

The \bfi{degree of polarization} is the quotient 
\[
p:=|\Sb|/S_0\in[0,1].
\]
It is a number between $0$ and $1$ since
\[
0\le \det\rho=(S_0^2-S_3^2)-(S_1^2+S_2^2)=S_0^2-\Sb^2.
\]
The fully polarized case $p=1$, i.e., $S_0=|\Sb|$, is equivalent to
$\det\rho=0$, hence holds iff the rank of $\rho$ is $0$ or $1$. In this
case we say that the state is \bfi{pure}. Thus the pure states
correspond precisely to fully polarized beams. In a pure state, the
coherence matrix can be written in the form $\rho=\psi\psi^*$
with a 2-dimensional \bfi{state vector} $\psi\in\Cz^2$ determined up to 
a phase, and the intensity of the beam is
\[
I=\<1\>=|\psi|^2=\psi^*\psi.
\]
Thus a positive definite Hermitian $\rho$ describes the state of an
arbitrary source, the trace of $\rho$ is the intensity of the source,
and certain Hermitian operators, here the components of $S$, represent
key quantities whose quantum value is observable.

Thus we modeled the simplest quantum phenomenon, the qubit in terms of
notions related to a complex vector space of dimension 2.

In the following, and with the hindsight of nearly a century of
successful quantum mechanics, we generalize from the polarization
experiments to experiments involving arbitrary emissions of quantum 
sources. Examples are alpha, beta, and/or gamma rays emitted by 
radioactive material, the products of collision experiments in a 
particle collider, and more complex situations some of which are 
mentioned later.

On the formal level, we model a \bfi{quantum system} using (in place of
$\Cz^2$) a \bfi{state space} $\Hz$, a complex vector space\footnote{
In conventional terms, the state space is a dense subspace of a Hilbert
space, a common domain of all operators used to describe the quantum 
system. In much of quantum mechanics, especially when phrased in terms 
of Dirac's bra-ket formalism, the Hilbert space itself is never needed.
} 
whose elements are called \bfi{state vectors}. $\Hz$ carries a 
positive definite Hermitian form antilinear in the first argument that 
assigns to $\phi,\psi\in\Hz$ the \bfi{inner product} $\phi^*\psi\in\Cz$.
A linear operator $A\in\Lin\Hz$ has an \bfi{adjoint} $A^*\in\Lin\Hz$ if
\[
(A\phi)^*\psi=\phi^*(A^*\psi) 
\]
for all $\phi,\psi\in\Hz$; if it exists,
the adjoint is unique. $A\in\Lin\Hz$ is \bfi{Hermitian} if $A^*=A$, and
\bfi{positive semidefinite} if $\psi^*A\psi\ge0$ for all $\psi\in\Hz$.
A \bfi{finite rank operator} has the form $A=B^*C$, where
$B,C\in\Lin(\Hz,\Cz^m)$ for some $m$. In this case, $CB^*$ is an
$m\times m$ matrix, and we can define the \bfi{trace} of $A$ to be
$\tr A=\tr CB^*$, where the second trace is the matrix trace, the sum
of the diagonal elements. The trace is independent of the decomposition
$A=B^*C$ of $A$. By taking limits, the trace can be extended to be
defined for all \bfi{trace class} operators.

We characterize a \bfi{quantum source} (in the following simply called
a \bfi{source}) -- one of the emitters in the quote by Peres -- by a
positive semidefinite Hermitian \bfi{density operator} (or 
\bfi{density matrix} if $\Hz=\Cz^n$) $\rho\in\Lin\Hz$.

The \bfi{state} of the source is\footnote{
similar as in the algebraic approach to quantum mechanics
(\sca{Haag} \cite[p.122]{Haa})
} 
the positive linear mapping $\<\cdot\>$ that assigns to each 
$X\in \Lin \Hz$ its \bfi{quantum value}
\lbeq{e.qEx}
\<X\>:=\tr\rho X.
\eeq
(Read ''$\<X\>$'' as ''brackets $X$'' or ''the quantum value of $X$''.)
More generally, given a fixed state and a vector $X\in(\Lin \Hz)^m$
with operator components $X_j\in\Lin \Hz$, we call the vector
$\ol X=\<X\>\in\Cz^m$ with components $\ol X_j=\<X_j\>$
the \bfi{quantum value} of $X$. As customary, we also refer to $\rho$ as
the \bfi{state}, since states and density operators are in a 
1-to-1 correspondence.

Note that the above definitions are sensible even without specifying 
what a source (or later a detector) ''is''. They allows us to recognize 
sources (and later detectors) when we study their behavior, and find 
that this behavior is consistent with the definitions imposed here. 
Indeed, this is what \bfi{quantum tomography} is about. 

To say that a source is characterized by its density operator amounts
to the claim (to be verified by experiment) that every property of the
source relevant for quantum detection is encoded in the density
operator, hence in the state. (This need not include other properties
of the source, such as its age or its manufacturer.)
The collection of all quantum values $\<X\>$ completely determines the
state of the source, and hence its density operator. Indeed, by 
linearity, it suffices to use for $X$ the elements of a particular 
basis. If $\Hz=\Cz^n$, a suitable basis consists of the $n^2$ matrices
\lbeq{e.Eee}
E_{jk}:=e_je_k^*,
\eeq
where $e_j$ is the column vector with zero entries except for a $1$ in 
the $j$th component. Another useful basis consists of the identity and 
$n^2-1$ suitably normalized trace zero Hermitian $n\times n$ matrices 
with only two nonzero entries. For $n=2$, we get the 3 Pauli matrices, 
and the coefficients in this basis correspond to the entries of the 
Stokes vector. For $n=3$, we get the infinitesimal $SU(3)$ symmetries 
describing, for example, the 8 gluons in quantum chromodynamics.

Thus any property of the source relevant for quantum detection can be 
expressed as a function of quantum values. We call the number
\[
I:=\<1\>=\tr\rho
\]
the \bfi{intensity} of the source. Note the 
slight\footnote{\label{f.norm}
In contrast to our convention to normalize the trace of $\rho$ to the
intensity, it is customary in the literature to normalize the intensity 
to be $1$. This can be achieved by dividing $\rho$ and all quantum 
values by the intensity. A disatvantage of this normalization is that 
$\rho$ loses one degree of freedom and the intensity of the source is 
no longer represented by the state.
} 
difference to conventional density operators, where the trace is instead
fixed to be one. The intensity is nonnegative since $\rho$ is positive 
semidefinite. $\rho=0$ defines the \bfi{dark state}; it is the only 
state with zero intensity.

A source is called \bfi{stationary} if its properties of interest are
time-independent, and \bfi{nonstationary} otherwise. A stationary
(nonstationary) source is described by a time-independent
(time-dependent) density operator.

A source and its state are called \bfi{pure} if the density operator
has rank 1, and hence is given by $\rho=\psi\psi^*$ for some
\bfi{state vector} $\psi$; if the source or state is not pure it is
called \bfi{mixed}. In a pure state, the quantum value takes the form
\[
\<X\>=\tr\psi\psi^* X =\psi^*X\psi.
\]

\subsection{Measurement devices and quantum measures}\label{ss.meas}

\nopagebreak
\hfill\parbox[t]{10.8cm}{\footnotesize

{\em
One would naturally like to know what is being measured in a
measurement.}

\hfill Jos Uffink, 1994 \cite[p.205]{Uff1994}
}

\bigskip

\nopagebreak
\hfill\parbox[t]{10.8cm}{\footnotesize

{\em
The theorist's problem is to predict the probability of response of this
or that detector, for a given emission procedure. Detectors are
represented by positive operators $E_\mu$, where $\mu$ is an arbitrary
label whose sole role is to identify the detector. The probability that
detector $\mu$ be excited is $\tr(\rho E_\mu)$. A complete set of
$E_\mu$, including the possibility of no detection, sums up to the unit
matrix and is called a positive operator valued measure (POVM).
\\
The various $E_\mu$ do not in general commute, and therefore a
detection event does not correspond to what is commonly called the
''measurement of an observable''. Still, the activation of a particular
detector is a macroscopic, objective phenomenon. There is no
uncertainty as to which detector actually clicked. [...]}

\hfill Asher Peres, 2003 \cite[p.1545]{Per2003}
}

\bigskip

\nopagebreak
\hfill\parbox[t]{10.8cm}{\footnotesize

{\em
The only form of ''interpretion'' of a physical theory that I find
legitimate and useful is to delineate approximately the ensemble of
 natural phenomena the theory is supposed to describe and to construct
something resembling a ''structure-preserving map'' from a subset of
mathematical symbols used in the theory that are supposed to represent
physical quantities to concrete physical objects and phenomena
(or events) to be described by the theory. Once these items are
clarified the theory is supposed to provide its own ''interpretation''.
}

\hfill J\"urg Fr\"ohlich, 2021 \cite[p.238]{Fro2021}
}

\bigskip

To relate states and quantum values to experimental practice,
we employ carefully defined rules, thereby providing a clear formal
notion of measurement.

In optics, combining two independent sources leads to the addition of
the intensities, and hence the corresponding densities. Similarly,
changing the intensity amounts to a scalar multiplication of the
corresponding densities. When measuring light with a photodetector, a
stationary light source produces at sufficiently high intensity a
steady photocurrent whose strength is proportional to the intensity.
As long as the current is steady we speak of \bfi{classical} 
measurements.

At sufficiently low intensity it is observed that the photocurrent 
degenerates into a random sequence of current pulses whose mean rate 
is proportional to the intensity. In this case we speak of 
\bfi{quantum} measurements. Now the detector response can be described 
as a sequence of ''quantized'' elementary \bfi{detection events}, and 
the intensity must be obtained by a statistical average goverened by
propbabilistic laws. Quantum electrodynamics -- see, e.g.,
\sca{Weinberg} \cite[Section 3.4]{WeiI} or
\sca{Mandel \& Wolf} \cite[Chapter 14]{ManW} -- allows one to predict
these mean rates.

The linearity of typical detector responses, described here for optical 
sources, is a feature observed in broad generality.\footnote{
but not in general. See the remark at the end of Subsection 
\ref{ss.mixed}. 
} 
For example, decay rates of radioactive sources have the same
additivity properties. The following definitions turn the insight from
our optical example system into a formal requirement for quantum
measurement processes.

On the formal level, a \bfi{quantum measurement device} (short a 
\bfi{measurement device}) is characterized by a collection of
\bfi{detection elements} labelled by labels $k$ from a finite\footnote{
Experimentally realizable detectors always produce only a finite number
of possible results; see the examples in Section \ref{s.ex}.
Idealizations violating this finiteness condition are not discussed in 
this paper.
} 
set $K$ satisfying the following postulate.

\bfi{(DRP)}: \bfi{Detector response principle}.
{\em A detection element $k$ responds to an incident stationary source 
with density operator $\rho$ with a nonnegative mean rate $p_k$ 
depending linearly on $\rho$. The mean rates sum to the intensity of 
the source. Each $p_k$ is positive for at least one density operator 
$\rho$.}

The detector response principle (DRP) and the abundant existence of 
quantum measurement devices are extremely well established basic 
empirical facts. The theoretical explanation of the DRP consitutes the 
so-called \bfi{quantum measurement problem},\index{measurement problem}
discussed later in Subsection \ref{ss.measP}.

Note that in order that a measured mean rate has a sensible operational
meaning, the source must be reasonably stationary at least during the
time measurements are taken. For nonstationary sources (discussed later 
in Subsection \ref{s.nonstat}), one only gets
time-dependent empirical rates of limited accuracy. Assuming
stationarity allows us to ignore all dynamical issues, including the
dynamical differences between isolated systems and open systems.
In particular, we may proceed independently of quantum mechanical models
for the measurement process itself, which would involve the dynamics of
microscopic degrees of freedom of the measurement device.

The key result for the theory of quantum measurements is the 
following \bfi{detector response theorem}:

\begin{thm}\label{t.pPOVM} 
For every measurement device, there is a unique discrete quantum
measure $P_k$ ($k\in K$) whose quantum values $\<P_k\>$ determine,
for every source with density operator $\rho$, the mean rates 
\lbeq{e.BornPOVM}
p_k=\<P_k\>=\tr\rho P_k  \for k\in K.
\eeq
\end{thm}

\bepf
For simplicity, we first assume a state space with finite dimension $d$;
the finiteness restriction is lifted later. By linearity, the mean 
rates satisfy
\lbeq{e.pk}
p_k=\sum_{i,j} P_{kji}\rho_{ij}
\eeq
for suitable complex numbers $P_{kji}$. If we introduce the matrices
$P_k$ with $(j,i)$ entries $P_{kji}$, \gzit{e.pk} can be written in the
concise form \gzit{e.BornPOVM}.

To find the properties of the matrices $P_k$, we first note that the
$p_k$ are nonnegative. Since $p_k>0$ for at least one density operator
$\rho$, \gzit{e.BornPOVM} implies that all $P_k$ are nonzero. Since
$p_k$ is real for all density operators $\rho$, we have
\[
\tr\rho P_k^*=\tr(P_k\rho)^*=\ol{\tr P_k\rho}=\ol p_k=p_k=\tr\rho P_k.
\]
This holds for all density operators $\rho$, hence $P_k^* = P_k$. Thus
the $P_k$ are Hermitian.
Picking arbitrary pure states with $\rho=\psi\psi^*$ shows that $P_k$
is positive semidefinite. Summing the mean rates gives
\[
\tr\rho=I=\sum p_k=\sum\tr\rho P_k=\tr\rho\sum P_k.
\]
Since this must hold for all positive semidefinite $\rho$, we conclude
that $\sum P_k=1$ is the identity.

Thus, in the case where the state space of the system measured is
finite-dimensional, the $P_k$ form a discrete quantum measure. It can
be shown (\sca{de Muynck} \cite[p.41]{deMuy2002}) that the same holds
in the
infinite-dimensional case, but the argument is considerably more
abstract. Instead of the matrix argument one uses the fact that each
bounded linear functional on the state space of Hilbert--Schmidt
operators can be represented as an inner product, resulting in $P_k$s
with $p_k=\tr\rho P_k$. By the above arguments, they are then found to
be positive semidefinite bounded Hermitian operators.
\epf

The detector response theorem characterizes the response of a quantum
measurement device in terms of a quantum measure. But what is being
measured? A quantum measurement device produces in the low intensity
case a stochastic sequence of \bfi{detection events}, but makes no
direct claims about values being measured. It just says which one of the
detection elements making up the measurement device responded at which 
time.
The quantum effects are in the response of the detection elements, not
in possible values assigned to the $k$th detection event. The latter
are provided by adding to the device a scale that annotates the
detection elements with appropriate values. These values can be
arbitrary numbers or vectors $a_k$ (or even more complex mathematical 
entities from a vector space) --
whatever has been written on the scale a pointer points to, or whatever
has been programmed to be written by an automatic digital recording
device.

To interpret the responses numerically, we therefore use a \bfi{scale},
formally an assignment of distinct complex numbers or vectors $a_k$ to
the possible detection elements $k$. In concrete settings, the scale is
part of the detector. Thus a
\bfi{quantum detector}\footnote{
A quantum detector may be considered as a technically precise version
of the informal notion of an \bfi{observer} that figures prominently in
the foundations of quantum mechanics. It removes from the latter term
all anthropomorphic connotations.
} 
(in the following simply called a \bfi{detector}) consists of a
measuring device with detection elements $k\in K$ characterized by a
quantum measure and a scale $a_k$ ($k\in K$).

\newpage
\section{The statistical interpretation of quantum mechanics}
\label{s.stat}

\nopagebreak
\hfill\parbox[t]{10.8cm}{\footnotesize

{\em [...] atomistics as such was of no great use without another 
fundamental idea, namely that the observable properties of matter are 
not intrinsic qualities of its smallest parts, but averages over 
distributions governed by the laws of chance.}

\hfill Max Born, 1949 \cite[p.46]{Bor.Nat}
}

\bigskip

\at{Intro missing}

We also discuss projective measurements, the class of measurements 
covered by the traditional foundations, and how the traditional 
eigenvalue setting of the founding fathers fits into the new 
approach taken.

\subsection{Statistical expectations}\label{ss.qExp}

\nopagebreak
\hfill\parbox[t]{10.8cm}{\footnotesize

{\em In a statistical description of nature only expectation values
or correlations are observable.
}

\hfill Christof Wetterich, 1997 \cite{Wet}
}
\bigskip

In this subsection, we normalize the intensity, hence 
the trace of the density matrix, to one; cf. 
Footnote ${}^{\mbox{\footnotesize\ref{f.norm}}}$. This has the effect 
that we now have $\sum p_k=1$, so that the normalized mean rate $p_k$ 
can be interpreted as the \bfi{response probability} of detector 
element $k$ given some response, or as the \bfi{detection probability} 
for the $k$th detection event. This gives an intuitive meaning for the 
$p_k$ in case of low intensity measurements. Formula \gzit{e.BornPOVM}, 
derived here\footnote{\label{f.3}
The usual practice is to assume Born's rule for projective
measurements as a basic premise, with a purely historical justification.
Later (if at all), the more general quantum measure (POVM) setting is 
postulated and justified in terms of Born's rule in an artificially 
extended state space defined using an appropriate ancilla. This 
justification is based on Naimark's theorem (\sca{Naimark} \cite{Nai}) 
-- also called Neumark's theorem, using a different transliteration of 
the Russian originator.
} 
from very simple first principles, becomes von Neumann's well-known 
extension of Born's probability formula. It gives the theoretical 
quantum values $\<P_k\>$ a statistical interpretation as response 
probabilities $p_k$ of a quantum measurement device.

The results of a detector in a sequence of repeated events define a
random variable or random vector $a_k$ (with a dummy index $k$ labeling
the particular detector element responding) that
allows us to define the \bfi{statistical expectation}
\lbeq{e.statEx}
\E(f(a_k)):=\sum_{k\in K} p_kf(a_k)
\eeq
of any function $f(a_k)$. As always in statistics, this statistical 
expectation is operationally approximated by finite sample means of 
$f(a)$, where $a$ ranges over a sequence of actually measured values. 
However, the exact statistical expectation is an abstraction of this;
it works with a nonoperational probabilistic limit of infinitely
many measured values, so that the replacement of relative sample
frequencies by probabilities is justified. Clearly, $\E$ is linear in
its argument. 

For any family $x_k$ ($k\in K$), we introduce the operators
\lbeq{e.Px}
P[x_k]:=\sum_{k\in K} x_kP_k.
\eeq
Since $\<P_k\>=p_k$ for all $k\in K$ by \gzit{e.BornPOVM}, this allows 
us to write \gzit{e.statEx} as
\lbeq{e.statExP}
\E(f(a_k))= \<P[f(a_k)]\>.
\eeq
We say that a detector defined by the quantum measure $P_k$ ($k\in K$)
and the scale $a_k$ ($k\in K$) \bfi{measures} the
\bfi{quantity}\footnote{
In traditional terminology (e.g., \sca{Schroeck}
\cite{Schr1985}, \sca{de Muynck} \cite[p.360]{deMuy2002}), one would
say that the detector measures the observable represented by the scalar
or vector operator $A$. Since there is a tradition for using the word
'observable' synonymous with POVMs in the form mentioned in Subsection
\ref{ss.tomoMeas} below, and since observables such as spectral widths 
or spectral intensities cannot be represented in this way, we use a 
more neutral terminology.
} 
\lbeq{e.obs}
A:=P[a_k]=\sum_{k\in K} a_kP_k.
\eeq
When the scale consists of general complex numbers, the operator
corresponding to the measurement is bounded. Indeed, since the $P_k$ are
positive semidefinite and add to $1$, we have $\|P_k\|\le 1$ in the
spectral norm, hence
\lbeq{e.bounded}
\|A\|\le \sum_{k\in K} |a_k|=\|a\|_1.
\eeq
When the scale consists of vectors, the operator corresponding
to the measurement is a vector with operator components.
When the scale consists of real numbers only, the operator corresponding
to the measurement is Hermitian.

From \gzit{e.BornPOVM} and \gzit{e.obs}, we find the formula
\lbeq{e.BornEx}
\E(a_k)=\tr\rho A=\langle A\rangle
\eeq
for the statistical expectation of the measurement results $a_k$
obtained from a source with density operator $\rho$. Comparing with
\gzit{e.qEx}, we see that the statistical expectation of measurement
results coincides with the theoretical quantum value of $A$ evaluated
in the state $\rho$ of the source. This is
\bfi{Born's rule in expectation form}, discovered by
\sca{Landau} \cite[(4a),(5)]{Landau1927}, but formulated in the context
of projective measurements first by \sca{von Neumann}
\cite[p.255]{vNeu1927}. For general quantum measures, the formula is in
\sca{Ali \& Emch} \cite[(2.14)]{AliE.meas}.

Born's rule gives the purely theoretical notion of a quantum value an
operational statistical interpretation in terms of expectations of
measurement results of a quantum detector. If we call the quantum value
$\<A\>$ the \bfi{quantum expectation} of $A$, we may express Born's rule
\gzit{e.BornEx} in the following suggestive form:

\bfi{(BR)}: \bfi{Born's rule}. 
{\em The statistical expectation of the measurement results equals the
quantum expectation of the measured quantity.}

A thorough investigation of this correspondence is given in my book
{\em Coherent Quantum Physics} \cite{Neu.CQP}. It includes a detailed
history of the various forms of Born's rule (\sca{Neumaier}
\cite[Section 3]{Neu.Ifound}, \cite[Chapter 14]{Neu.CQP}).

\subsection{Projective measurements}\label{ss.proj}

\nopagebreak
\hfill\parbox[t]{10.8cm}{\footnotesize

{\em The standard von Neumann description of quantum measurements
applies when all the detector outcomes are well-defined and correspond
to precise classical measurement values that unambiguously reflect the
state of the system. [...]
When measuring a quantum light state, this ideal situation corresponds,
for example, to the case of a perfect photon-number detector
with unit efficiency and no dark counts. }

\hfill Zavatta and Bellini, 2012 \cite[p.350]{ZavB}
}

\bigskip

\at{New intro needed} 
Complementing the preceding discussion of concrete measurement
arrangements, we now discuss a conceptually important class of discrete
quantum measures and the associated measurements.
We call a discrete quantum measure \bfi{projective} if the $P_k$
satisfy the \bfi{orthogonality relations}
\lbeq{e.orthP}
P_jP_k=\delta_{jk}P_k \for j,k\in K.
\eeq
We say that a detector measuring $A$ performs a
\bfi{projective measurement} of $A$ if its quantum measure is
projective.
Such projective measurements are unstable under imperfection in the
detector. Therefore they are realistic only under special circumstances.
Examples are two detection elements behind
polarization filters perfectly polarizing in two orthogonal directions,
or the arrangement in an ideal Stern--Gerlach experiment for spin
measurement. Most measurements, and in particular all measurements of
quantities with a continuous spectrum, are not projective.

The orthogonality relations imply that $AP_k=a_kP_k$. Since the $P_k$
sum up to $1$, any $\psi\in\Hz$ can be decomposed into a sum
$\psi=\sum P_k\psi =\sum\psi_k$ of vectors $\psi_k:=P_k\psi$
satisfying the equation $A\psi_k=AP_k\psi=a_kP_k\psi=a_k\psi_k$.
Therefore $\psi_k$ (if nonzero) is an eigenvector of $A$ (or of each
component of $A$ in case the $a_k$ are not just numbers) corresponding
to the eigenvalue $a_k$ of $A$. Since $P_k^2=P_k=P_k^*$, the $P_k$ are
orthogonal projectors to the eigenspaces of the $a_k$.

When the $a_k$ are numbers,
this implies that $A$ is an operator with a finite spectrum. Moreover,
$A$ and $A^*$ commute, i.e., $A$ is a normal operator, and in case the
$a_k$ are real numbers, a Hermitian, self-adjoint operator. (This is the
setting traditionally assumed from the outset.) When the
$a_k$ are not numbers, our analysis implies that the components of $A$
are mutually commuting normal operators with a finite joint spectrum,
and if all $a_k$ have real components only, the components of $A$ are
Hermitian, self-adjoint operators.
Thus the projective setting is quite limited with respect to the
kind of quantities that it can represent.

For projective measurements, \gzit{e.obs} implies
\[
f(A^*,A)=\sum_k f(\ol a_k,a_k) P_k
\]
for all functions $f$ for which the right hand side is defined.
Therefore the modified scale $f(\ol a,a)$ measures $f(A^*,A)$, as we
are accustomed from classical measurements, and defines a projective
measurement of it. But when the components of $A$ are not normal or do
not commute, this relation does not hold.

\subsection{The spectral version of Born's rule}\label{ss.Born}

\nopagebreak
\hfill\parbox[t]{10.8cm}{\footnotesize

{\em This work naturally leads to the question of whether POVMs ought to
be considered, in some sense, more fundamental than standard
projection-valued measures.}

\hfill Caves, Fuchs, Manne, and Renes, 2004 \cite[p.208]{CavFMR}
}

\bigskip

We are now ready to connect our developments to the traditional
spectral view of quantum measurements that goes back to Born, Dirac, 
and von Neumann.

From the discussion in Subsection \ref{ss.proj}, we may conclude that in
a projective measurement of a Hermitian, self-adjoint operator $A$, 
the possible values are precisely the finitely many eigenvalues of 
$A$ (or joint eigenvalues of the components), measured with a 
probability of $p_k=\tr \rho P_k$. This \bfi{spectral version} of 
\bfi{Born's rule} is valid only\footnote{
In a general measurement as discussed in Subsection \ref{ss.meas}, the
measurement results are usually unrelated to the eigenvalues of $A$.
} 
for projective measurements of quantities represented by mutually
commuting normal operators with a finite joint spectrum.

In the special case where the spectrum of $A$ is \bfi{nondegenerate},
i.e., all eigenspaces have dimension 1, the orthogonal projectors have
the special form $P_k=\phi_k\phi_k^*$, where $\phi_k$ are normalized
eigenstates corresponding to the eigenvalue $a_k$. In this case, the
probabilities take the form
\[
p_k=\phi_k^*\rho\phi_k
\]
appearing in every textbook on quantum mechanics.
If, in addition, the source is pure, described by $\rho=\psi\psi^*$ with
the normalized state vector $\psi\in\Hz$, this can be written in the
more familiar squared amplitude form
\lbeq{e.BornSquare}
p_k=|\phi_k^*\psi|^2.
\eeq
In practice, the orthogonality relations \gzit{e.orthP} can be
implemented only approximately (due to problems with efficiency, losses,
inaccurate preparation of directions, etc.). Thus the present
derivation shows that measurements satisfying the spectral version of 
Born's rule (i.e., projective measurements) are always idealizations.

Whenever one simultaneously measures quantities corresponding to 
noncommuting operators, Born's rule in textbook form does not apply, 
and one needs a quantum measure that is not projective.
The operators corresponding to most measurements discussed in Section
\ref{s.ex} do not commute; therefore such joint measurements cannot
even be formulated in the textbook setting of projective measurements.

In general, the quantum measure description of a real device cannot
simply be postulated to consist of orthogonal projectors. The correct
quantum measure must be
found out by quantum tomography, guided by the theoretical model of the
measuring equipment, then ultimately calibrating it using the formula
\gzit{e.BornPOVM} for probabilities. This formula is a proper extension
of Born's rule for probabilities of projective measurements.
It cannot be reduced to the latter unless one adds to the description
ancillas not represented in the physical state space of the system plus 
its environment, but in an unphysical artificial state space formally 
constructed on the basis of Naimark's theorem
(cf. Footnote ${}^{\mbox{\footnotesize\ref{f.3}}}$).

Thus we recover the traditional spectral setting as a consequence of 
the general approach, restricted to the special case where the 
components of $A$ are commuting self-adjoint (or at least normal) 
operators, hence have a joint spectral resolution with real (or complex)
eigenvalues $a_k$, and the $P_k$ are the projection operators to the 
eigenspaces of $A$. 

Rather than postulating Born's rule for projective measurements, as done
in standard textbooks, we derived it together with all its ramifications
and its domain of validity, from simple, easily motivated definitions.
In particular, the spectral notions are not postulated input as in
traditional expositions. Instead they appear as consequences of a
derivation which shows under which (rather special) circumstances Born's
rule is valid.

By \gzit{e.bounded}, quantities measurable by quantum detectors are
necessarily described by bounded linear operators. This is in agreement
with experimental practice. For example, measuring arbitrarily large
values of the position or momentum of a particle would need infinitely
many detection elements distributed over an unbounded region of space.
However, the definition \gzit{e.qEx} of the quantum value extends
without problems to many unbounded linear operators $A$. Indeed,
unbounded operators and their quantum expectations appear routinely in
the theoretical part of quantum mechanics. In quantum field theory,
quantum values of non-Hermitian operators are needed in the definition
of so-called $N$-point functions as quantum expectations of products of
field operators.

Thus the operators corresponding to observable quantities need not be
Hermitian. Indeed, Dirac's 1930 quantum mechanics textbook 
(\sca{Dirac} \cite[p.28]{Dir1}), which introduced the name  
''observable'' for operators, used this terminology for arbitrary 
linear operators. 

It is therefore interesting to note that all later editions of Dirac's 
textbook and {\em all} later textbooks on quantum mechanics require 
the restriction to Hermitian operators possessing a real spectral 
resolution, i.e., in modern terminology, to self-adjoint Hermitian 
operators. The probable reason is that  when $X$ is not normal (and in 
particular when it is defective, hence has not even a spectral 
resolution), it is impossible to give the quantum value $\<X\>$ of 
$X$ a statistical interpretation in the traditional spectral sense.

\subsection{The approximate nature of the spectral version}
\label{ss.appBorn}

\nopagebreak
\hfill\parbox[t]{10.8cm}{\footnotesize

{\em A student has read such and such a number on his thermometer.
He has taken no precautions. It does not matter; he has read it, and if
it is only the fact which counts, this is a reality [...]
Experiment only gives us a certain number of isolated points. They must
be connected by a continuous line, and this is a true generalisation.
But more is done. The curve thus traced will pass between and near the
points observed; it will not pass through the points themselves.
Thus we are not restricted to generalising our experiment, we correct
it; and the physicist who would abstain from these corrections, and
really content himself with experiment pure and simple, would be
compelled to enunciate very extraordinary laws indeed.}

\hfill Henri Poincar\'e, 1902 \cite[p.142f]{PoiScH}
}

\bigskip

\nopagebreak
\hfill\parbox[t]{10.8cm}{\footnotesize

{\em The following 'laboratory report' of the historic Stern-Gerlach
experiment stands quite in contrast to the usual textbook 'caricatures'.
A beam of silver atoms, produced in a furnace, is directed through an
inhomogeneous magnetic field, eventually impinging on a glass plate.
[...]
Only visual measurements through a microscope were made. No statistics
on the distributions were made, nor did one obtain 'two spots' as is
stated in some texts. The beam was clearly split into distinguishable
but not disjoint beams. [...]
Strictly speaking, only an unsharp spin observable, hence a POV measure,
is obtained.}

\hfill Busch, Grabowski and Lahti, 1995 \cite[Example 1, p.7]{BusGL}
}

\bigskip

In the standard formulation of \bfi{Born's statistical interpretation}
of quantum mechanics, based on projective measurements, the measurement
results of an observable represented by a bounded operator $A$ are 
\bfi{quantized}: the measured 
result will be one of the eigenvalues $\lambda_k$ of $A$.
Multiple repetition of the measurement results in a random sequence of
values $\lambda_k$, with probabilities computed from \gzit{e.BornSquare}
if the system is in a pure state. In the limit of arbitrarily many
repetitions, the mean value of this sequence approaches $\ol A$ and the
standard deviation approaches $\sigma_A$.

According to Born's statistical interpretation in the standard
formulation (i.e., for projective measurements), each actual measurement
result $\lambda$ is claimed to be one of the eigenvalues, which is 
exactly (according to the literal reading\footnote{\label{f.proj}
The formulation appearing in \sca{Wikipedia} \cite{Wik.Born} is
''the measured result will be one of the eigenvalues''.
\sca{Griffiths \& Schroeter} \cite[p.133]{GriS} declare, ''If you
measure an observable [...] you are certain to get one of the
eigenvalues''. \sca{Peres} \cite[p.95]{Peres} defines, ''each one of
these outcomes corresponds to one of the eigenvalues of $A$; that
eigenvalue is then said to be the result of a measurement of $A$''.
The only exceptions seem to be textbooks such as
\sca{Nielsen \& Chuang} \cite[p.84f]{NieC} that start the formal
exposition with the POVM approach rather than Born's interpretation.
But in their informal introduction of qubits, even they give priority 
to projective measurements!
} 
of most formulations) or approximately (in a more liberal reading) 
measured, with probabilities computed from $A$ and the density
operator $\rho$ by the probability form of Born's rule. 

Let us consider the measurement of an unknown quantity 
$A\in\Cz^{2\times 2}$ of a qubit. In a quantum tomography context, 
$A$ has unknown eigenvalues $\lambda_1,\lambda_2$. The 
prediction made by Born's rule is that the observed distribution is  
bimodal; in the literal form of Born's rule it even has a two-point 
support at the eigenvalues $\lambda_1$ and $\lambda_2$. But whenever 
these eigenvalues are irrational or cannot be exactly measured 
-- the prevailing situation in spectroscopy --, deviations from the 
eigenvalues are unavoidable. 

Similarly, the eigenvalues of the spin operator allegedly measured in
a textbook Stern--Gerlach experiment, are the two discrete values 
$\pm\hbar/2$. Taking the spectral form of his rule literally, Born 
could have deduced in 1927 from the Stern--Gerlach experiment the exact 
value of Planck's constant! But the original Stern-Gerlach experiment 
produced on the screen two overlapping lips of silver, and not -- as 
its textbook caricature -- two well-separated tiny spots from which 
one could get a high precision value for $\hbar/2$.

Therefore Born's rule cannot be exactly valid; it only describes 
measurements up to some approximation error. This means that Born's 
rule is not, as generally claimed, about actual measurements. Instead 
it is a postulate about idealized true values representing the results 
of hypothetical measurements whose observations are theoretical numbers,
not actual, inaccurate results. 

Thus {\em the statistical interpretation based on the spectral form of 
Born's rule paints an inadequate, idealized picture whenever 
eigenvalues are only approximately known and must therefore be 
inferred experimentally}.

\subsection{The statistical interpretation} 
\label{s.prob}

\nopagebreak
\hfill\parbox[t]{10.8cm}{\footnotesize

{\em We discuss a mean-field quantum-mechanical model which describes 
the dynamics of a homogeneously broadened system of two-level atoms 
contained in a pencil-shaped resonant cavity and driven by a coherent
resonant field. [...] Bistability is shown to be a consequence of 
atomic cooperation.}

\hfill Bonifacio and Lugiato, 1978 \cite[p.20]{BonL} 
}

\bigskip

\nopagebreak
\hfill\parbox[t]{10.8cm}{\footnotesize

{\em The Born measure is a mathematical construction; what is its 
relationship to experiment? This relationship must be the source of the 
(alleged) randomness of quantum mechanics, for the Schr\"odinger 
equation is deterministic.}

\hfill Klaas Landsman, 2019 \cite[p.20]{Lan2019} 
}

\bigskip

On the other hand, the statistical interpretation based on Born's rule 
in the expectation form (BR) derived from the DRP is impeccable. It
gives a correct account of the actual experimental situation since it 
gives projective measurements (and hence eigenvalues) no special status.
Measurements are defined as quantum values of macroscopic pointer 
variables whose statistical mean agrees with the quantum value of the 
system quantity measured, as discussed in Section \ref{s.measP}. 
They are regarded as approximate readings of these detector properties, 
fluctuating from measurement to measurement.

In a realistic measurement, the possible values obtained when measuring
a particular quantity $A$ depend -- according to our discussion in 
Section \ref{s.measP}  -- on the decomposition $A=P[a_k]$ used to 
construct the scale. 
That this decomposition is ambiguous follows from Subsection
\ref{ss.tomoStates}, where it is shown that the scale is not determined 
by the quantity $A$ measured. Since the scale determines the measurement
results, this means that one can with equal right ascribe different
results to the measurement of the same quantity $A$. Thus, in general,
different detectors measuring the same quantity $A$ have different sets
of possible measurement results. In other words, the same quantity $A$ 
can be measured by detectors with different mathematical 
characteristics, and in particular different measurement results that 
generally have nothing to do with the eigenvalues of $A$.

Thus the new approach is in full agreement with the standard 
recipes for drawing inferences from inaccurate measurement results.
The situation is precisely the same as in classical metrology, where 
observable quantities always have a true value determined by the 
theoretical description, and all randomness in measurements is assumed
to be due to measurement noise.

In contrast to Born's statistical interpretation, which needs 
probabilities in the very foundations, due to the multi-valuedness of 
the true values discussed in Subsection \ref{ss.appBorn}, the 
single-valuedness of the true values in the new setting established here
implies that probabilities are not intrinsic to quantum physics. Just 
like in classical mechanics, they are emergent imperfections due to the 
experimental conditions in which the measurements are made. 

For example, the experimental arrangement in a Stern--Gerlach experiment
together with a simple, essentially classical argument leads to two 
more or less focussed entangled beams, which fully accounts for 
the two fuzzy spots of the response at the screen. Consistent with the 
experimental record, the response of a sufficiently good detector to a 
Stern--Gerlach setup produces both in theory and in practice, 
measurement results concentrated more or less close to two spots (or 
parallel lines) of the detector. This results in a bimodal distribution 
with two more or less sharp peaks, with details depending on the 
detection method used and its resolution -- just as what one gets when 
measuring a classical diffusion process in a double-well potential 
(see, e.g., \sca{Hongler \& Zheng} \cite{HonZ}).
Thus the experimental setup together with a semiclassical theoretical 
analysis is a sufficient explanation of the experimental results. There 
is no need to attribute them to a hypothetical true, discrete value 
$\pm\hbar/2$ of the spin. 

On the other hand, the discrete responses on the screen of a double 
slit experiment or in a Geiger counter are taken to be due to the 
discreteness of the recording device. The latter introduces a systematic
error of the same nature as the errors discussed in Subsection 
\ref{ss.appBorn}. 
Analogous to the bistability that give rise to binary responses in 
classical coin tossing, binary responses of the macroscopic detector 
elements may be explained as in \sca{Neumaier} 
\cite[Chapter 11]{Neu.CQP} by the bistability of their coarse-grained 
microscopic dynamics. While this has apparently not been done in detail 
in the detector setting, environment-induced randomness and 
environment-induced dissipation lead in related contexts (see, e.g., 
\sca{Bonifacio \& Lugiato} \cite{BonL}, 
\sca{Gevorgyan} et al. \cite{GevSCK}) to a reduced semiclassical 
description yielding a bimodal distribution, in full analogy with a 
classical, environment-induced diffusion process in a double-well 
potential. For a discussion of the dynamical aspects of the quantum 
chaos responsible for this see, e.g., 
\sca{Ingraham \& Acosta} \cite{IngA},
\sca{Zhang \& Feng} \cite{ZhaF},
\sca{Belot \& Earman} \cite{BelE},
\sca{Gomez} et al. \cite{GomLL}, 
and Chapter 11 of  \sca{Neumaier} \cite{Neu.CQP}.

\newpage
\section{Quantum tomography}\label{s.qTomo}

The operational meaning of both states and quantum measures is given by
Theorem \ref{t.pPOVM}, which says how to get predictions for the
response probabilities when a state given by an arbitrary density matrix
is measured by a measurement device with an arbitrary quantum measure.
These predictions can be compared with experiment, and the agreement
has been overwhelmingly good.

This section discusses the various forms of quantum tomography for 
determining quantum states for sources, quantum measures for detectors, 
and completely positive maps for quantum processes and quantum 
instruments. The type of the quantum system to be measured and hence 
its state space is here considered to be fixed. 

Because of computational limitations that become apparent below, 
quantum tomography is done in practice only for quantum systems, where a 
low-dimensional state space reproduces the experimentally relevant 
features sufficiently well. The behavior of more complicated systems
is determined from theoretical models and prior knowledge assumed for 
the behavior of the building blocks of the various devices used in
real experiments.

It is interesting to note that the duality between quantum state
tomography and quantum detector tomography has a parallel in the
analogy between Theorem \ref{t.pPOVM} and a result by \sca{Busch}
\cite{Bus} (see also \sca{Caves} et al. \cite{CavFMR}) dual to this
theorem. Busch assumes properties of states in terms of quantum measures
and proves the existence of a corresponding density operator $\rho$
satisfying \gzit{e.BornPOVM}.

\subsection{Quantum measurement tomography and POVMs}\label{ss.tomoMeas}

\nopagebreak
\hfill\parbox[t]{10.8cm}{\footnotesize

{\em How should one mathematically describe an observable in the
quantum setting?}

\hfill Franklin Schroeck, 1989 \cite[p.247]{Schr1989}
}

\bigskip

\nopagebreak
\hfill\parbox[t]{10.8cm}{\footnotesize

{\em In the most standard picture of quantum mechanics the statistics 
of every measurement are governed by the projection of the system state 
on the orthogonal eigenstates of commuting self-adjoint operators 
representing the measured observables. This implicitly asumes that the 
measurement is performed on a closed system.\\
A more complete and realistic picture must encompass the possibility of 
controllable as well as unpredictable couplings of the observed system 
with external agents. This is to say that the measurement is performed, 
in general, on an open system. Among other consequences, this extends 
the idea of observables beyond self-adjoint operators, introducing
 generalized measurements described by positive operator measures.\\
{}[...] To some extent, the real measurement differs from the intended 
one in an unpredictable way.}

\hfill Luis and S\'anchez-Soto, 1998 \cite[p.3573]{LuiS}
}

\bigskip

The proof of Theorem \ref{t.pPOVM} shows how to get a quantum measure
for an arbitrary experimentally realized measurement device. If the
experimentally accessible part of the state space has dimension $d$,
one has to find stationary sources that prepare at least $d^2$ states
with linearly independent density matrices, giving the $\rho_{ij}$.
Then one has to collect for each of theses states enough statistics to
determine the rates of response, giving for each state the corresponding
$p_k$. Finally, one has to solve, for each $k$, the linear system
\gzit{e.pk} for the $P_{kij}$ to get the quantum measure.
This procedure for determining quantum measures, called
\bfi{quantum measurement tomography} or 
\bfi{quantum detector tomography}, constitutes the quantum
tomography principle for the characterization of measurement devices,
first formulated by \sca{Luis \& S\'anchez-Soto} \cite{LuiS}. 
Conversely, one can in principle design detectors whose behavior is 
arbitrarily close to that of a given quantum measure; see \sca{Lundeen} 
et al. \cite{LunFCP}.

No idealization is involved in the quantum tomography principle, except
for the limited accuracy necessarily inherent in statistical estimation,
due to the limitations imposed by the law of large numbers. Of course, 
to do all this, one needs sources with known density operator.
In optical applications, textbook quantum optics is used for these.
In practice, one knows the density operators only approximately, and 
can only get approximations for the rates, leading to approximate 
quantum measures only. How to create and extract the optimal
amount of information from finite statistics is beyond the present
paper. For some entry points to the large literature see, e.g.,
\sca{D'Ariano} et al. \cite{DArMP}.

For error estimation in quantum tomography see, e.g., 
\sca{Faist \& Renner} \cite{FaiR} and \sca{Flammia} et al. 
\cite{FlaGLE}.

\bigskip

In our discussion of quantum measures we essentially followed the view
of \sca{Schroeck} \cite[p.252]{Schr1989}. It is slightly different
from the more traditional point of view as presented, e.g., in
\sca{Busch} et al. \cite{BusGL,BusLPY} or \sca{Holevo}
\cite{Hol1982,Hol2001,Hol2012}.
There one uses in place of the discrete quantum measure a discrete
POVM, a family of finitely many positive operators $\Pi(\Delta)$ with
suitable properties, where $\Delta$ ranges over the subsets of $\Cz^m$
(or even only of $\Rz$). In the notation of the current setting, these
operators are given by
\[
\Pi(\Delta):=\sum_{k\in K: a_k\in\Delta} P_k.
\]
Since this implies $\Pi(\{\xi\})=P_k$ if $\xi=a_k$ and
$\Pi(\{\xi\})=\emptyset$ otherwise, the traditional POVM
encodes both the detection elements and the scale, and hence fully
specifies the detector.

When restricted to the commutative case (representing classical 
statistical physics in the state space formulation by 
\sca{Koopman} \cite{Koo}), the terminology of Busch amounts to treating 
all real random variables as observables, and that of Holevo to treating
all real random variables as generalized observables. But this is not 
the view of classical metrology, where (cf. \sca{Rabiniwitz} \cite{Rab})
classical observables always have a true value determined by the
theoretical description, and all randomness in measurements is assumed
to be due to measurement noise.

\bigskip

One can extend the concept of a discrete POVM
by dropping the discreteness requirement. This allows one to describe
idealized detectors with infinitely many possible detection elements
leading to infinitely many outcomes. This is needed, e.g., to discuss
thought experiments with arbitrarily accurate position measurements or
measurements of arbitrarily large momentum. The price for this is the
increased abstraction of the mathematical machinery needed.
In particular, the introduction of these more general POVMs requires 
measure theoretic concepts; details can be found in the books mentioned 
in the introduction. This added abstractness and formal complexity makes
general POVMs less suitable for introductory courses on quantum 
mechanics or quantum information theory.

\subsection{Informationally complete quantum measures} \label{ss.ic}

\nopagebreak
\hfill\parbox[t]{10.8cm}{\footnotesize

{\em
This would mean that one can measure all observables of a system in a
single experiment, merely by relabeling the outcomes. This would,
indeed, offer a radical new solution to the joint measurement problem.}

\hfill Jos Uffink, 1994 \cite[p.207]{Uff1994}
}

\bigskip

\nopagebreak
\hfill\parbox[t]{10.8cm}{\footnotesize

{\em
In this new approach we have a nonuniqueness in places which the old
theory accepted as physically significant. This means that, for
self-adjoint operators with a well-defined physical meaning like spin,
energy, position, momentum, etc., we have now many mathematical formulas
(many nonorthogonal resolultions of the identity). The natural question
arises: what this means, and how to remove such an ambiguity.
Awareness of this nonuniqueness existed in the early papers, but the
question was not worked out.}

\hfill Marian Grabowski, 1989 \cite[p.925]{Grabow}
}

\bigskip

The same measurement device allows one to measure many different
quantities, depending on which values are assigned to the various
responses of its detection elements.

We call a measurement device and its associated quantum measure $P_k$
($k\in K$) \bfi{informationally complete}
(\sca{Prugove\v{c}ki} \cite{Pru}) if the $P_k$ span the real vector
space $\Pc$ of Hermitian operators with finite trace.
By choosing the scale appropriately, a {\em single} informationally
complete measurement device allows the measurement of {\em arbitrary}
vector quantities $A$. Indeed, by definition, equation \gzit{e.obs} can
be solved componentwise for the components of the $a_k$.

If the state space $\Hz$ has finite dimension $d$, and there are
$|K|>d^2$ detection elements, there is a nontrivial relation
\[
\sum_{k\in K} \alpha_kP_k=0
\]
with real coefficients $\alpha_k$ that do not all vanish. Then
\gzit{e.Px} implies $P[a_k+\alpha_k\xi]=P[a_k]=A$ for all $\xi\in \Rz^m$
(or even $\xi\in \Cz^m$). Therefore  the scale is not uniquely
determined by the detection elements and the quantity $A$ measured.
In particular, a quantum measure can be informationally complete only
when the state space $\Hz$ has finite dimension $d$, and then requires
$|K|\ge \dim \Pc =d^2$ detection elements.
A \bfi{minimal} informationally complete quantum measure has
$|K|=d^2$. Together with the quantity $A$ measured, a minimal
informationally complete quantum measure uniquely determines the scale.
Of course, many measurements with a single informationally complete
measurement device must be averaged to get a good accuracy.

\subsection{Quantum state tomography}\label{ss.tomoStates}

\nopagebreak
\hfill\parbox[t]{10.8cm}{\footnotesize

{\em Whether a given device indeed does prepare the system in the state
indicated by an analysis of its operation must be verified through
agreement between the results of subsequent observations and the
theoretical predictions.}

\hfill Carl Helstrom, 1976 \cite[p.66]{Hel}
}

\bigskip

In generalization of the notion of a beam of light we say that a
\bfi{quantum beam} passes through the medium between a source and a
detector, carrying information (such as mass, energy, momentum, charge,
or spin) that somehow causes the detector responses. Strictly speaking,
what we measure are not properties of the source, but properties of the
end of the beam -- the part directly incident with the detector.

The properties of a stationary quantum beam can be inspected wherever
we place a detector -- just as a beam of light is invisible unless it
is scattered into a detector -- whether a photodiode or an eye.

Similar to quantum measurement tomography, the density operator of a
stationary source or the associated quantum beam can also be found 
operationally. In the simplest case, this is done by measuring the mean 
rates with which a detector with a known informationally complete 
quantum measure responds to the source. Then \gzit{e.pk} is a linear 
system for $\rho$ with a unique solution if sufficiently many and 
sufficiently diverse mean rates are measured. Solving this linear 
system gives the density operator $\rho$.

More precisely, to find the density matrix for an arbitrary
experimentally realized and sufficiently stationary source one has to
collect statistics for the response of a quantum measure corrsponding 
to at least $d^2$ linearly independent detection elements, giving the 
$P_{kij}$ and the corresponding response probabilities $p_k$. Then one 
solves the linear system \gzit{e.pk} for the $\rho_{ij}$ to get the 
density matrix. This procedure for determining density matrices, called 
\bfi{quantum state tomography}, constitutes the quantum tomography 
principle for the characterization of states. This process is essential 
for calibrating sources and is called \bfi{quantum state tomography}.

Of course, to do all this one needs detectors having a known quantum
measure. In optical applications, one uses for these arrangements of
photodetectors and optical devices modifying the source in a known way.
In practice, one knows the quantum measures only approximately and can
only get approximations for the mean rates, leading to approximate 
quantum states only. How to create and extract the optimal amount of 
information from a given finite amount of statistics is beyond the 
present paper. For some entry points to the large literature see, e.g., 
\sca{Je\v zek} et al. \cite{JezFH} or \sca{Granade} et al. \cite{GraFF}.

In particular, if the state space has finite dimension $d$, the
knowledge of the probabilities $p_k$ of an informationally complete
quantum measure determines the associated state, and hence its density
operator. For further discussion of this see, e.g., \sca{Caves} et al.
\cite{CavFS} and \sca{Allahverdyan} et al. \cite{AllBN0}.

Quantum systems with more complex properties have higher-dimensional 
state spaces than simple quantum systems such as qubits. The bigger the 
state space dimension $d$ the more difficult it is to create the $d^2$ 
detection elements needed. A single detector suffices only if the 
measurement device is informationally complete, but by an appropriate 
combination of equipment this can be relaxed, using the concept of 
composite detectors; see Subsection \ref{ss.compos} below.

\subsection{Completely positive maps and quantum process tomography}
\label{ss.tomoProc}

\nopagebreak
\hfill\parbox[t]{10.8cm}{\footnotesize

{\em The most arbitrary transfer function of a quantum black box is to
map one density matrix into another, $\rho_\iin\to\rho_\out$, and this
is determined by a linear mapping $\cal E$ which we shall give a method
for obtaining.}

\hfill Chuang and Nielsen, 1997 \cite[p.2455]{ChuN7}
}

\bigskip

\nopagebreak
\hfill\parbox[t]{10.8cm}{\footnotesize

{\em Parametric amplifiers and phase-conjugating mirrors are active 
devices -- they require a source of energy, mostly light of a higher 
frequency. Yet these active devices share a key property with the 
passive instruments -- they are linear devices in the sense that the 
input modes are linear transformations of the output modes. 
}

\hfill Ulf Leonhardt, 2003 \cite[p.1213]{Leo.i}
}

\bigskip

In addition to sources and detectors, experimental equipment for
studying quantum systems includes \bfi{filters} that modify a source.
As in the case of states and detectors, we generalize the notion from
our discussion of optical equipment in Subsection \ref{ss.pol}.
As discussed there, a linear, non-mixing (not depolarizing) optical
filter is characterized by a complex $2\times 2$ transmission matrix
$T$. The filter transforms an in-going beam with state $\rho$ into an
out-going beam with state $\rho'=T\rho T^*$. The intensity of a
beam after passing the filter is
\[
I'=\tr \rho'=\tr T\rho T^*=\tr \rho T^*T.
\]
If the filter is lossless, the intensities of the in-going and the
out-going beam are identical. This is the case if and only if the
transmission matrix $T$ is unitary.

More generally, a linear, mixing (depolarizing) optical filter
transforms $\rho$ instead into a sum of several terms of the form
$T\rho T^*$. This is a linear expression in $\rho$, hence can be
described by a real $4\times 4$ matrix acting on the Stokes vector,
called the \bfi{Mueller matrix} (\sca{Perez \& Ossikovski} \cite{PerO}).
Equivalently, it is described by a completely positive linear map
(\sca{Kossakowski} \cite{Kos}) on the space of $2\times 2$ matrices,
acting on the coherence matrix.

Completely positive maps are special cases of \bfi{superoperators},
linear maps $E\in\Lin\Lin \Hz$ that map the space $\Lin \Hz$ of linear
operators $X:\Hz\to\Hz$ into itself. $\Lin \Hz$ has a natural Hermitian 
inner product
\[
\<X|Y\>:=\tr X^* Y.
\]
In the special case where $\Hz=\Cz^n$, operators $X\in\Lin \Hz$ are
$n\times n$ matrices whose entries $X_{jk}$ are indexed by two indices
$j,k$ running from $1$ to $n$. Therefore a superoperator $E$ maps $X$
into a matrix $E(X)$ whose components are
\[
E(X)_{ab}:=\sum _{j,k} E_{abjk} X_{jk},
\]
with an array of components $E_{abjk}$ indexed by four indices. Giving
this array specifies the superoperator; the case $n=2$ corresponds to
the optical situation just discussed. By transposing the middle indices,
we get from this array another array of components
\[
\wh E_{abjk}:=E_{ajbk}.
\]
This array defines another superoperator $\wh E$ that we call the
\bfi{Choi transform} of $E$. It was first used by \sca{Choi} \cite{Choi}
to characterize completely positive maps. Applying the Choi transform
twice clearly recovers the original superoperator.

\begin{thm}\label{t.CP}~\\
(i) If $E(X)=TXT^*$ then $\wh E(X)=T\tr T^*X$, and we have
\lbeq{e.op}
\<\wh E(X)|Y\>=\<X|\wh E(Y)\>,
\eeq
\lbeq{e.CP}
\<X|\wh E(X)\>\ge 0 \for X\in\Lin \Hz  \mbox{ of finite rank},
\eeq
and hence for all bounded $X$ for which the left hand side is defined.

(ii)\footnote{
Part (ii) is due to \sca{Sudarshan} et al. \cite{SudMR} (see also
\sca{Jagadish \& Petruccione} \cite{JagP}), but their derivation of 
\gzit{e.CP} from a positivity condition is defective. 
}
The Choi transform of a superoperator $E$ satisfies \gzit{e.op}
and \gzit{e.CP} iff $E$ can be written in the form of an
\bfi{operator sum expansion} (also called \bfi{Kraus representation})
\lbeq{e.OSE}
E(X)=\sum_\ell T_\ell X T_\ell^*.
\eeq
(iii) In terms of an arbitrary orthonormal basis $e_1,e_2,\ldots$ of
$\Hz=\Cz^n$, and the associated basis \gzit{e.Eee} of $\Lin \Hz$,
the Choi transform maps $X\in \Lin \Hz$ to
\lbeq{e.Choi}
\wh E(X)=\sum_{j,k}E(e_je_k^*)Xe_ke_j^*.
\eeq
\end{thm}

A superoperator $E$ whose Choi transform satisfies \gzit{e.op} and the
\bfi{positivity condition} \gzit{e.CP} is called a
\bfi{completely positive map} on $\Lin\Hz$. 

\bepf
(i) Since $\D E(X)_{ab}=\sum_{j,k}T_{aj}X_{jk}\ol T_{bk}$, we have
$E_{abjk}=T_{aj}\ol T_{bk}$, hence
\[
\wh E(X)_{ab}=\sum_{j,k} \wh E_{abjk} X_{jk}
=\sum_{j,k} E_{ajbk} X_{jk} =\sum_{j,k} T_{ab}\ol T_{jk}X_{jk}
=T_{ab}\tr T^*X.
\]
This proves $\wh E(X)=T\tr T^*X$. Now \gzit{e.op} follows from
\[
\bary{lll}
\<\wh E(X)|Y\>
&=&\tr \wh E(X)^*Y=\tr (T\tr T^*X)^*Y=(\tr X^*T)\tr T^*Y\\
&=&\tr X^*\wh E(Y)=\<X|\wh E(Y)\>,
\eary
\]
and then \gzit{e.CP} follows from
\[
\<\wh E(X)|X\>=|\tr T^*X|^2\ge 0.
\]
(ii) If $E$ has the operator sum expansion \gzit{e.OSE} then \gzit{e.op}
and \gzit{e.CP} hold by (i) applied to each term in the sum. Conversely,
if \gzit{e.op} and \gzit{e.CP} hold, this says that $\wh E$ is Hermitian
positive semidefinite as an operator on $\Lin \Hz$.
Any such operator can be written as a product
\lbeq{e.Efact}
\wh E=\wh R^*\wh R
\eeq
of a superoperator $\wh R$ and its adjoint. Therefore
\[
\bary{lll}
\<X|\wh E(Y)\>&=&\<X|\wh R^*(\wh R(Y)\>
=\<\wh R(X)|\wh R(Y)\>
=\tr \wh R(X)^*\wh R(Y)\\[2mm]
&=&\D\sum_{c,d} \ol{\wh R(X)}_{cd}\wh R(Y)_{cd}
=\sum_{c,d} \Big(\ol{\sum_{jk}\wh R_{cdjk}X_{jk}}\Big)
            \Big(\sum_{j'k'}\wh R_{cdj'k'}Y_{j'k'}\Big) \\[6mm]
&=&\D\sum_{jk}\ol X_{jk}\sum_{c,d,j',k'}\ol R_{cjdk}R_{cj'dk'}Y_{j'k'},
\eary
\]
hence 
\[
\wh E(Y)_{jk}=\D\sum_{c,d,j',k'}\ol R_{cjdk}R_{cj'dk'}Y_{j'k'},~~~
\wh E_{jkj'k'}=\sum_{c,d}\ol R_{cjdk}R_{cj'dk'}.
\]
If we introduce the matrices $T_{cd}$ with entries
\[
(T_{cd})_{aj}:=\ol R_{cadj},
\] 
we find
\[
\bary{lll}
E(X)_{ab}&=&\D\sum_{j,k} E_{abjk}X_{jk}
=\sum_{j,k}\wh E_{ajbk}X_{jk}
=\sum_{j,k}\sum_{c,d}\ol R_{cadj}R_{cbdk}X_{jk}\\[6mm]
&=&\D\sum_{c,d}\sum_{j,k}T_{cdaj}X_{jk}\ol T_{cdbk}
    =\sum_{c,d}\sum_{k}(T_{cd}X)_{ak}\ol T_{cdbk}
=\sum_{c,d}(T_{cd}XT_{cd}^*)_{ab}.
\eary
\]
This implies the operator sum expansion
$E(X)=\D\sum_{c,d}T_{cd}XT_{cd}^*$.
\\
(iii) The Choi transform maps $X$ to the operator $\wh E(X)$ with 
components
\[
\wh E(X)_{ab}=\sum _{\ell,d} \wh E_{ab\ell d} X_{\ell d}
=\sum _{\ell,d} E_{a\ell bd} X_{\ell d}.
\]
On the other hand, since
$E(e_je_k^*)_{a\ell}=\sum_{c,d} E_{a\ell cd}e_{jc}\ol e_{kd}$,
the $(a,b)$ entry of the right hand side $S$ is
\[
\bary{lll}
S_{ab}&=&\D\sum_{j,k}\Big(E(e_je_k^*)Xe_ke_j^*\Big)_{ab}
= \D\sum_{j,k,\ell,m}
    E(e_je_k^*)_{a\ell}X_{\ell m}(e_ke_j^*)_{mb}\\[6mm]
&=&\D\sum_{j,k,\ell,m,c,d}
    E_{a\ell cd}e_{jc}\ol e_{kd}X_{\ell m}e_{km}\ol e_{jb}
= \D\sum_{\ell,d} E_{a\ell bd}X_{\ell d}=\wh E(X)_{ab},
\eary
\]
and we conclude that $\wh E(X)=S$. Here $e_{jc}$ is the $c$th component 
of $e_j$. In the second last step, we summed over $k$ and $j$ using 
that for an orthonormal basis
\[
\sum_k \ol e_{kd}e_{km}=\delta_{dm},~~~
\sum_j e_{jc}\ol e_{jb}=\delta_{cb},
\]
where $\delta$ is the Kronecker symbol. Thus only the terms with 
$m=d$ and $c=b$ give a nonzero contribution. 
\epf

The operator sum expansion is not uniquely determined since
\gzit{e.Efact} remains valid when we multiply $\wh T$ on the left by an
arbitrary unitary matrix. The simplest choice for $\wh T$ is a Cholesky
factor (\sca{Golub \& van Loan} \cite{GolvL}), where $\wh T$ is upper
triangular and the diagonal entries are real and nonnegative. This
determines $\wh T$ uniquely, given the basis and some fixed ordering of
the index pairs.

As a consequence of \gzit{e.OSE}, a completely positive map maps a
density operator $\rho$ into another density operator. Indeed,
$\rho':=E(\rho)$ is Hermitian and positive semidefinite since
\[
\rho'^*=E(\rho)^*=E(\rho^*)=E(\rho),
\]
\[
\psi^*\rho'\psi=\psi^*E(\rho)\psi
=\sum_\ell \psi^*T_\ell\rho T_\ell^*\psi
=\sum_\ell (T_\ell^*\psi)^*\rho T_\ell^*\psi\ge 0.
\]
It was proved by \sca{Kraus} \cite{Kra} that \gzit{e.OSE} is equivalent
to the traditional definition of complete positivity traditionally used
to define quantum processes (called \bfi{operations} in the pioneering
paper by \sca{Hellwig \& Kraus} \cite{HelK}). Tradition also requires in
addition that the trace is preserved, a condition too strong in our
setting where density operators are not normalized.

As in the case of states, we may generalize the optical situation from
the state space $\Cz^2$ to arbitrary separable state spaces $\Hz$.
We may use formula \gzit{e.Choi} to define the Choi transform in
general. Theorem \ref{t.CP} remains valid with essentially the same
proof. In the infinite-dimensional case, one can use either
a discrete basis or a continuous representation; the operator sum
expansion turns into the \bfi{Stinespring factorization}
(\sca{Stinespring} \cite{Sti}), with the sum replaced by an
infinite sum when a discrete basis is used, or by an integral when a
continuous representation is used.

As sources are characterized by their state and measurement devices are
characterized by their quantum measure, so linear filters are
characterized by a bounded completely positive map $E$ specifying the
input-output behavior of the filter.

A \bfi{linear quantum filter} (also called a \bfi{quantum process})
incident with an input source whose state is given by a density
operator $\rho\in \Lin \Hz $ produces an output whose state is given by
the density operator $E(\rho)\in \Lin \Hz $, with a completely positive
map $E\in\Lin\Lin \Hz $ specifying the details of the input-output
behavior. If the completely positive map is given as an operator sum
expansion \gzit{e.OSE}, the intensity of the outpot beam is
\[
I_\out=\tr E(\rho)=\sum_\ell \tr T_\ell \rho T_\ell^* 
=\sum_\ell \tr T_\ell^* T_\ell \rho  =\tr \pi(E)\rho,
\]
where
\lbeq{e.piE}
\pi(E):=\sum_\ell \tr T_\ell^* T_\ell
\eeq
is Hermitian and positive semidefinite. 

A filter is called \bfi{lossless} if it preserves the intensity of any
beam, thus if 
\lbeq{e.ll}
\pi(E)=\sum_\ell  T_\ell^* T_\ell=1.
\eeq
A filter is called \bfi{active} if it increases  the intensity of some
beam, and \bfi{passive} otherwise; in particular, lossless filters are
passive. Most filters discussed in the literature are passive; active
filters need an input of external energy. In quantum information theory,
a linear passive filter is called a \bfi{quantum gate} or 
\bfi{quantum channel}.

Important examples of filters are \bfi{linear, nonmixing quantum
filters}, whose completely positive map is given by
\[
E(X)=TX T^*
\]
with an operator $T\in\Lin \Hz $ called the \bfi{transmission operator},
or \bfi{transmission matrix} if the input basis and the output basis
are fixed. Thus their operator sum expansion \gzit{e.OSE} has only
a single term. 
Linear quantum filters that cannot be represented in the form
$E(X)=TX T^*$ are called \bfi{mixing} since they turn most pure states
into mixed states. 

The transmission operator of a nonmixing filter may be viewed as a 
generalization of the S-matrix. Indeed, by \gzit{e.ll}, a nonmixing 
quantum filter is lossless iff its transmission operator $T$ is unitary.
A lossless nonmixing filter is essentially a scattering process, and 
$T$ is the S-matrix. Multiple scattering introduces losses. It also 
introduces nonlinearities, but as long as these are small they can be 
neglected. 

For the idealized case of beams in a pure state $\psi$, further
development is possible since nonmixing quantum filters map a pure
state with state vector $\psi$ into a pure state with state vector
$T\psi$. For example, the perfect polarizer with $T=\phi\phi^*$ where
$\phi^*\phi=1$ reduces the intensity of a pure state $\psi$ to
\[
S_0'=\<T^*T\>=\psi^*T^*T\psi=\psi^*\phi\phi^*\psi=|\phi^*\psi|^2.
\]
For the special case of polarized light, this is \bfi{Malus' law} from
1809 (\sca{Malus} \cite{Mal}). Reinterpreted in terms of detection
probabilities, it gives Born's squared amplitude formula for quantum
probabilities.

In principle, the completely positive map characterizing a linear filter
can be determined by \bfi{quantum process tomography} (\sca{Poyatos}
\cite{PoyCZ}). In the conceptually simplest way, one has to prepare the
source in sufficiently many different states $\rho_\ell$, passing the
resulting quantum beams through the filter, and using quantum state
tomography to measure the resulting states $\rho'_{\ell}$. Then $E$ is
determined by solving the linear equations
\[
E(\rho_\ell)=\rho'_{\ell} \Forall\ell
\]
for the array $E_{abjk}$ defining the superoperator $E$.
This is possible unique if the input states span the space of all
Hermitian operators. For details and efficient finite precision schemes
see, e.g., \sca{Chuang \& Nielsen} \cite{ChuN7} and
\sca{Mohseni} et al. \cite{MohRL}.

\bigskip

From a sufficiently diverse collection of states and measurement devices
and statistics for all state-device pairs, one can, in principle, even
determine both density matrices and quantum measures together; see,
e.g., \sca{Mogilevtsev} \cite{Mog}. This
involves the solution of \gzit{e.pk} for both the $P_{kij}$ and the
$\rho_{ij}$. But now this is a nonlinear system for the joint collection
of unknowns, and checking for a unique solution is difficult. However,
known approximate solutions may be improved in this way, leading to a
simultaneous calibration of the states used and the measurement devices
used. This eliminates a possible source of circularity in the relation
between the theoretical concepts and their phenomenological expression.

In principle, one can also determine the properties of both sources and
filters simultaneously. This is called
\bfi{self-calibrating tomography}. The state $\rho_{k\ell}$ of the
output of the $k$th filter when fed with an input beam from the $\ell$th
source can be determined by quantum state tomography. Then one has to
solve the nonlinear equations
\[
F_k(\rho_\ell)=\rho_{k\ell} \Forall k,\ell
\]
simultaneously for the transmission operators $F_k$ and the density
operators $\rho_\ell$. For more practical schemes with finite precision
see \sca{Quesada} et al. \cite{QueBJ}, \sca{Sim} et al. \cite{SimSNE},
and \sca{Lepp\"aj\"arvi \& Sedl\'ak} \cite{LepS}.

\subsection{Nonstationary sources}\label{s.nonstat}

\nopagebreak
\hfill\parbox[t]{10.8cm}{\footnotesize

{\em It is always good to know which ideas cannot be checked directly, 
but it is not necessary to remove them all. It is not true that we can 
pursue science completely by using only those concepts which are 
directly subject to experiment. In quantum mechanics itself [...] there 
are many constructs that we cannot measure directly. The basis of a 
science is its ability to predict. To predict means to tell what will 
happen in an experiment that has never been done.}

\hfill Richard Feynman, 1971 \cite[p.2-9]{FeyLS}
}

\bigskip

\nopagebreak
\hfill\parbox[t]{10.8cm}{\footnotesize

{\em These experiments have a special fascination, as, in the language 
of modern quantum mechanics, they allow one to watch the reduction of 
the wave function by the measurement process on the oscilloscope 
screen.}

\hfill Nagourney, Sandberg and Dehmelt, 1986 \cite[p.2787]{NagSD}
}

\bigskip

\nopagebreak
\hfill\parbox[t]{10.8cm}{\footnotesize

{\em The environment of the qubit is not stationary. Moreover, initial 
correlations between SQD and qubit -- both microscopic systems -- might 
exist. When integrating out the environment, the factorizability and the
stationarity of the environment are, however, often invoked to eliminate
the so-called slip of the initial condition for the subsystem coming 
from the initial state of the environment and its short-time transient
evolution.}

\hfill Hell, Wegewijs and DiVincenzo, 2016 \cite[p.3]{HelWD}
}

\bigskip

Implicit in the exposition so far are idealizations that are 
not always appropriate. One must be aware of the limitations these
idealization impose on our quantum models and on the power of quantum 
tomography.

An idealization we used most of the time was stationarity. 
So far, we only considered {\em stationary} beams since their 
statistical properties are independent of time, hence can be probed in 
principle with arbitrary accuracy. While stationary quantum systems
constitute the most important building blocks for the theoretical and 
experimental analysis of quantum phenomena, they only describe the 
short-time behavior of real systems. In the real world, everything is 
subject to change, and therefore everything is nonstationary on scales 
where this change becomes observable. 

We therefore need to extend the concepts introduced to nonstationary 
systems. When a source is nonstationary its quantum state $\rho(t)$ 
depends on time $t$. Since each measurement takes time, the reachable 
accuracy in determining the state is limited by the amount of 
smoothness of the state in time and by the amount of information that 
can be obtained during a time interval where the form of the time 
dependence can be assumed to be simple enough to determine the 
coefficients. Moreover, to get the state to higher accuracy, fainter 
sources must be measured for much longer timse. This requires that the 
system is much closer to stationarity.

The situation is essentially the same as for classical measurements of 
time-dependent signals, where there is a tension between the resolution 
in the time domain and in the frequency domain, given by 
\bfi{Nyquist's theorem}, a classical analogue of 
\bfi{Heisenberg's uncertainty relation} for quantum systems.
Low frequency phenomena are slowly varying at short times, hence can be 
reliably resolved by repeated sampling in time. But sufficiently high 
frequency phenomena are varying so fast that sampling with a fixed 
time resolution cannot capture their details. 

This problem is more pronounced in the quantum domain since most 
quantum systems have a continuous spectrum extending to infinity, 
implying the presence of arbitrarily large oscillation frequencies. 
Since the microscopic domain cannot be described in classical terms,
experiments directly based on stationarity assumptions can probe only
the low frequency, semiclassical behavior of microscopic systems.
However, higher frequencies can be probed in a stationary setting by 
spectroscopic means, using the fact that eigenfrequencies of physical 
systems  -- discussed quantitatively in Subsection \ref{ss.Espec} below 
-- can be monitored. This is done by studying their resonance behavior 
in interactions with systems whose spectral properties are known. 
In particular, the probing by visible light produces the visible 
spectra whose explanation was one of the first decisive successes of 
modern quantum mechanics. 

The extrapolation of the concepts of state, filters, and detectors to
the nonstationary, and in the microscopic domain has proved 
extraordinarily productive, without needing any changes in the 
mathematical foundations. Today it is possible to perform experiments 
studying the temporal behavior of single microscopic quantum systems 
such as a single atom in an ion trap and to quantitatively predict 
their stochastic behavior. These are highly nonstationary systems where
the simple quantum tomography methods discussed in this paper no longer 
work.

In the nonstationary case, the measuring frequency must be low enough 
to avoid that each measurement probes a significantly different quantum 
state. The law of large numbers therefore limits the accuracy with which
a nostationary state can be determined. This is analogous to the limit 
imposed by the well-known Nyquist bound for sampling periodic signals. 

Thereforeour understanding and analysis of  nonstationary processes 
must rely more on theory and interpretation. Temporal quantum 
tomography -- the quest for measuring time-dependent quantum states 
(\sca{Nokkala} \cite{Nok}, \sca{Tran \& Nakajima} \cite{TraN}) --  
has barely been studied. 
However, using statistical tools from time series analysis for 
nonstationary systems described by suitable stochastic models 
(\sca{Bertolotti} et al. \cite{BerSF}), measurements can still be used 
get quantitative information of limited accuracy about the 
time-dependent state.

\subsection{Quantum tomography of instruments}\label{ss.tomoInstr}

\nopagebreak
\hfill\parbox[t]{10.8cm}{\footnotesize

{\em The state of the system after the observation must be an
eigenstate of [the observable] $\alpha$, since the result of a
measurement of $\alpha$ for this state must be a certainty.}

\hfill Paul Dirac, 1930 \cite[p.49]{Dir1}
}

\bigskip

\nopagebreak
\hfill\parbox[t]{10.8cm}{\footnotesize

{\em It is very important to notice that the functions $\Psi_n(q)$ do
not, in general, coincide with the functions $\phi_n(q)$; the latter are
in general not even orthogonal, and do not form a set of eigenfunctions
of any operator. If the electron was in a state $\Psi_n(q)$, then a
measurement of the quantity $f$ carried out on it leads with certainty
to the value $f_n$. After the measurement, however, the electron is in
a state $\phi_n(q)$ different from its initial one, and in this state
the quantity $f$ does not in general take any definite value.}

\hfill Landau and Lifschitz, 1958, 1977 \cite[p.23]{LL.3}
}

\bigskip

\nopagebreak
\hfill\parbox[t]{10.8cm}{\footnotesize

{\em Such an operation could, for example, describe the state
reduction from a measurement apparatus for a given fixed
outcome, which occurs with probability $\tr(\rho A^*A)\le 1$.}

\hfill D'Ariano and Presti, 2001 \cite[p.49]{DArP}
}

\bigskip

\nopagebreak
\hfill\parbox[t]{10.8cm}{\footnotesize

{\em Von Neumann's quantum postulates successfully predict the
probability distribution of the outcome of the measurement. However, 
they are not sufficient to predict the joint probability distribution 
of the outcomes of successive measurements in time. [...] 
For the statistics of successive measurements, we need the notion of 
state changes caused by measurements, often called quantum state 
reductions, which has been considered one of the most difficult notions 
in quantum mechanics. In order to deal with quantum systems to be 
measured sequentially, we need a general notion of quantum state 
reductions and mathematical methods to calculate them. For this purpose,
we have to mathematically characterize all of the possible state 
changes caused by the most general type of measurements.
}

\hfill Masanao Ozawa, 2021 \cite[p.2]{Oza2021}
}

\bigskip

In quantum mechanical experiments, the simplest kind of a nonstationary 
situation is described by a quantum instrument. (More general 
nonstationary situations will be discussed in Subsection 
\ref{s.nonstat} below.) Quantum instruments generalize quantum filters 
and quantum measurement devices. Like quantum measurement devices, 
they produce random measurable results. But, in addition, they produce 
like filters a random output perfectly correlated with these results. 
Examples are the wire detectors in particle colliders, where particles 
passing a wire are deflected while depositing a measurable amount of 
energy.

A \bfi{quantum instrument}\footnote{
also called a \bfi{quantum operation}, but this terminology is ambiguous
and not recommended since in the literature, this term is also used for 
what we call here quantum filters. 
} 
(\sca{Davies \& Lewis} \cite{DavL}) consists
of a collection of detection elements indexed by $j\in J$ (where $J$ 
does not contain the label $0$ for a null response), and transforms an 
input beam in state $\rho$ into an output beam whose state $\rho'$ is 
\[
\rho'=\cases{E_j(\rho) & when detection element $j$ responded,\cr
             E_0(\rho) & when no detection element responded,}
\]
with completely positive maps $E_j$ ($j\in J$) and $E_0$. These maps 
specify not only the output state but also the response rate 
\lbeq{e.pj}
p_j:=\tr E_j(\rho)=\tr \pi(E_j)\rho=\<\pi(E_j)\>
\eeq
of the $j$th detection element, expressed in terms of \gzit{e.piE}.
This definition assumes that time is split into intervals associated to 
a particular detector response and times where nothing happens. This
usually makes experimental sense, but is conceptually a little fuzzy --
as often when experiments and theoretical descriptions must be matched.
The special case where $J$ is empty (no detection elements, hence no 
measurement result) just describes a filter, but if $J$ is nonempty, the
output beam is nonstationary since it depends on what happened at the 
detector elements. 

Another important special case is that of a \bfi{projective instrument},
where each $E_j$ ($j\in J\cup\{0\}$) is nonmixing and the corresponding 
transition operators $P_j$ with
\[
E_j(\rho)=P_j\rho P_j^*
\]
are orthogonal projectors to the eigenspaces of a Hermitian operator 
$A$ with distinct eigenvalues $a_j$  ($j\in J\cup\{0\}$). Thus
\[
P_j^2=P_j=P_j^*,~~~P_jP_k=0 \for j\ne k.
\]
In this case, the instrument acts as a detector performing a projective 
measurement of $A$. For a pure input state with state vector $\psi$, its
output is the projection $P_j\psi$ of the input state to the eigenspace 
of $A$ corresponding to the eigenvalue $a_j$ obtained as measurement 
result. This is the Dirac/von Neumann reduction of the state vector, 
often referred to as \bfi{collapse}. 

Thus quantum instruments capture in a generally valid, precise fashion 
the operationally relevant generalized version of the notion of
collapse (of the state vector, wave function, or wave packet) or
state reduction that are found in most introductory textbooks
on quantum mechanics.

The determination of the completely positive maps $E_j$ 
($j\in J':=J\cup\{0\}$) and for a given instrument is called 
\bfi{quantum instrument tomography}; cf. \sca{Sugiyama} \cite{Sug}.
Though very little systematic work has been done on it, the basic 
principle is again quite easy. The behavior of an instrument is 
experimentally probed through so-called \bfi{conincidence measurements}.
Here the output beam of the instrument is observed with a second 
measurement device whose detection elements are labelled by $k\in K$. 
Allowing for null detections, we again put $K':=K\cup\{0\}$. One then 
records the pairs $(j,k)\in J'\times K'$ of events where the $j$th 
detection element of the instrument and the $k$th detection element of 
the second measurement device respond simultaneously (within some 
temporal resolution bound accounting also for beam transmission times). 
Here one records $(j,0)$ when the $j$th detection element of the 
instrument responds, but not the second measurement device, and $(0,k)$ 
when the second measurement device responds, but no instrument detection
element. By \bfi{postselection}, one extracts the statistics of the 
subsample of events $(j,k)$ with fixed $j\in J\cup\{0\}$. This subsample
is distributed according to the completely positive map $E_j(\rho)$, 
hence the latter can be determined by quantum process tomography. 
The response rates can be obtained from the complete sample togeher 
with the time stamps of the events, and \gzit{e.pj} serves for a check
either of consistency, or of the assumptions underlying the whole 
setting.

\subsection{Composite sources, filters, and detectors}\label{ss.compos}

\nopagebreak
\hfill\parbox[t]{10.8cm}{\footnotesize

{\em One can construct, in a controlled way, optical networks, also 
called multiports, from the basic building blocks such as beam splitters
and mirrors, networks with interesting quantum properties.}

\hfill Ulf Leonhardt, 2003 \cite[p.1213]{Leo.i}
}

\bigskip

In practice, sources, detectors, filters, and instruments often have
controls that modify their behavior. The controls can be modelled by
parameters on which the states, quantum measures, and transmission
operators depend.The dependence of these on the controls can also be
elicited by quantum tomography, by performing these with different
values of the controls. Doing this with a limited amount of
observations and maximal accuracy involves nontrivial optimization
techniques that are beyond the present exposition.

We get composite sources by combining a source with one or more appended
filters, and regarding the output of the last filter as the output  of
the composite source when the first filter is fed with the source. 
Similarly, we get composite filters by combining  several filters in 
series. We also get composite detectors by combining a detector with 
one or more prepended filters, and measuring a beam entering the first 
filter by measuring the output of the last filter. 
As in the optical case discussed in Subsection \ref{ss.pol}, this 
allows one to create an almost unlimited variety of sources and 
detectors. Their behavior can be predicted by applying repeatedly the 
transfromation rules discussed above.

Combinations of controllable sources, filters including splitters,
instruments, and detectors in a similar way as discussed in the optical
case in Subsection \ref{ss.pol} allow one to design (within the
experimentally realizable accuracy) complex quantum circuits with a
prespecified input-output behavior.

\newpage
\section{Quantum dynamics} \label{s.qDyn}

\nopagebreak
\hfill\parbox[t]{10.8cm}{\footnotesize

{\em For we have frequent need of the general view, but not so often of 
the detailed exposition. Indeed it is necessary to go back on the main 
principles, and constantly to fix in one's memory enough to give one 
the most essential comprehension of the truth. And in fact the accurate 
knowledge of details will be fully discovered, if the general principles
in the various departments are thoroughly grasped and borne in mind; for
even in the case of one fully initiated the most essential feature in 
all accurate knowledge is the capacity to make a rapid use of 
observation and mental apprehension, and this can be done if everything 
is summed up in elementary principles and formulae. For it is not 
possible for anyone to abbreviate the complete course through the whole
system, if he cannot embrace in his own mind by means of short formulae 
all that might be set out with accuracy in detail.}

\hfill Epicurus, ca. 300 BC \cite[p.1]{Epi}
}

\bigskip

Equipped with the concepts and results from quantum tomography, we now 
obtain in a deductive fashion the standard machinery of quantum 
mechanics, including everything traditionally simply assumed at the 
very beginning.

The notion of a quantum phase space and the associated Poisson algebra 
brings conservative quantum dynamics and conservative classical dynamics
under a common dynamical framework.

The basic dynamical equations of quantum mechanics -- the conservative 
time-dependent Schr\"odinger equation for pure states in a nonmixing 
medium, the Ehrenfest equation for the dynamoics of quantum values, and 
the dissipative Lindblad equation for states in mixing linear media 
-- are derived by a continuum limit of a sequence of quantum filters. 

The more complex case of mixing nonlinear media gives rise to the 
phenomenon of metastability, which mediates between the continuous 
quantum dynamics and the discrete behavior of detection elements.

\subsection{Quantum beams and quantum Liouville equation}\label{ss.qDyn}

\nopagebreak
\hfill\parbox[t]{10.8cm}{\footnotesize

{\em In physics, causal description, originally adapted to the problems
of mechanics, rests on the assumption that the knowledge of the state
of a material system at a given time permits the prediction of the
state at any subsequent step.}

\hfill Niels Bohr, 1948 \cite[p.51]{Bohr}
}

\bigskip

The passage of a quantum beam through an inhomogeneous, nonmixing linear
medium can be modeled by means of many slices consisting of very thin
nonmixing filters with transmission operators close to the identity,
hence of the form
\lbeq{e.TK}
T(t)=1+\Delta t K(t)+O(\Delta t^2),
\eeq
where $\Delta t$ is the very short time needed to pass through one
slice and $K(t)$ is an operator specifying how $T(t)$ deviates from the
identity. If $\rho(t)$ denotes the density operator at time $t$ then
$\rho(t+\Delta t) = T(t)\rho(t)T(t)^*$, so that
\[
\frac{d}{dt}\rho(t)
= \frac{\rho(t+\Delta t)-\rho(t)}{\Delta t}+O(\Delta t)
= K(t)\rho(t)+\rho(t)K(t)^*+O(\Delta t).
\]
In the continuum limit $\Delta t\to 0$, we obtain the
\bfi{quantum Liouville equation}
\lbeq{e.qLiou}
\frac{d}{dt}\rho(t) = K(t)\rho(t)+\rho(t)K(t)^*.
\eeq
It is customary to scale $K(t)$ by \bfi{Planck's constant} $\hbar$ to
obtain a quantity with the units of energy, and to express
\lbeq{e.KH}
i\hbar K(t)=H(t)-iV(t)
\eeq
in terms of two Hermitian operators, the \bfi{Hamiltonian} $H(t)$ and
a \bfi{dissipative potential} $V(t)$. For a passive medium,
\[
0\le \tr \rho - \tr T(t)\rho T(t)^*=\tr \rho -\tr T(t)^*T(t)\rho
=\Delta t\ \tr \Big(K(t)+K(t)^*+O(\Delta t)\Big)\rho.
\]
In the limit $\Delta t\to 0$, multiplication by $\hbar/\Delta t$ gives 
for $\rho=\psi\psi^*$ the relation
\[
0\le\hbar \tr(K(t)+K(t)^*)\rho=\tr V(t)\rho=\tr V(t)\psi\psi^*
=\psi^*V(t)\psi.
\]
Thus for a passive medium, $V(t)$ is positive semidefinite. 

In the even more special case of a lossless medium, equality holds in 
the above argument, hence the dissipative potential $V(t)$ vanishes.
Therefore,  for a lossless medium, the quantum Liouville equation takes 
the special commutator form
\lbeq{e.vNeu}
i\hbar\frac{d}{dt}\rho(t) = [H(t),\rho(t)]~~~(H(t) \mbox{~Hermitian})
\eeq
of the \bfi{von Neumann equation}.

\subsection{Quantum phase space}\label{s.phase}

\nopagebreak
\hfill\parbox[t]{10.8cm}{\footnotesize

{\em The general structures of classical and quantum mechanics are 
usually regarded as essentially different and the relation between them 
has been the subject of several investigations. As a matter of fact, 
the connection between the usual formulation of classical mechanics
and quantum mechanics is not immediate. [...]
A better understanding of the relation between the two theories may be 
obtained by a formal theory of generalized dynamics which includes 
classical and quantum mechanics as special cases.}

\hfill Franco Strocchi, 1964 \cite[p.36]{Stro}
}

\bigskip

A consequence of the present approach is the natural emergence of a 
phase space structure for quantum dynamics. To motivate the concept, 
we first review the basics of classical Hamiltonian mechanics.

{\bf Conservative classical mechanics.}
In the Hamiltonian formulation of conservative classical mechanics, $N$ 
distinguishable particles are modelled in terms of their position and 
momentum. These constitute phase space variables $q_k$ and $p_k$ 
($k=1,\ldots,N$) which take values in $\Rz^3$ and are usually combined 
into vectors $(p,q)$ with values in the phase space $\Rz^{6N}$. 
Their value at a given time $t_0$, togegther with the Hamiltonian 
$H(p,q)$, a twice continuously differentiable function of the phase 
space variables, uniquely determine, through the ordinary differential 
equations
\lbeq{e.HamiltonDyn}
\frac{d}{dt} q(t)=\frac{\partial}{\partial p} H(p,q),~~~
\frac{d}{dt} p(t)=-\frac{\partial}{\partial q} H(p,q),
\eeq
their value $p(t)$ and $q(t)$ at any time in the past or future, 

For a system of free particles, these phase space variables are in 
principle measurable to arbitrary accuracy by probing their motion with 
appropriate measurement devices. For a system of interacting particles, 
measurement accuracy is restricted to an accuracy depending on how fast 
the phase space variables change with time. Classical observables are
all sufficiently nice complex-valued functions $f=f(p,q)$ of the phase 
space variables, and include the Hamiltonian $H$, the observable 
describing the total energy of the system. Classical observables change 
in time according to the equations 
\lbeq{e.LieDyn}
\frac{d}{dt} f=H\lp f
\eeq
with the \bfi{Lie product} (the negative Poisson bracket) 
\lbeq{e.LieClass}
f\lp g:=\partial_p f \cdot \partial_q g -\partial_p g\cdot \partial_q f.
\eeq
The Lie product is antisymmetric. It satisfies the Jacobi identity and 
the Leibniz rule, hence turns $C^\infty(\Rz^{6N})$ into a commutative 
Poisson algebra.
Applied to the phase space variables themselves (wich are simple linear 
functions of $p$ and $q$), we recover \gzit{e.HamiltonDyn} and find for 
their Lie product the well-known canonical relations 
\[
p_{ka}\lp q_{ka}=1,~~~ q_{ka}\lp p_{ka}=-1~~~(k=1,\ldots,N,~a=1,2,3),
\]
and vanishing other Lie products, defining the Heisenberg Lie algebra 
$h(3N)$. The Poisson algebra $C^\infty(\Rz^{6N})$ is canonically 
associated to this Heisenberg algebra as its Lie--Poisson algebra.
For details see, e.g., \sca{Neumaier \& Westra} \cite[Chapter 12]{NeuW}.

Many other classical conservative systems have a formulation in terms 
of commutative Poisson algebras constructed from phase spaces forming 
Poisson manifolds; see, e.g., \sca{Marsden \& Ratiu} \cite{MarR}. 
Poisson algebras form an even more comprehensive framework than 
phase space manifolds. For example, indistinguishable particles cannot 
be modelled in terms of observable phase space variables since the only 
observables are the permutation invariant functions in 
$C^\infty(\Rz^{6N})$. However, these form again a commutative Poisson 
algebra. Based on this, classical equilibrium statistical mechanics and 
equilibrium thermodynamics can be developped in a purely algebraic way; 
see \sca{Neumaier \& Westra} \cite[Part II]{NeuW}.

{\bf Conservative quantum mechanics.}
We now show that conservative quantum mechanics can also be cast into 
the framework of commutative Poisson algebras. As we have seen, quantum 
systems can be  modelled in terms of their normalized density operators 
$\rho$. For a stationary quantum system, the components of the density 
operator in a fixed basis (and from these by computation all nice 
functions of the density operator) are in principle measurable to 
arbitrary accuracy through quantum tomography, by probing the system 
with appropriate measurement devices. For a nonstationary quantum 
system, measurement accuracy is restricted to an accuracy depending on 
how fast the density operator changes with time.

This close analogy to the classical situation just discussed suggests 
that we may treat the components of the normalized density operator in 
a fixed basis as the phase space variables of a quantum system.
Following this analogy, the 
\bfi{quantum phase space} consists of all Hermitian positive 
semidefinite operators $\rho$ of trace $1$, and the quantum observables 
are all sufficiently nice complex-valued functions $f=f(\rho)$ of the 
phase space variables. In particular, this includes the linear 
functionals
\[
\<A\>=\tr \rho A.
\]
Thus all quantum values and all sufficiently nice functions of quantum 
values are observable. The commutator Lie product for linear operators
\lbeq{e.LieOp}
A\lp B:=\iota(AB-BA),~~~\iota:=\frac{i}{\hbar}
\eeq
defines a Lie algebra whose Lie--Poisson algebra is the algebra of all 
infinitely often differentiable functions$f=f(\rho)$  of a compact 
operator $\rho\in\Lin\Hz$, with induced Lie product 
\lbeq{e.LieQuant}
f\lp g:=\tr\rho(\partial_\rho f \lp \partial_\rho g).
\eeq
If we restrict the arguments $\rho$ to phase space variables, i.e., 
Hermitian positive semidefinite operators $\rho$ of trace $1$, we 
obtain a commutative Poisson algebra of quantum observables, which 
equips our quantum phase space with the structure of a Poisson manifold.
The Poisson structure of the space of normalized density operators 
justifies treating this space as quantum phase space. This was first 
recognized by \sca{Man'ko} et al. \cite{ManMSZ};
symplectic reductions to coadjoint orbits are discussed, e.g., in 
\sca{Requist} \cite{Req} and \sca{Bonet-Luz \& Tronci} \cite{BonT},
This is quite different from the far more common approach, pioneered by
\sca{Strocchi} \cite{Stro}, to treat the projective space of wave 
functions up to phase as a symplectic quantum phase space manifold, 
which covers only pure states.

\subsection{The Ehrenfest picture}\label{s.Ehren}

\nopagebreak
\hfill\parbox[t]{10.8cm}{\footnotesize

{\em Es ist w\"unschenswert, die folgende Frage m\"oglichst elementar 
beantworten zu k\"onnen: Welcher R\"uckblick ergibt sich vom Standpunkt
der Quantenmechanik auf die Newtonschen Grundgleichungen der klassischen
Mechanik?
\\
(It is desirable to be able to answer the following question in 
the most elementary way: Which hindsight do we get from the point of 
view of quantum mechanics on Newton's basic equations of classical 
mechanics?)}

\hfill Paul Ehrenfest, 1927 \cite[p.455]{Ehr}
}

\bigskip

The Lie product \gzit{e.LieQuant} induces a Lie product
\lbeq{e.Lie}
\<A\> \lp \<B\> := \<A\lp B\>
\eeq
for the quantum values. The quantum Hamiltonian $H$ is an operator, 
hence not a quantum observable in the present sense. However, its 
quantum value
\[
\ol H:=\<H\>
\]
qualifies as a quantum observable, and describes the total 
energy of the system. Quantum observables change in time according to 
the equations 
\lbeq{e.LieDynQ}
\frac{d}{dt} f=\ol H\lp f
\eeq
Applied to time-dependent quantum values 
\[
\<A\>_t:=\tr \rho(t) A,
\]
we find 
\lbeq{e.Ehrenfest}
\frac{d}{dt} \<A\>_t=\<H\lp A\>_t
\eeq
as differential equation for arbitrary quantum values.
We call \gzit{e.Ehrenfest} the \bfi{Ehrenfest equation} since 
\sca{Ehrenfest} \cite{Ehr} found in 1927 the special case of this 
equation where $A$ is a quantum position or momentum 
variable, and 
\[
H=\D\frac{p^2}{2m}+V(q)
\]
is the sum of kinetic and potential energy. Indeed, due to the 
canonical commutation rules, we have in this case
\lbeq{e.EhrenfestOrig}
\frac{d}{dt} \<q\>_t=\<H\lp q\>_t = \frac{\<p\>_t}{m},~~~
\frac{d}{dt} \<p\>_t=\<H\lp p\>_t = \<-\nabla V(q)\>_t.
\eeq
The Ehrenfest equation implies that the collection of all quantum values
at a given time $t_0$ uniquely determines all quantum values at any 
time in the past or future. In the literature, this dynamics for 
quantum values is referred to as the \bfi{Ehrenfest picture}
(\sca{Man'ko} et al. \cite{ManMSZ}). 
The Ehrenfest picture is more thoroughly discussed in \sca{Neumaier} 
\cite[Section 2.2]{Neu.CQP}.

By specializing $A$ to a basis of the space of operators, we see that 
the Ehrenfest equation is equivalent to the conservative quantum 
Liouville equation
\lbeq{e.Lio}
    i \hbar \dot\rho(t) = [H,\rho(t)].
\eeq
Thus the phase space setting for the quantum dynamics is equivalent 
to the traditional setting for conservative quantum dynamics.

\subsection{Pure state idealization and the Schr\"odinger equation}
\label{s.pure}

\nopagebreak
\hfill\parbox[t]{10.8cm}{\footnotesize

{\em The basis of the mathematical formalism of quantum mechanics lies 
in the proposition that the state of a system can be described by a
definite (in general complex) function $\Psi(q)$ of the coordinates.
[...] The function $\Psi$ is called the wave function of the system.}

\hfill Landau and Lifschitz, 1977 \cite[p.6]{LL.3}
}

\bigskip

A filter with transmission matrix $T$
transforms a beam in the pure state $\psi$ into a beam in the pure
state $\psi'=T\psi$. If $\psi(t)$ denotes the pure state at time $t$,
the same slicing scenario as above gives
$\psi(t+\Delta t) = T(t) \psi(t)$. Therefore
\[
\frac{d}{dt}\psi(t)
= \frac{\psi(t+\Delta t)-\psi(t)}{\Delta t}+O(\Delta t)
= K(t) \psi(t)+O(\Delta t).
\]
In the continuum limit $\Delta t\to 0$, we obtain the differential
equation
\[
\frac{d}{dt}\psi(t)= K(t) \psi(t).
\]
In the special case of a lossless medium, this becomes the
\bfi{time-dependent Schr\"odinger equation}
\lbeq{e.tSchr}
i\hbar\frac{d}{dt}\psi(t) = H(t) \psi(t).
\eeq
Pure states are experimentally relevant only if they can be prepared. 
This is the case if they come from the ground state of a Hamiltonian 
for which the ground state is nondegenerate with energy $E_0$, and the 
first excited state with energy $E_1$ has a large energy gap 
$\Delta:=E_1-E_0$. 

We assume for simplicity that $H$ has a discrete spectrum. In an 
orthonormal basis of eigenstates $\phi_k$, functions $f(H)$ of the 
Hamiltonian $H$ are defined by
\[
   f(H) = \sum_k f(E_k) \phi_k \phi_k^*
\]
whenever the function $f$ is defined on the spectrum. 

Dissipation through contact with an environment at 
\bfi{absolute temperature} $T$ brings and keeps the system close to 
equilibrium. The equilibrium density is that of the 
\bfi{canonical ensemble},
\[
\rho_T = Z(T)^{-1} e^{(E_0-H)/\kbar T}
= Z(T)^{-1}\sum_k e^{(E_0-E_k)/T}\phi_k\phi_k^*;
\]
here $\kbar>0$ is the \bfi{Boltzmann constant}, and the so-called 
\bfi{partition function}
\[
   Z(T) := \sum_k e^{(E_0-E_k)/\kbar T},
\]
is determined by the requirement that the trace equals $1$. For low
enough temperatures, namely when the energy gap $\Delta$ exceeds a 
small integer multiple of $E^* := \kbar T$, all terms $ e^{(E_0-E_k)/T}$
with $k>0$ are of the order $O(e^{-\Delta/\kbar T})$ or smaller. 
Therefore 
\[
\rho_T = \phi_1 \phi_1^*+O(e^{-\Delta/\kbar T}).
\]
This implies that for sufficiently low temperatures, the equilibrium 
state is approximately pure.

States of sufficiently simple systems (i.e., those with only a few 
experimentally accessible energy levels) can therefore be prepared in 
nearly pure states by realizing a source whose dynamics is governed by 
a Hamiltonian in which the first excited state has a much larger energy 
than the ground state. Thermal dissipation then brings the system into 
equilibrium, and as seen above, the resulting equilibrium state is 
nearly pure. 

Low lying excited states for which a selection rule 
suppresses the transition to a lower energy state can be made nearly 
pure in the same way. 

If the ground state of a quantum system is finitely degenerate, 
thermal sources produce mixed states in the subspace of minimal energy, 
and one needs more care to prepare a pure state. Now a filter is needed 
that temporarily creates by an interaction a modified Hamiltonian that 
removes the degeneracy. In the purification of electron beams through 
idealized Stern--Gerlach magnets (\sca{Busch} et al.\cite{BusGL}), 
the twofold spin degeneracy of electrons is removed by utilizing the 
different motion of different spin states in magnetic fields. This acts 
as a beam splitter, and upon projection to a single beam, the state 
collapses to a pure state. The twofold helicity degeneracy of light is 
removed through a similar process in the purification of light through 
beam-splitting polarization filters, while in absorptive polarization 
filters, the degeneracy is removed through lossy thermal convergence to 
the ground state of the modified Hamiltonian.

\subsection{Quantum dynamics in mixing linear media}\label{ss.mixed}

\nopagebreak
\hfill\parbox[t]{10.8cm}{\footnotesize

{\em It seems that the only possibility of introducing an irreversible
behaviour in a finite system is to avoid the unitary time evolution
altogether by considering non-Hamiltonian systems.}

\hfill G\"oran Lindblad, 1976 \cite[p.119]{Lin}
}

\bigskip

\nopagebreak
\hfill\parbox[t]{10.8cm}{\footnotesize

{\em It is fairly clear that although these measurements can be
predicted by quantum theory they do not fall within the usual
description of measurement theory.}

\hfill Brian Davies, 1969 \cite[p.278]{Dav69}
}

\bigskip

We can also find the correct dynamical laws under assumptions where the
Schr\"odinger equation or the von Neumann equation is inappropriate.
It is well-known that in practice most physical processes are
\bfi{dissipative}. This means that depending on the situation, energy
decreases or entropy increases, due to interaction with the medium
through which a quantum system passes. This cannot happen for processes
described by the Schr\"odinger equation or the von Neumann equation,
hence a more complicated description is needed.

We may repeat the derivation of the quantum Liouville equation in the
more general situation of a mixing linear medium. We assume again that
the medium is modeled by means of many slices consisting of very thin
filters. But these are now mixing, hence described by a multi-term
operator sum expansion. Only one of the terms can be close to the 
identity; the others are small. The assumption that leads to a 
sufficiently general dynamics for the state is that 
\[
\rho(t+\Delta t)
=T(t)\rho(t)T(t)^*+\sum_\ell \gamma_\ell T_\ell(t)\rho(t)T_\ell(t)^*,
\]
with a primary transmission operator $T(t)$ of the form \gzit{e.TK},
secondary transmission operators $T_\ell(t)$ of the form
\[
T_\ell(t)=\Delta t^{1/2}L_\ell(t)+O(\Delta t),
\] 
and nonnegative constants $\gamma_\ell$. 
(As in \sca{Jagadish \& Petruccione} \cite[Section 7]{JagP}, whose 
argument we follow, these could be absorbed into the $L_\ell(t)$ by 
rescaling the latter; but it is more convenient in the applications to 
have them separately.) Then 
\[
\frac{d}{dt}\rho(t)
= \frac{\rho(t+\Delta t)-\rho(t)}{\Delta t}+O(\Delta t)
= K(t)\rho(t)+\rho(t)K(t)^*
+\sum_\ell \gamma_\ell L_\ell(t)\rho(t)L_\ell(t)^*+O(\Delta t^{1/2}).
\]
Suppressing the time argument, and taking the continuum limit 
$\Delta t\to 0$, we obtain the \bfi{quantum master equation}
\lbeq{e.qMaster}
\frac{d}{dt}\rho 
= K\rho+\rho K^*+\sum_\ell \gamma_\ell L_\ell\rho L_\ell^*.
\eeq
In the lossless case, the intensity $I=\tr \rho(t)$ remains constant, 
hence
\[
0=\tr \frac{d}{dt}\rho 
= \tr K\rho+\tr \rho K^*+\sum_\ell \gamma_\ell \tr L_\ell\rho L_\ell^*
=\tr\Big(K+K^*+\sum_\ell \gamma_\ell L_\ell^*L_\ell\Big)\rho.
\]
This must hold for all positive semidefinite $\rho$. By linearity, 
it must therefore hold for all $\rho$, giving
\[
K+K^*+\sum_\ell \gamma_\ell L_\ell^*L_\ell=0.
\]
We use again the decomposition \gzit{e.KH}, and find that 
\[
\frac{2}{\hbar}V=-K-K^*=\sum_\ell \gamma_\ell L_\ell^*L_\ell,
\]
\[
K=-\frac{i}{\hbar}H-\frac{1}{\hbar}V
=-\frac{i}{\hbar}H+\half \sum_\ell \gamma_\ell L_\ell^*L_\ell.
\]
Inserting this into the quantum master equation \gzit{e.qMaster}
results in the \bfi{Lindblad equation}
\lbeq{e.Lindblad}
\frac{d}{dt}\rho = \frac{i}{\hbar}[\rho,H] 
 +\sum_\ell \gamma_\ell\Big(
  L_\ell\rho L_\ell^*-\half L_\ell^*L_\ell\rho-\half\rho L_\ell^*L_\ell
                       \Big)
\eeq
(\sca{Lindblad} \cite{Lin}, \sca{Gorini} et al. \cite{GorKS}), the 
standard equation for describing dissipative, memoryless quantum
 systems.

\subsection{Nonlinear dissipative quantum dynamics}\label{s.nonlDyn}

\nopagebreak
\hfill\parbox[t]{10.8cm}{\footnotesize

{\em In quantum theory of many-body systems one expects nonlinear
quantum evolution equations. [...] The canonical forms of the
nonlinear generators are derived and discussed for classes of open 
quantum mean-field models.}

\hfill Alicki \& Messer, 1983 \cite[p.300]{AliM}
}

\bigskip

\nopagebreak
\hfill\parbox[t]{10.8cm}{\footnotesize

{\em Often quantum systems are not isolated and interactions with their 
environments must be taken into account. In such open quantum systems 
these environmental interactions can lead to decoherence and 
dissipation, which have a marked influence on the properties of the 
quantum system. In many instances the environment is well-approximated 
by classical mechanics, so that one is led to consider the dynamics of 
open quantum-classical systems.}

\hfill Raymond Kapral, 2015 \cite[Abstract]{Kap}
}

\bigskip

The situation is quite different for the passage through \bfi{nonlinear
media} characterized by multiple scattering, which destroys the 
linearity. The thin slices would now have to be described by nonlinear 
filters, for which the theory developed here is inadequate. Instead one
needs to start from quantum field theory and derive the appropriate 
nonlinear properties by perturbation theory; see, e.g., \sca{Akhmanov} 
et al. \cite{AkhCDK} for the case of nonlinear quantum optics.

Because of this nonlinearity, measurements through nonlinear media
no longer follow the detector response principle (DRP) that we assumed 
at the outset for the present investigations. 
In place of a linear Lindblad dynamics for the density operator, one 
obtains a dynamics described by nonlinear quantum dynamical semigroups 
(\sca{Alicki \& Messer} \cite{AliM}). A comprehensive view of the 
resulting general theory of open quantum systems is given in 
\sca{Breuer \& Petruccione} \cite{BreP}.

Nonlinearities are easily incorporated in conservative (nonmixing) 
quantum phase space dynamics by choosing a Hamiltonian that is a 
nonlinear function of quantum values. In the mixing case, one needs 
to augment the conservative phase space approach as in the classical 
case, where an appropriate dissipative bracket allows a phase space 
description of complex dissipative fluids; see, e.g., 
\sca{Beris \& Edwards} \cite{BerE} and \sca{\"Ottinger} \cite{Oet}.
\sca{Taj \& \"Ottinger} \cite{TajO} discuss a generic approach to 
nonlinear dissipative quantum dynamics via the Markov approximation.

A \bfi{semiclassical description} of a quantum system is appropriate if 
the system can be described sufficiently well in terms of an effective 
dynamics of a limited number of local quantum values. In terms of the
Poisson algebra, the projection to a suitable manifold of quantum 
values of relevant quantities produces the semiclassical approximation. 
The local equilibrium description of a macroscopic system is 
semicalssical in this sense.

In most cases of interest, the reduced dynamics can again be described 
in a phase space setting. If the reduced Poisson algebra is a tensor 
product of a quantum Poisson algebra with Lie product \gzit{e.LieQuant} 
and a classical Poisson algebra with Lie product \gzit{e.LieClass}, the
reduced system is commonly called \bfi{quantum-classical}. In essence,
a quantum-classical description treats the fast degrees of freedom by 
quantum mechanics while the slower ones are approximately handled as 
classical quantities. A prototypical example is the Born--Oppenheimer 
approximation on which much of chemistry is based.
The Ehrenfest picture discussed in Subsection \ref{s.phase} plays a 
significant role in these nonlinear quantum-classical approachesn to 
chemical and physical systems. 
For a thorough discussion of quantum-classical dynamics see, e.g., 
\sca{Kapral \& Ciccotti} \cite{KapC}, \sca{Marx \& Hutter} \cite{MarH}, 
and \sca{Kapral} \cite{Kap}.

\newpage
\section{Spectroscopy and system identification}\label{s.systemId}

In this section, we discuss the implications of our developments for
applications in spectroscopy and high precision measurements.

An analysis of the oscillations of quantum values of states satisfying
the Schr\"odinger equation produces the Rydberg--Ritz combination 
principle underlying spectroscopy, whose prediction marked the onset
of modern quantum mechanics. Today, spectroscopy is a huge field 
providing high quality experimental data for a host of physical 
and chemical systems. 

The accuracy with which resonance measurements can determine 
frequencies is the origin of applications to high precision 
measurements. We study in some detail single ion spectroscopy with a 
Penning trap, as used for the determination of the gyromagnetic ratio 
of the electron to 12 decimal digits of accuracy.

Together with a discussion of the relation between classical stochastic 
processes and individual classical systems, this provides insight into 
what quantum mechanics can say about individual quantum systems.

\subsection{Energy spectrum and quantum fluctuations}\label{ss.Espec}

\nopagebreak
\hfill\parbox[t]{10.8cm}{\footnotesize

{\em Wohl aber kann dem Elektron auch in der Quantentheorie eine 
Aus\-strahlung zugeordnet werden; diese Strahlung wird beschrieben 
erstens durch die Frequenzen, die als Funktionen zweier Variablen 
auftreten, quantentheoretisch in der Gestalt:\\
(However, even in quantum theory it is possible to 
ascribe to an electron the emission of radiation. In order to 
characterize this radiation we first need the frequencies which appear 
as functions of two variables. In quantum theory these functions are of 
the form)\\
\hspace*{2cm} $\nu(n,n-a)=\frac{1}{j}[W(n)-W(n-a)]$.}

\hfill Werner Heisenberg, 1925 \cite[p.881]{Hei1925}
}

\bigskip

We return to the derivation of the Schr\"odinger equation \gzit{e.tSchr}
in Section \ref{s.pure}.
In the special case of passage through a homogeneous medium, the filters
are independent of their location, hence $T(t)$ is time-independent.
This implies that the Hamiltonian $H$ is also time-independent.
In particular, the Schr\"odinger equation $i\hbar\dot\psi=H\psi$ for 
pure states in a lossless medium can be solved by a standard 
exponential ansatz. This gives
\lbeq{e.psi}
\psi(t)=\sum_k e^{-i t E_k/\hbar}\psi_k
\eeq
as a superposition of \bfi{eigenvectors} $\psi_k$ of $H$ satisfying the
\bfi{time-independent Schr\"o\-dinger equation}
\lbeq{e.ESchr}
H\psi_k = E_k \psi_k
\eeq
with appropriate boundary conditions. All real \bfi{eigenvalues} $E_k$ 
together form the \bfi{energy spectrum} of $H$. 

In quantum information theory, only finite-dimensional state spaces 
figure. This is possible since in this context the setting is such that 
for its building blocks, only a few states are energetically accessible.
Thus the state space is a tensor product of very low-dimensional 
state spaces (most typical of qubits). 
If the state-space is finite-dimensional, the spectrum is finite.

But most quantum systems need for their description an 
infinite-dimensional state space. This is already the case for an 
optical beam. While a beam of classical monochromatic light oscillates 
with a fixed frequency, the associated quantum beam oscillates with 
infinitely many overtones of this classical frequency, though in dim 
light most of them are hardly excited. 
After the qubit (which ignores in its optical realization the 
excited states of a photon), the quantum oscillator is the second 
simplest quantum system of fundamental importance. The operators on 
this state space are generally expressed in terms of basic operators 
satisfying canonical commutation relations. This points to the
fundamental importance of Lie algebras in quantum physics. See the books
by \sca{Neumaier \& Westra} \cite{NeuW} and \sca{Woit} \cite{Woi} for
Lie algebras and Lie groups in general quantum physics, and more 
specifically \sca{Leonhardt \& Neumaier} \cite{LeoN} for Lie algebra 
techniques for the design of quantum optical circuits.
 
In general, each classical degree of freedom of a quantum system 
determines a corresponding quantum oscillator with an associated 
infinite-dimensional state space. The observation by
\sca{Dirac} \cite{Dirac1925} that the Poisson bracket is the
classical limit of the scaled commutator implies that coupled quantum
oscillators are described by a tensor product of oscillator state 
spaces. This is easily generalized to arbitrary composite systems, and 
is relevant even in quantum information theory, which exploits the 
exponential complexity of tensor products of many qubits. 

If the state space is infinite-dimensional, the energy spectrum is
always bounded from below, but usually not bounded from above. The
number of eigenvalues is usually infinite and the sum in \gzit{e.psi} 
has infinitely many terms. In the common case where part or all of the 
energy spectrum is continuous, the sum in \gzit{e.psi} must be 
augmented by or replaced by an integral to make sense. 

In a pure time-dependent state $\psi(t)$, the quantum values take the 
form
\[
\<A\>_t=\psi(t)^*A\psi(t)
=\sum_{j,k} e^{-i t (E_j-E_k)/\hbar}\psi_j^*A\psi_k.
\]
For mixed states, a similar expression 
\[
\<A\>_t=\sum_{j,k} a_{jk}e^{-i t (E_j-E_k)/\hbar}
\]
can be derived from the von Neumann equations, using the spectral 
theorem. This shows that every quantum system oscillates with angular 
frequencies
\[
\omega_{jk}=|E_j-E_k|/\hbar
\]
proportional to the energy differences.

An unbounded energy spectrum implies the presence of arbitrarily high 
frequencies, forming \bfi{quantum fluctuations} observable statistically
as noise (even with the human eye, \sca{Stiles} \cite{Stil}). Lower 
frequency oscillations can be probed by spectroscopy, and produce 
absorption or emission lines with frequencies 
$\nu_{jk}=\omega_{jk}/2\pi$. With $h:=2\pi\hbar$ we find the 
\bfi{Rydberg--Ritz combination principle} that the observed spectral
lines are given by differences
\lbeq{e.RR}
\nu_{jk}=|\nu_j-\nu_k|,
\eeq
of frequencies $\nu_j$ determined from Planck's relation $E_k=h\nu_k$
between energy and frequency. This relation was the starting point of 
Heisenberg's matrix mechanics (\sca{Heisenberg} \cite{Hei1925}), which 
lead to modern quantum physics.

\subsection{The relevance of eigenvalues}\label{s.eig}

\nopagebreak
\hfill\parbox[t]{10.8cm}{\footnotesize

{\em Wir werden also an dem Bohrschen Bilde festhalten, dass ein
atomares System stets nur in einem stationaren Zustand ist. Wir k\"onnen
sogar annehmen, dass wir gelegentlich durch Beobachtung in einem 
Augenblick mit Sicherheit wissen, dass das Atom im $n$-ten Zustand ist
(Fussnote: Z.B. wissen wir, dass in einem sich selbst \"uberlassenen 
Gase bei hinreichend tiefer Temperatur alle Atome mit verschwindender 
Ausnahme im Normalzustand sind); 
im allgemeinen aber werden wir in einem Augenblick nur wissen, dass auf
Grund der Vorgeschichte und der bestehenden physikalischen Bedingungen
eine gewisse Wahrscheinlichkeit daf\"ur besteht, dass das Atom im 
$n$-ten Zustand ist.
\\
(Thus we shall stick to Bohr's picture that an atomic system always is 
in a stationary state only. We may even assume that through observation 
we know occasionally with certainty in some moment that the atom is in 
the $n$th state
(Footnote: E.g., we know that in a gas left to itself at sufficiently 
low temperature, all atoms with vanishingly few exceptions are in the 
ground state);
but in general we will know in some moment only that due to the prior 
history and the prevailing physical conditions there is a certain 
probability that the atom is in the $n$th state.)
}

\hfill Max Born, 1927 \cite[p.170]{Bor1927}
}

\bigskip

Though in the present setting eigenvalues no longer \at{} play the 
fundamental role they traditionally have in quantum measurement theory, 
they continue to be essential in several different areas -- to analyze 
time-independent quantum dynamics, for spectroscopy, for equilibrium 
thermodynamics, and for high precision measurements. 

In particular, the eigenvalues of the Hamiltonian are most relevant, 
since they specify the possible energy levels of a stationary quantum 
system. This is relevant not only on the theoretical level, where a 
spectral representation of the Hamiltonian allows the explicit solution 
of the quantum Liouville equation and the time-dependent Schr\"odinger 
equation. 

The eigenvalues of the Hamiltonian are also important in many contexts 
with experimentally relevant consequences: We have seen the 
Rydberg--Ritz formula \gzit{e.RR} from spectroscopy, which can access 
a large number of energy levels. 
For the reasons discussed in Subsection \ref{s.pure}, in much of 
quantum chemistry, only the electronic ground state is considered 
relevant. Its energy, parameterized by the nuclear 
coordinates, determines the potential energy surface whose properties 
are sufficient to describe the shape of all molecules and their 
chemical reactions. In many other cases, only few energy levels are 
experimentally accessible, in 
which case the quantum system can be modeled as a few level system, 
drastically simplifying the task of quantum state tomography.

The energy spectrum is also prominent in thermal equilibrium physics 
since it determines the partition function from which all other 
thermodynamic properties can be derived. For a treatment of thermal 
equilibrium quantum physics based on quantum expectations without 
making reference to statistical assumptions -- see the online textbook 
by \sca{Neumaier \& Westra} \cite{NeuW}.

For high precision measurements, one must be able to prepare states 
with tiny quantum uncertainties, so that their measurement can be 
performed with very high accuracy. An exampole is the 
\bfi{quantum Hall effect} and its variations, relevant for preparing 
calibration states for metrology standards (\sca{Lindeley} (\cite{Lin}, 
\sca{Kaneko} et al. \cite{KanNO}), 

The possibility of achieving tiny uncertainties is 
linked to the spectrum. Let $X$ be a scalar or vector quantity whose
$m$ components are defined on a common dense domain.
We call the set of $\xi\in\Cz^m$ for which no linear operator $R(\xi)$ 
exists such that $R(\xi)|X-\xi|^2$ is the identity the \bfi{spectrum} 
$\spec X$\index{$\spec X$, spectrum} of $X$. The spectrum
is always a closed set, but it may be empty, as, e.g., for the vector
formed by position and momentum of a particle. The following result 
implies that the preparation of states such that $X$ has arbitrarily 
small uncertainty is possible precisely when $\ol X$ belongs to the 
spectrum of $X$. 

\begin{thm}
$\xi\in\Cz^m$ belongs to the spectrum of a scalar or vector quantity $X$
with $m$ components iff states exist that have arbitrarily small 
positive $\<|X-\xi|^2\>$.
\end{thm}

\bepf
By definition, the operator $B:=|X-\xi|^2$ is Hermitian and positive 
semidefinite, hence essentially self-adjoint with nonnegative spectrum 
$\Sigma$. Note that $\xi$ belongs to the spectrum of $X$ iff 
$0\in\Sigma$. We consider the spectral projector $P(s)$ to the invariant
subspace corresponding to the spectrum in $[0,s]$. If $0\not\in\Sigma$
then $P(s)=0$ for some $s>0$, and $\<|X-\xi|^2\>=\<B\>\ge s$ cannot be 
arbitrarily small. On the other hand, if $0\in\Sigma$ then, for all 
$s>0$, the projector $P(s)$ is nonzero. Thus we can find vectors 
$\phi(s)$ such that $\psi(s):=P(s)\phi(s)$ is nonzero. By scaling 
$\phi(s)$ appropriately, we can ensure that $\psi(s)$ has norm one. 
Then in the corresponding pure state, $\<|X-\xi|^2\>=\<B\>\le s$. 
Since this works for any $s>0$, the claim follows.
\epf

\subsection{Spectroscopy and high precision measurements}
\label{s.specHigh}

\nopagebreak
\hfill\parbox[t]{10.8cm}{\footnotesize

{\em Never in the history of science has a subject sprung so suddenly 
from a state of complete obscurity and unintelligibility to a condition 
of full illumination and predictability as has the field of 
spectroscopy since the year 1913. [...]\\
It has now become possible to make a nearer approach than heretofore to 
presenting a physical interpretation of a group of remarkable 
spectroscopic rules developed, largely empirically, within the past two 
years by Russell, Heisenberg, Pauli and Hund, and embracing in an 
altogether remarkable way, most of the facts of spectroscopy known up 
to the present. The success with which these empirical rules describe 
the facts of spectroscopy is little less than magical. These rules 
naturally all start with, and grow out of, the fundamental assumption 
underlying all quantum theory, namely, that all periodic motions must 
be quantized.}

\hfill Robert A. Millikan, 1927 \cite[pp.211,225]{Mil}
}

\bigskip

Physical systems are described in theoretical physics mathematically,
by means of models that quantitatively reflect the essence of the 
systems. These models are formulated within a given theoretical 
framework, which allows one to fix the model assumptions. Typically 
one specifies the phase space (including a reference coordinate system 
for fixing the collections of the observables), the causal rules 
(Galilean in nonrelativistic physics, Minkowski in special relativity, 
local Minkowski in general relativity), and the parameterized 
Hamiltonian in conservative mechanics, more elaborate equations of 
motion in dissipative mechanics.

A physical model is a template which contains parameters that are 
usually not fixed but to be determined by \bfi{system identification}, 
the process of matching key properties of the model with experimental 
results. 

For example, a classical quartic oscillator is a model defined 
by a Hamiltonian 
\[
H=p^2/2m+kq^2/2 +gq^2/4
\] 
together with the specification of classical Hamiltonian dynamics for 
the conjugate position and momentum variables $q$ and $p$. The 
coefficients in the Hamiltonian, the mass $m$ and the coupling constant 
$g$ are the parameters. The claim that, within a given accuracy, a 
particular oscillator is well described by this model can be decided 
by making experiments on the oscillator and comparing it with the
predictions of the model.
In particular, the spectroscopic information gathered by experimenters
can be used to infer and partially reconstruct the interaction 
potentials of two atoms close to their equilibrium position; see, e.g.,
\sca{Varshni} \cite{Var}.

If a physical model is correct to some accuracy, there will be a 
parameter choice matching the predictions to this accuracy. The best 
set of parameters will usually be determined from experiments using 
standard statistical parameter estimation techniques. In the limit of 
arbitrarily many and arbitrarily accurate measurements, the statistical 
error can be made arbitrarily small. Thus the parameters are 
operationally quantifiable, independent of an observer. 

In system identification one has no interest in the determination of a 
particular state of the system investigated, only in its time invariant 
properties -- the parameters that go into the description of the state 
space and the system dynamics. Thus no accuracy restrictions apply, 
unlike those discussed in Subsection \ref{s.nonstat} for the accuracy 
with which nonstationary states can be obtained. This means that 
parameters are measurable to in principle unlimited accuracy.

This is the basis of high precision measurement schemes enabled by
quantum spectroscopy. 
Since isolated frequency differences \gzit{e.RR} of quantum systems can 
be determined with extremely high accuracy through resonance phenomena, 
they are used for metrology purposes. Measuring accuracies are so high 
that spectroscopic standards figure prominently in the basic definitions
of the international system of units \cite{SIunits}: A \bfi{second} 
is defined -- since 1964 -- in terms of the hyperfine transition 
frequency of an caesium 133 atom (later made more precise to be 
isolated at rest at a temperature of 0 K, unperturbed by an electric 
field).

\subsection{Example: The gyromagnetic ratio of the electron}
\label{s.gyro}

\nopagebreak
\hfill\parbox[t]{10.8cm}{\footnotesize

{\em One of the experiments now possible is to observe the 'quantum
jumps' to and from a metastable state in a single atom by monitoring of 
the resonance fluorescence of a strong transition in which at least one 
of the states is coupled to the metastable state. When the atomic 
electron moves to the metastable state, the fluorescence from the 
strongly driven transition vanishes. When the electron drops back into 
the strongly driven transition, the fluorescence abruptly returns. 
Thus the strong  transition fluorescence is a monitor of the quantum 
state of the atom.}

\hfill Bergquist, Hulet, Itano and Wineland, 1986 \cite[p.1699]{BerHIW}
}

\bigskip

\nopagebreak
\hfill\parbox[t]{10.8cm}{\footnotesize

{\em 
The dream of the spectroscopist is to be able to study a single atom or 
ion under constant conditions for a long period of time. In recent 
years, this dream has to a large extent been realized. The basic tool 
is here the {\em ion trap}, which was introduced in the 1950s by 
another of this year's laureates, Wolfgang Paul in Bonn. His technique 
was further refined by the third laureate, Hans Dehmelt, and his 
co-workers in Seattle into what is now known as ion-trap spectroscopy.

Dehmelt and his associates used this spectroscopy primarily for 
studying electrons, and in 1973 they succeeded for the first time in 
observing a single electron in an ion trap, and in confining it 
there for weeks and months. One property of the electron, its magnetic 
moment, was measured to 12 digits, 11 of which have later been 
verified theoretically. This represents a most stringent test of the 
atomic theory known as quantum electrodynamics (QED).
%
}

\hfill Award ceremony speech for the Nobel prize in physics 1989 
\cite{Nobel1989s}
}

\bigskip

\nopagebreak
\hfill\parbox[t]{10.8cm}{\footnotesize

{\em Quantum mechanics has often been classified as a merely
statistical ensemble theory, with not much bearings on the individual
members of the ensembles. Yet there is an increasing variety of
experiments exhibiting individual quantum processes which were
conceived, devised and explained on the basis of this very theory.
}

\hfill Busch and Lahti, 1996 \cite[p.5899]{BusL}
}

\bigskip

\nopagebreak
\hfill\parbox[t]{10.8cm}{\footnotesize

{\em In order to obtain some insight into the time evolution of possible
individual measurement records of $\<I^\out\>$ in the subsequent 
discussions, the quantum state diffusion (QSD) model of state reduction 
will be used.
This model of state reduction has been introduced as a general approach 
to continuous quantum measurement processes in which the time evolution 
of an individual quantum system, i.e., a  single member of a statistical
ensemble, is represented explicitly. In this respect the QSD model
transcends the framework of traditional quantum mechanics and its 
significance for the quantum-mechanical measurement problem still 
remains an open question.}

\hfill Rigo, Alber, Mota-Furtado and O'Mahony, 1998 \cite[p.482]{RigAM}
}

\bigskip

\nopagebreak
\hfill\parbox[t]{10.8cm}{\footnotesize

{\em For atomic and molecular spectroscopists it is clear that quantum 
jumps exist; this is part of their daily bread. Most spectroscopists 
are also aware that the time involved in a transition is finite but 
very short.
}

\hfill  de la Pe\~na, Cetto and Vald\'es-Hern\'andez, 2020 
\cite{dlPenCV}
}

\bigskip

Part of the Nobel prize in physics 1989 was given for the ion trap 
technique which has made it possible to study a single electron or a 
single ion with extreme precision. The experimental details of the 
Nobel prize winning experiment are in \sca{Van Dyck} et al. 
\cite{vDyckSD}. In more detail, \sca{Brown \& Gabrielse} \cite{BroG} 
discuss the theoretical and experimental aspects of the determination 
of the gyromagnetic ratio $g$ of the electron from measurements of two 
or three frequency differences in the spectrum of a \bfi{geonium atom} 
made of a single electron in a Penning trap, a nonstationary quantum 
source controlled by a photocurrent. They assign a time-dependent 
density operator $\rho(t)$ to the single electron at time $t$. There is 
no ensemble of systems to which the traditional probabilistic 
interpretation could be applied. However, the setting of the present 
paper, where we work throughout with single quantum systems, applies 
without problems.

The basic formula in \cite{BroG}  is (2.65) on p.243, which expresses 
$g/2$ as the ratio of the spin precession frequency and the cyclotron 
frequency. This formula is obtained by assuming the standard magnetic 
interaction term for a spin. 
Brown and Gabrielse get a highly accurate value of $g-2$ by measuring 
with very high accuracy both the cyclotron frequency and the difference 
of the spin precession frequency to it, or in a refinement discussed 
later closely related frequencies that are easier to prepare with low 
experimental error). These frequencies are found by resonant coupling 
of the ion trap to an external circuit described in Fig.9 on p.248 and
scanning the response to different frequencies to be able to observe a 
resonance curve (Fig.10), from which the value of the resonance 
frequency can be obtained by simple curve fitting.

Brown and Gabrielse first discuss (p.248f) the axial motion $z(t)$ of 
the electron in the ion trap coupled to the external circuit: 
{\em ''The oscillating charged particle induces alternating image 
charges in the electrodes, which in turn cause an oscillating current 
$I$ to flow through a resistor $R$.''} The amplitude of this current is 
measured as a function of the frequency of the driving force of the 
circuit. Brown and Gabrielse derive, using purely classical reasoning, 
the ordinary differential equation (3.9) -- a driven and damped harmonic
oscillator equation, later (p.255) refined by anharmonic terms -- for 
$z(t)$. Then they apply classical linear response theory (on p.251) to 
obtain from time correlations the spectral properties of the noise 
voltage in dependence on the circuit frequency.

On pp.269--271, Brown and Gabrielse derive, using the quantum dynamics 
of a driven spin in a magnetic field, the coupled ordinary differential 
equations (5.24ab) for the diagonal matrix elements $\rho_{++}(t)$ and
 $\rho_{--}(t)$ of the spin density operator. (Note that (5.24ab) 
contain spurious factors 4, as can be seen by comparing with the 
solution given in (5.26ab), but this does not effect the remaining 
analysis.) By (5.25), $\rho_{++}(t)$ and $\rho_{--}(t)$ sum to one and
 can be expressed in terms of the quantum expectations 
\[
S_z:=\<\sigma_z\>
\]
as 
\[
\rho_{++}(t)=\half(1+ S_z),~~~ 
\rho_{--}(t)=\half(1- S_z). 
\]
The spin dynamics (5.24ab) is therefore equivalent to the dynamics 
\[
\frac{d}{dt}S_z=-2RS_z
\]
for the quantum expectation $S_z$ of the spin in $z$ direction.
Brown and Gabrielse discuss on p.246 a quantum ensemble to justify their
classical decay equation (2.94) for radiation damping by equating 
quantum expectations with their classical analogue. But no attempt is 
made to justify treating the axial motion $z(t)$ of the electron as 
classical. Probably $z(t)$ must also be considered as a quantum 
expectation -- this time that of the $z$-component of the electron 
position. 

\sca{Rigo} et al. \cite{RigAM} give a somewhat more rigorous model for 
a single electron in a Penning trap by means of a Lindblad master 
equation. The dynamics of the density operator is represented in terms 
of differential equations for the quantum expectations of Pauli 
matrices for the spin and bosonic creation and annihilation operators 
$a_+$, $a_z$ for oscillator describing the cyclotron motion and the 
axial motion $z(t)$, respectively. 

In both \sca{Brown \& Gabrielse} \cite{BroG} and \sca{Rigo} et al. 
\cite{RigAM}, everything theoretical is deduced from these differential 
equations and variations that take into account relativistic 
corrections and refinements in the experimental procedure. The only 
measurements made are those of the currents produced in the external 
circuit. This is enough to obtain the two frequencies whose quotient 
gives $g-2$. In this way, the latter can be determined by experiment to 
extraordinary accuracy. The analysis requires the objectivity of the 
states and their dynamics, from which the gyromagnetic ratio is derived.

We conclude that the current analysis techniques for nonstationary 
single quantum systems exhibiting quantum jumps -- of which an electron 
in a Penning trap is just the most prominant example -- require that 
{\em each individual quantum system has at each time an objective, 
measurement-independent state represented by a time-dependent density 
operator.} 

While this flatly contradicts the traditional interpretations of quantum
mechanical states purely in terms of ensembles or knowledge, it is fully
consistent with the present tomography approach to quantum mechanics,
which ''transcends the framework of traditional quantum mechanics'',
since the quantum expectations of an imagined ensemble have an 
interpretation as quantum values of a single quantum system.


\subsection{The physics of imagined ensembles}\label{s.imagined}

\nopagebreak
\hfill\parbox[t]{10.8cm}{\footnotesize

{\em We may imagine a great number of systems of the same nature, but
differing in the configurations and velocities which they have at a
given instant, and differing not merely infinitesimally, but it may be
so as to embrace every conceivable combination of configuration
and velocities. [...] The first inquiries in this field were indeed
somewhat narrower in their scope than that which has been mentioned,
being applied to the particles of a system, rather than to independent
systems.
}

\hfill Josiah Willard Gibbs, 1902 \cite[pp. vii--viii]{Gib}
}

\bigskip

\nopagebreak
\hfill\parbox[t]{10.8cm}{\footnotesize

{\em So aufgefasst, scheint die Gibbssche Definition geradezu 
widersinnig. Wie soll eine dem K\"orper wirklich eignende Gr\"osse 
abh\"angen nicht von dem Zustand, den er hat, sondern den er 
m\"oglicherweise haben k\"onnte? [...] Es wird eine Gesamtheit 
mathematisch fingiert. [...] erscheint es schwierig, wenn nicht 
ausgeschlossen, dem Begriffe der kanonischen Gesamtheit eine 
physikalische Bedeutung abzugewinnen.
\\
(Viewed in this way, the definition by Gibbs seems downright absurd.
How can a quantity really belonging to a body depend not on the 
state that it has, but on that it could possibly have? [...] 
An ensemble is mathematically feigned. [...] It seems difficult if not
impossible, to extract a physical meaning from the concept of a 
canonical ensemble.)
}

\hfill Paul Hertz, 1910 \cite[p.226f]{Her} 
}

\bigskip

\nopagebreak
\hfill\parbox[t]{10.8cm}{\footnotesize

{\em To include the effect of the noise, we imagine performing the 
experiment a large number of times, so as to produce an ensemble 
average, the average we have previously used in deriving the 
correlation function.
}

\hfill Brown and Gabrielse, 1986 \cite[p.252]{BroG} 
}

\bigskip

\nopagebreak
\hfill\parbox[t]{10.8cm}{\footnotesize

{\em It seems hardly necessary to stress that every electromagnetic 
field found in nature has some fluctuations associated with it. Even 
though these fluctuations are, as a rule, much too rapid to be observed 
directly, one can deduce their existence from suitable experiments that 
provide information about correlations between the fluctuations at two 
or more space-time points. [...] \\
We assume that the light is [...] macroscopically steady. 
(Footnote: By 'macroscopically steady' we mean that it does not exhibit 
fluctuations on a macroscopic time-scale. In the more precise language 
of stochastic processes, the fluctuations can be represented as as 
stationaryu random process, whose mean period and correlation time are
 much shorter than the average intervasl needed to make an observation.)
}

\hfill Mandel and Wolf, 1995 \cite[pp.147ff]{ManW}
}

\bigskip

\nopagebreak
\hfill\parbox[t]{10.8cm}{\footnotesize

{\em Random functions of time frequently have the property that the 
character of the fluctuations does not change with time, even though 
any realization of the ensemble changes continually in time.[...]\\
So far the means, or expectations, have been calculated by 
averaging over the ensemble of all realizations. However, sometimes 
only a single realization of the ensemble is available. [...]\\
Ergodicity holds if correlations of the random process die out 
sufficiently rapidly in time. In that case a sufficiently long record 
of a single realization of the random process can be divided up into 
sections of shorter length which are uncorrelated, so that an ensemble 
can be constructed from a single realization. The average over this 
ensemble then equals the average over time, because a single 
realization of sufficient length already contains all the information 
about the ensemble. If the process is stationary and ergodic, all 
realizations of the random process look somewhat similar and differ 
only in detail.
}

\hfill Mandel and Wolf, 1995 \cite[pp.45.47,49f]{ManW}
}

\bigskip

To derive the spectral properties of the Penning trap, both Brown and 
Gabrielse and Rigo et al. need density operators, traditionally 
ascribed to ensembles rather than individual quantum systems. While 
Rigo et al. simply state that their use ''transcends the framework of 
traditional quantum mechanics'', Brown and Gabrielse justify the use 
of density operators by an imagined ensemble average. The ensemble on 
which they base their noise analysis is completely fictitious 
(''imagined'') since in reality they do the experiment with a single 
electron which does not form an ensemble. 
(Note that Brown and Gabrielse do not distinguish between classical 
ensemble averages, quantum ensemble averages, and statistical averages, 
as can be seen by comparing corresponding statements on p.252, p.268 
before (4.63), p.270 before (5.24a), and p.271!)

The association of an imagined ensemble with a {\em single} physical 
system goes back to Gibbs, the founder of the ensemble approach to 
classical statistical mechanics. He was very aware that thermodynamics, 
and hence statistical mechanics, applies to single systems, although his
statistical mechanics handles them in terms of a probability density 
that expresses the properties of a population of identical systems.
In his famous statistical mechanics textbook, \sca{Gibbs} \cite{Gib} 
introduced, in the preface (from which the above quote is taken) ,
imagined ensembles to bridge the conceptual gap.

To deduce properties of macroscopic materials, Gibbs uses -- in 
contrast to Boltzmann, who introduced statistical mechanics for gases 
by using ensembles of microscopic atoms -- an ensemble of classical 
macroscopic systems. Treating a collection of particles in a gas as an 
ensemble (as Boltzmann did) makes statistical sense as there are a huge 
number of them. But the Gibbs formalism is applied to single
macroscopic systems such as a brick of iron rather than to its many
constituents. Treating a single system as part of an imagined ensemble
was a very bold step taken by Gibbs. It allowed him to extend 
Boltzmann's work from ideal gases to arbitrary chemical systems, in a 
very robust way. His statistical mechanics formalism, as encoded in the 
textbook (\sca{Gibbs} \cite{Gib}), survived the quantum revolution 
almost without change -- the book reads almost like a modern book on 
statistical mechanics!

Though exceedingly successful, Gibbs' imagined ensemble raised in
his time severe objections in the physics community, such as the 
response by Hertz quoted above, who complained that the ensemble is 
feigned mathematically. Of course Gibbs was aware that imagined systems 
have no physical implications. But these were needed at a time where
mathematics had not yet the abstract character that it has today. Today 
we are accustomed to the fact that the same mathematical concept -- of 
a vector, say -- may have totally different physical interpretations 
as long as they satisfy the same axioms. 

What holds for vectors also 
holds for expectations: Imagined ensembles are simply mathematical 
entities that satisfy the axioms for expectation (\sca{Whittle} 
\cite{Whi}) although there is no underlying ensemble of things whose 
expectation is implied. Thus using an imagined ensemble simply expresses
the idea of borrowing the terminology and notation of ensembles but 
ignoring their original physical meaning. The expectations in the 
imagined ensemble are just the quantum values of our present setting, 
which have an objective meaning independent of any ensemble. 

To understand why one can sometimes get correct inferences for a single 
system using an imagined ensemble, we consider, like 
\sca{Mandel \& Wolf} \cite[Chapter 4]{ManW}, a single, classical 
deterministic system with unobservable high frequency fluctuations, 
e.g., a beam of light. 
It is common to treat small, irregular fluctuations as noise
of an associated random stationary process. This means that one treats
the -- on very short time scales highly nonstationary -- deterministic 
system as a particular realization of a stationary ensemble of 
stochastic systems. \sca{Mandel \& Wolf} \cite[Section 2.2]{ManW}
justify how under certain conditions (stationarity and ergodicity), the 
stochastic ensemble averages have a mathematically valid interpretation 
as time averages of deterministic systems. The effect of this 
interpretation is that one may treat time correlations as additional 
effective classical observables. If the correlations of the random 
process die out sufficiently rapidly in time, short time averages are 
sufficient to replicate ensemble averages. Then it is enough that the 
deterministic system is macroscopically steady, at time scales long 
compared to the fluctuations but short compared to measurement times.
(This condition is a classical analogue of the condition discussed in
Subsection \ref{s.nonstat} for measuring instationary quantum systems.)

This kind of explanation is the standard rationalization for the use 
of stochastic methods applied to single systems in classical physics. 
Thus we may say that in classical mechanics, the correlations of an 
imagined ensemble describe unmodelled high frequency fluctuations of 
the single system represented by this imagined ensemble.

The ergodic rationalization has no analogue in traditional quantum 
mechanics where states for individual quantum systems have no objective 
meaning. In the present approach, where individual quantum systems have 
a density matrix, such a rationalization is indeed superfluous.

\newpage
\section{Measurement uncertainty}\label{s.unc}

\nopagebreak
\hfill\parbox[t]{10.8cm}{\footnotesize

{\em Some hypotheses are dangerous, first and foremost those which are
tacit and unconscious. And since we make them without knowing them, we
cannot get rid of them. Here again, there is a service that mathematical
physics may render us. By the precision which is its characteristic, we
are compelled to formulate all the hypotheses that we would
unhesitatingly make without its aid.}

\hfill Henri Poincar\'e, 1902 \cite[p.151]{PoiScH}
}

\bigskip

This more technical section -- which can be skipped on first reading 
without loss of continuity -- gives a thorough, precise discussion of 
various aspects of uncertainty in quantum measurements.

\subsection{Statistical uncertainty}\label{ss.unc}

\nopagebreak
\hfill\parbox[t]{10.8cm}{\footnotesize

{\em Aus diesen Gr\"unden ist eine gleichzeitige genaue Beobachtung
von $q$ und $p$ prinzipiell ausgeschlossen. [...]
Man kann aber auch beide Gr\"o{\ss}en in einer einzigen Beobachtung
messen, also wohl gleichzeitig, aber nur mit beschr\"ankter Genauigkeit.
Bei einer solchen Beobachtung fragt man in der klassischen Theorie nach
dem 'Fehler' des gemessenen Wertes. [...]
Die 'Beobachtungsfehler' erscheinen in der neuen Theorie als
mit der statistischen Unbestimmtheit selbst zusammengeschmolzen.\\
(For these reasons, a simultaneous exact observation of $q$ and $p$ is 
impossible in principle. [...] However one can measure both quantities 
in a single measurement, hence simputaneously, but only with limited
accuracy. In such an observation one asks in the classical theory for 
the 'error' of the value measured. [...] 
The 'observation errors' appear in the new theory as being melted 
together with the statistical uncertainty itself.)}

\hfill Earle Kennard 1927 \cite[p.340f]{Ken}
}

\bigskip

\nopagebreak
\hfill\parbox[t]{10.8cm}{\footnotesize

{\em
Results of measurements cannot be absolutely accurate. This
unavoidable imperfection of measurements is expressed in their
inaccuracy.}

\hfill Semyon Rabinovich, 2005 (\cite[p.2]{Rab})
}

\bigskip

We write $|x|:=\sqrt{x^*x}$ for the Euclidean norm of a vector
$x\in \Cz^m$, and generalize it to vectors $A\in(\Lin\Hz)^m$  with
operator components by defining\footnote{
Actually, we never need $|A|$, but only the notation $|A|^2$ for $A^*A$,
especially when $A$ is some composite formula. 
} 
the operator
\[
|A|:=\sqrt{A^*A}.
\]
In this section, we add the adjective ''quantum''
to all theoretical quantum notions that suggest by their traditional 
name a statistical interpretation, and hence might confuse the borderline
between theory and measurement. This is in the spirit of the convention 
of \sca{Allahverdyan} et al. \cite{AllBN2} who add instead the prefix 
''q-''. 
Since we use statistical language, we adopt the trace 1 normalization 
of states, so that mean rates reduce to probabilities. Thus the state 
of the source is the positive linear mapping that assigns to each 
$X\in \Lin \Hz$ its quantum expectation
\[
\ol X=\<X\>:=\Tr\rho X.
\]
More generally, given a fixed state, the \bfi{quantum expectation}
of a vector $X\in(\Lin \Hz)^m$ with operator components is the vector
$\ol X=\<X\>\in\Cz^m$ with components $\ol X_j=\<X_j\>$. 
Its \bfi{quantum uncertainty} is the nonnegative 
number
\lbeq{e.sigmaXv}
\sigma_X:=\sqrt{\<(X-\ol X)^*(X-\ol X)\>}=\sqrt{\<X^*X\>-|\ol X|^2}.
\eeq
We may also define the \bfi{quantum covariance matrix}
\[
C_X:=\<(X-\ol X)(X-\ol X)^*\>\in\Cz^{n\times n},
\]
in terms of which
\[
\sigma_X=\sqrt{\tr C_X}.
\]

\begin{prop} (\sca{Robertson} \cite{Rob29})~\\
Any two Hermitian operators $A,B\in\Lin\Hz$ satisfy the 
\bfi{uncertainty relation}
\lbeq{e.uncRel}
\sigma_A\sigma_B\ge \half |\<[A,B]\>|
\eeq
with the \bfi{commutator}  
\[
[A,B]:=AB-BA.
\]
\end{prop}

\bepf
The relation remains unchanged when subtracting from $A$ and
$B$ its quantum expectation, hence it suffices to prove it for the case
where both quantum expectations vanish. In this case, 
$\<A^2\>=\sigma_A^2$ and $\<B^2\>=\sigma_B^2$, and the Cauchy--Schwarz 
inequality gives
\[
|\<AB\>|^2\le \<A^2\>\<B^2\>=\sigma_A^2\sigma_B^2.
\]
Hence $|\<AB\>|\le\sigma_A\sigma_B$. On the other hand, one easily 
checks that
$i\im\<AB\>=\half\<[A,B]\>$, so that
$\half |\<[A,B]\>|=|\im\<AB\>|\le |\<AB\>|$. Combining both inequalities
gives the assertion.
\epf

We now relate these theoretical quantum notions to the statistical 
notions introduced in Subsection \ref{ss.qExp}.

\begin{prop}
The inequality
\lbeq{e.Esigma}
\min_{\xi\in\Cz^m}\E(|a_k-\xi|^2)=\E(|a_k-\ol A|^2)
\ge \<|A-\ol A|^2\>=\sigma_A^2
\eeq
bounds the statistical uncertainty of the measurement results $a_k$
in terms of the theoretical quantum uncertainty of the quantity $A$ 
measured.
\end{prop}

For $m=1$, this is due to \sca{Holeveo} \cite[(9.8), p.88]{Hol1982}; it
was later redicovered by \sca{de Muynck \& Koelman} \cite{deMuyK} and
\sca{Werner} \cite[Proposition 3(2)]{Wer}.

\bepf
This follows by observing that
\lbeq{e.Eaxi}
\E(|a_k-\xi|^2)-\E(|a_k-\ol A|^2)=\E(|a_k-\xi|^2-|a_k-\ol A|^2)
=\E(|\ol A-\xi|^2)=|\ol A-\xi|^2
\eeq
is minimal for $\xi=\ol A$. The claim now follows from the following 
proposition.
\epf

\begin{prop}\label{p.Esigma}
Given a detector measuring $A=P[a_k]\in\Hz^m$. Then:

(i) For every Hermitian positive semidefinite $G\in\Cz^{m\times m}$, the
operator $P[a_k^*Ga_k]-A^*GA$ is positive semidefinite.

(ii) For any $\xi\in\Cz^m$ and any state $\<\cdot\>$,
\lbeq{e.Exi}
\E(|a_k-\xi|^2)\ge \<|A-\xi|^2\>.
\eeq
\end{prop}

The case $m=1$ is already in \sca{Holevo} \cite[Lemma 13.1]{Hol1973}.

\bepf
Using the quantum measure $P_k$ ($k\in K$) of the detector, we write
\[
\Delta:=\sum_{k\in K} (A-a_k)^*GP_k(A-a_k).
\]
(i) For any $\psi\in\Hz$, we define the vectors $\psi_k:=(A-a_k)\psi$,
and find $\psi^*N\psi=\sum \psi_k^* GP_k\psi_k\ge 0$. Hence $\Delta$ is
positive semidefinite. Now the first part follows from
\[
\bary{lll}
\Delta&:=&\D\sum\Big(A^*GP_kA-a_k^*GP_kA-A^*GP_ka_k+a_k^*GP_ka_k\Big)\\
&=&A^*GP[1]A-A^*GP[a_k]-P[a_k]^*GA+P[a_k^*Ga_k]\\
&=&A^*GA-A^*GA-A^*GA+P[a_k^*Ga_k]=P[a_k^*Ga_k]-A^*GA.
\eary
\]
(ii) In the special case where $G$ is the identity matrix, we find from
\gzit{e.statExP} that
\[
\E(|a_k|^2)=\<P[|a_k|^2]\>=\<P[a_k^*a_k]\>=\<N+A^*A\>
=\<\Delta:\>+\<A^*A\>\ge \<A^*A\>=\<|A|^2\>.
\]
Since $P[a_k-\xi]=P[a_k]-\xi=A-\xi$, we may apply this with
$a_k-\xi$ in place of $a_k$ and $A-\xi$ in place of $A$ and find
\gzit{e.Exi}.
\epf

In the special case of projective measurements, $A=P[a_k]$ satisfies
\[
P_k(A-a_k)=P_kA-a_kP_k=\sum_j a_jP_kP_j-a_kP_k=0,
\]
so that $\Delta:=0$ in the proof of Proposition \ref{p.Esigma}.
Therefore inequality \gzit{e.Esigma} holds in this case with equality 
for all states. A converse was proved in \sca{Kruszy\'nski \& De Muynck}
\cite[Proposition 2]{KruM}. Thus the difference in \gzit{e.Esigma}
describes the lack of projectivity. We may view it as a measure of
excess uncertainty of a real detector, as opposed to an idealized 
projective one. \sca{Busch} et al. \cite[(16)]{BusHL} view this 
difference as a measure of intrinsic noise.

\subsection{Imperfect measurements}\label{ss.imp}

\nopagebreak
\hfill\parbox[t]{10.8cm}{\footnotesize

{\em
The fact that actual measurements are always imprecise is well-known
and led Poincare to distinguish carefully the "mathematical continuum"
from the "physical continuum." In the mathematical continuum the notion
of identity satisfies the usual transitivity condition [...]
By contrast, this property cannot be assumed for the notion of
"indistinguishability" in the physical continuum attached to the raw
data of experiments. [...]
To pass from the physical continuum to the mathematical continuum
requires an idealization, namely that infinitely precise measurements
are in principle, if not in fact, attainable.}

\hfill Ali and Emch, 1974 \cite[p.1545]{AliE.meas}
}

\bigskip

\nopagebreak
\hfill\parbox[t]{10.8cm}{\footnotesize

{\em The discrete nature of the reading scale entails that a given
measuring apparatus allows only a measurement of a discrete version of
the observable under consideration. With this we do not, however, deny
the operational relevance of continuous observables. On the contrary,
their usefulness as idealisations shows itself in the fact that they
represent the possibility of indefinitely increasing the accuracy of
measurements by choosing increasingly refined reading scales.
}

\hfill Busch, Lahti and Mittelstaedt, 1996 (\cite[p.81]{BusLM})
}

\bigskip

The spectral norm of a quantity measured by an arbitrary detector is
bounded since the sum \gzit{e.obs} is finite and $\|P_k\|\le 1$ for all
$k$. Hence the components of $A$ are bounded linear
operators.\footnote{
To extend the notion of measurement to unbounded quantities such as
position or momentum -- which is outside the scope of the present paper
--, one would need to proceed in an idealized fashion, using continuous
POVMs for idealized measurements with infinite precision. Then
quantum expectations are defined only for sufficiently regular density
operators. For a proper treatment of the unbounded case see, e.g., the 
books mentioned at the beginning of this paper.
} 
In particular, \gzit{e.BornEx} is defined for all
density operators $\rho$. Note that the same bounded linear operator
$A$ can be decomposed in many ways into a linear combination of the
form \gzit{e.obs}; thus there may be many different detectors with
different scales measuring quantities corresponding to the same
operator $A$.

In practice, one is often interested in measuring a given (bounded or
unbounded) operator $X$ of interest. Designing realistic detectors that
allow a high quality measurement corresponding to theoretically
important operators is the challenge of high precision experimental
physics. Due to experimental limitations, this generally involves both
statistical and systematic errors.

If we approximate a (possibly vector-valued) quantity $X$ by a
measurable substitute quantity $A$ -- some such approximation is
unavoidable in practice --, we make a systematic error that may depend 
on the state of the system measured. \gzit{e.Eaxi} implies the formula
\lbeq{e.Emse}
\E(|a_k-\ol X|^2)=\E(|a_k-\ol A|^2)+|\ol A-\ol X|^2
\eeq
for the mean squared error of $a_k$ as an approximation of $\ol X$.
In particular, the term
\[
\Delta:=|\ol A-\ol X|=|\<X\>-\<A\>|,
\]
the \bfi{bias} due to the substitution of $A$ for $X$, is a lower bound
for the \bfi{root mean squared error} (\bfi{RMSE})
\[
\eps_X:=\sqrt{\E(|a_k-\ol X|^2)}.
\]
Unlike the RMSE, but like quantum expectations, the bias is a theoretical
property of a state, independent of measurement.
We say that $A$ is an \bfi{unbiased} approximation of $X$ in all states
such that the bias vanishes. If the bias vanishes in all states from an
open subset of the set of all states, we necessarily have $X=A$.
In particular, there are no everywhere unbiased approximations of
unbounded operators.\footnote{
For unbounded operators $A$ (for example the position operator vector
$q$ in a particular coordinate system), formula \gzit{e.BornEx} and
hence the bias is well-defined only for sufficiently regular density
operators $\rho$. This reflects a slight deficiency of the standard
textbook presentation of expectations. However,
measurement equipment for unbounded operators such as position, say,
only produces results in a bounded range. Hence it corresponds in fact
to a measurement device for a clipped version $X=F(q)$ of the
position operator $q$ with bounded $F$, resulting in a bounded $X$.
} 
Note that in practice, $\ol A \ne \ol X$ due to imperfections.
In particular, unbiased measurements are necessarily idealizations.

The RMSE $\eps_X$ measures the uncertainty in the value assigned
to $\ol X$. But because of the broken watch effect,\footnote{
Measuring time with a broken watch shows twice a day the exact time,
whereas a watch that is slow 1 second per day shows the correct time
at most once in a century.
} 
it cannot be regarded as the measured uncertainty in the value assigned
to $X$. Thus we need to add a systematic uncertainty correction that
corrects for the possibility that the uncertainty of $A$ is less than
the uncertainty of $X$. The latter has nothing to do with measurement
and hence must be a theoretical quantity computable from
quantum uncertainties. Now
\lbeq{e.qmse}
\<|A-\ol X|^2\>=\<|A-\ol A|^2\>+|\ol A-\ol X|^2
=\sigma_A^2+|\ol A-\ol X|^2
\eeq
in analogy to \gzit{e.Emse}, proved by expanding all squares. Comparing
this with the formula
\[
\<|X-\ol A|^2\>=\<|X-\ol X|^2\>+|\ol X-\ol A|^2
=\sigma_X^2+|\ol X-\ol A|^2
\]
obtained by interchanging the role of $A$ and $X$, we see that
the natural uncertainty correction to the mean squared error is
$(\sigma_X^2-\sigma_A^2)_+$, where $x_+:=\max(x,0)$ denotes the positive
part of a real number $x$. We therefore regard
\lbeq{e.measError}
\Delta_X[a_k]:=\sqrt{\E(|a_k-\ol X|^2)+(\sigma_X^2-\sigma_A^2)_+},
\eeq
the square root of the corrected mean squared error,
as the \bfi{measurement uncertainty} when measurements of $A$ are
performed in place of measurements of $X$.
Such \bfi{imperfect measurements} make sense even for unbounded
quantities $X$ in states with finite quantum uncertainty $\sigma_X$.
By \gzit{e.Exi} and \gzit{e.qmse},
\[
\bary{lll}
\Delta_X[a_k]^2&=&\E(|a_k-\ol X|^2)+(\sigma_X^2-\sigma_A^2)_+\\
&\ge& \<|A-\ol X|^2\>+\sigma_X^2-\sigma_A^2\\
&=&\sigma_X^2+|\ol A-\ol X|^2.
\eary
\]
Hence the measurement uncertainty is bounded from below by theoretical
error measures,
\[
\Delta_X(a_k)\ge \sqrt{\sigma_X^2+|\ol A-\ol X|^2}
=\sqrt{\<|X-\ol A|^2\>}.
\]
In particular, $\Delta_X(a_k)$ is always at least as large as the
quantum uncertainty of $X$, and larger if there is a nonzero bias.
This gives an operational interpretation to these theoretical terms.

By changing the scale of a detector we may define measurements of many
different quantities $A$ based on the same quantum measure. By picking 
the scale
carefully one can in many cases choose it such that $A$ approximates a
particular operator $X$ of interest with small measurement uncertainty
$\Delta_X[a_k]$ for the collection of states of interest. There are
several alternative ways to quantify what constitutes adequate
approximations; see, e.g., \sca{Appleby} \cite{App1998,App2016},
\sca{Barcielli} et al. \cite{BarGT},
\sca{Busch} et al. \cite{BusHL,BusLPY},
and references there.

Finding a good match of $A$ and $X$ by choosing a good scale $a_k$ is
the process called \bfi{tuning}. It corresponds to the classical
situation of labeling the scale of a meter to optimally match a desired
quantity. If the detector can also be tuned by adjusting parameters
$\theta$ affecting its responses, the operators $P_k=P_k(\theta)$
depend on these parameters, giving
\[
A=A(\theta)=\sum a_kP_k(\theta).
\]
Now both the labels $a_k$ and the parameters $\theta$ can be tuned to
improve the accuracy with which the desired $X$ is approximated by
$A(\theta)$, perfecting the tuning.

\subsection{Measurement errors}\label{ss.measErr}

\nopagebreak
\hfill\parbox[t]{10.8cm}{\footnotesize

{\em
Von dem neuen Standpunkt
schwebt nun zun\"achst dieser Fehlerbegriff in der Luft, denn er setzt
doch nicht nur den Begriff eines beobachteten, sondern auch den
Begriff eines 'wahren' Wertes voraus, und letzteren gibt es im
physikalischen Sinne nicht mehr.\\
(From the new point of view, theis notion of error floats at first in 
the air, bor it presupposes not only the notion of an observed but only 
that of a 'true' value, and the latter no longer exists in a physical 
sense.)
}

\hfill Earle Kennard 1927 \cite[p.340]{Ken}
}

\bigskip

\nopagebreak
\hfill\parbox[t]{10.8cm}{\footnotesize

{\em This mean value $x_0$ locates the wave packet in the crude sense
that an observing apparatus must be placed near $x_0$ if it is to have
a significant chance of interacting with the particle.}

\hfill Carl Helstrom, 1974 \cite[p.454]{Hel74}
}

\bigskip

\nopagebreak
\hfill\parbox[t]{10.8cm}{\footnotesize

{\em Measurement is the process of determinating the value of a physical
quantity experimentally with the help of special technical means called
measuring instruments. The value of a physical quantity [...] is found 
as the result of a measurement.
The true value of a measurand is the value of the measured physical
quantity, which, being known, would ideally reflect, both qualitatively
and quantitatively, the corresponding property of the object.}

\hfill Semyon Rabinovich, 2005 (\cite[p.1f]{Rab})
}

\bigskip

\nopagebreak
\hfill\parbox[t]{10.8cm}{\footnotesize

{\em Measurement errors are in principle unavoidable, because a
measurement is an experimental procedure and the true value of the
measurable quantity is an abstract concept. }

\hfill Semyon Rabinovich, 2005 (\cite[p.11]{Rab})
}

\bigskip

Measurement errors are ubiquitous in physical practice; their
definition requires, however, some care. A single measurement produces
a number, the \bfi{measurement result}. The splitting of the measurement
result into the sum of an intended result (the \bfi{true value}) and
the deviation from it (the \bfi{measurement error}) depends on what 
one declares to be the true value. What can be said about measurement 
errors depends, therefore, on what one regards as the true value of 
something measured.

In general, the true value is necessarily a theoretical construct, an
idealization arrived at by convention.
Since measured are only actual results, never the hypothesized true
values, there is no way to determine experimentally which convention is
the right one. Both the quantum formalism and the experimental record
are independent of what one declares to be the true value of a
measurement. Different conventions only define different ways of
bookkeeping, i.e., different ways of splitting the same actual
measurement results into a sum of true values and errors, in the
communication about quantum predictions and experiments. Nothing in the
bookkeeping changes the predictions and the level of their agreement
with experiment.

Thus the convention specifying what to consider as true values is
entirely a matter of choice, an \bfi{interpretation}. The convention
one chooses determines what one ends up with, and each interpretation
must be judged in terms of its implications for convenience and
accuracy. Like conventions about defining measurement units
\cite{SIunits}, interpretations can be adjusted to improvements in
theoretical and experimental understanding, in order to better serve
the scientific community.

In Subsections \ref{ss.unc} and \ref{ss.imp}, we did not
assume a notion of true value in the quantum case. However, 
\gzit{e.Esigma} implies that the least uncertain value $\xi$ is the 
quantum expectation $\ol A$. Thus in a statistical sense, the best
possible value that can be assigned is $\ol A$. This suggests that in
the unbiased case, $\ol A$ should be (in analogy to classical
statistics) considered as being the true value of $A$, which is 
measured approximately.

Alternatively, the problem of true values can be avoided if one regards 
not the quantum values of operators as the basic observables, but a 
vector of unknown parameters characterizing a quantum state. In this 
case, techniques of quantum estimation theory apply. For recent results,
see, e.g., \sca{Sidhu} et al. \cite{SidOCK}.

\newpage
\section{Quantum measures for particular detectors}\label{s.ex}

In this section, we give a number of relevant examples of detectors in 
the sense defined in Section \ref{ss.qExp}. Among them are exapmples 
from quantum optics, the joint measurement of noncommuting quantities, 
and the measurement of particle tracks in modern collision experiments.

Part of the challenge of experimental physics is to devise appropriate
preparation and measurement protocols in such a way that experiments
with desired properties are possible. Often this is the most difficult 
aspect of an experiment. On the other hand, it is also a difficult task 
to work out theoretically, i.e., in terms of statistical mechanics 
rather than quantum tomography, the right quantum measure for a given 
experiment described in terms of an -- even idealized -- microscopic 
model. In this paper, we do not discuss such microscopic models and 
refer instead to the literature; see, e.g., 
\sca{Breuer \& Petruccione} \cite{BreP.QC} and
\sca{Allahverdyan} et al. \cite{AllBN1}.

\subsection{Beam splitters and polarization state measurements}
\label{ss.pol}

In continuation of the introductory example in Subsection
\ref{ss.states}, we show here how quantum measures model the
simultaneous measurement of the polarization state $\rho$ of a
stationary optical source; cf. \sca{Brandt} \cite{Bra}. This can be done
in terms of \bfi{quantum optical networks} consisting of a combination
of optical filters and beam splitters. A $b$-port optical network
transforms $b$ input beams into the same number of output beams. The
input beams together (and the output beams together) form a quantum
system modelled on a state space $\Hz$ consisting of block vectors with 
$b$ components of size $2$. Hence the density matrices are block 
$b\times b$ matrices
\[
\rho=\pmatrix{\rho_{11} & \dots  & \rho_{1b} \cr
              \vdots    & \ddots & \vdots   \cr
              \rho_{b1} & \dots  & \rho_{bb}}
\]
with blocks $\rho_{jk}$ of size $2\times 2$. The diagonal block
$\rho_{kk}$ encodes the state of the $k$th beam, while the off-diagonal
blocks  $\rho_{jk}=\rho_{kj}^*$ encode correlations
between beams $j$ and $k$. The absence of the $k$th beam is modelled by
the dark state, $\rho_{kk}=0$; since $\rho$ is positive semidefinite,
$\rho_{kk}=0$ implies  $\rho_{jk}=\rho_{kj}=0$ \for all $j$.

If a beam of light is originally in state $\rho$, the state after 
having passed through a linear and non-mixing (non-depolarizing) 
optical filter is described according to 
\sca{Mandel \& Wolf} \cite[Section 6.2]{ManW} by $\rho'=T\rho T^*$,
where $T$ is a complex $2\times 2$ matrix, the \bfi{transmission matrix}
of the filter. In optics, the transmission matrix is called the
\bfi{Jones matrix}, after \sca{Jones} \cite{Jon}.
Jones matrices for relevant filters include $T=\gamma I$ with
$0<\gamma<1$, representing simple attenuation of the intensity by the
factor $\gamma^2$, and $T=\phi\phi^*$ with a vector $\phi\in\Cz^2$ of
norm $\le 1$, representing a \bfi{polarizer} with complete
polarization in the complex direction $\phi$. If $\phi$ is
\bfi{normalized}, i.e., $\phi^*\phi=1$, then the polarizer is
\bfi{perfect} in the sense that it preserves the intensity of light
that is already completely polarized in direction $\phi$.

A \bfi{beam splitter} is a half-silvered mirror in which two input beams
entering the mirror at 45 degrees from opposite sides combine to two
output beams exiting the mirror at 45 degrees. Thus its input and
output states are block $2\times 2$ matrices with blocks of size
$2\times 2$. Fully polarized beams of light are oscillations of the
classical electromagnetic field along some line segment, carrying
energy proportional to their intensity. If several beams are present at
the same time, they form a coherent superposition of the corresponding
oscillation modes. The phenomenological view of a beam splitter just
given therefore implies that the input field modes $\psi$ (a block
vector whose blocks describe the two beams) transforms into $T\psi$,
where
\[
T=\frac{1}{\sqrt{2}}\pmatrix{ 1 & 1 \cr 1 & -1 }
\]
is a $2\times 2$ block matrix composed of unit $2\times 2$ matrices $1$.
The density matrix $\rho=\psi\psi^*$ is therefore transformed into
$\rho'=(T\psi)(T\psi)^*=T\rho_B T^*$. It is empirically found (and can
be explained by stochastic electrodynamics) that this relation
continues to hold for mixed polarization states. Thus a perfect beam
splitter maps the input state $\rho_B$ to the output state
$\rho_B':=T\rho_B T^*$. In particular, if a beam in state
$\rho\in\Cz^{2\times 2}$ enters a beam splitter while the other input
port is unoccupied (represented by the dark state), the input state
is
\[
\rho_B=\pmatrix{\rho & 0 \cr 0 & 0},
\]
and the output state becomes
\[
\rho_B'=\half\pmatrix{\rho & \rho \cr \rho & \rho}.
\]
Thus the single input beam turns into two perfectly correlated output
beams, both attenuated by a factor $\half$.

The simultaneous measurement of the state of a primary beam now proceeds
by repeatedly splitting the primary beam emanating from the source
by means of a cascade of beam splitters into a finite number of
secondary beams labelled by a label set $K$, passing the $k$th secondary
beam through an optical filter, and detecting individual responses at
the $k$th detection element, defined as the screen at which the $k$th
secondary beam ends. We also introduce a null detector $0$ accounting
for the part of the beams absorbed at a filter, assumed to respond
whenever we expect an event but none of the screens shows one.

If we assume that the beam splitters are lossless, and that the filters
are linear and nonmixing, it is easy to see that this setting defines a
detector with quantum measure given by
\[
P_k=c_kT_k^*T_k,~~~P_0=1-\sum_k P_k,
\]
where $T_k$ is the Jones matrix of the $k$th filter and $c_k$ is the
attenuation factor incurred from the beam splitters. Indeed, the beam
splitters split a beam with density $\rho$ into multiple copies with
density $c_k\rho$, where $c_k=2^{-s}$ if the $k$th beam passed $s$
splitters. The subsequent filter produces the density
$\rho_k:=T_k(c_k\rho)T_k^*$. Therefore the $k$th detection element
responds proportional to the resulting intensity
\[
\tr\rho_k=\tr T_k(c_k\rho)T_k^*=\tr (c_kT_k^*T_k\rho)=\tr P_k\rho.
\]
For multiphoton states, the state space and hence the collection of
possible optical filters is much bigger. Using linear quantum-optical
networks (\sca{Leonhardt} \cite{Leo}) in place of simple filters and
Lie algebra techniques to encode their structural features
(\sca{Leonhardt \& Neumaier} \cite{LeoN}), one can design detectors
measuring arbitrary multiphoton states in essentially the same way.

\subsection{Joint measurements of noncommuting quantities}

\nopagebreak
\hfill\parbox[t]{10.8cm}{\footnotesize

{\em Aus diesen Gr\"unden ist eine gleichzeitige genaue Beobachtung
von $q$ und $p$ prinzipiell ausgeschlossen. [...]
Man kann aber auch beide Gr\"o{\ss}en in einer einzigen Beobachtung
messen, also wohl gleichzeitig, aber nur mit beschr\"ankter
Genauigkeit.\\
(For these reasons, a simultaneous exact observation of $q$ and $p$ is 
impossible in principle. [...] However one can measure both quantities 
in a single measurement, hence simputaneously, but only with limited
accuracy.)
}

\hfill Earle Kennard 1927 \cite[p.340]{Ken}
}

\bigskip

\nopagebreak
\hfill\parbox[t]{10.8cm}{\footnotesize

{\em The very meaning of the concepts involved -- the concept of a
simultaneous measurement of position and momentum, and the concept of
experimental accuracy -- continues to be the subject of discussion.
[...]
Ordinary laboratory practice depends on the assumption that it is
possible to make simultaneous, imperfectly accurate determinations of
the position and momentum of macroscopic objects.}

\hfill Marcus Appleby, 1998 \cite{App1998}
}

\bigskip

One of the basic and most well-known assertions of quantum mechanics
is Heisenberg's uncertainty relation. It asserts limitations in the
accuracy of joint measurements of position and momentum. Such
measurements are performed routinely. Indeed, every momentum measurement
in a lab is automatically accompanied by a simultaneous position
measurement that asserts at least that the particle measured is in a
position within the location of the measurement device. But Born's rule
cannot be applied to say anything about such a joint measurement.
In this context, the traditional foundations based on Born's rule are
therefore defective. Foundations based on quantum measures are free
from this defect.

Joint measurements of position and momentum are often described in
terms of a quantum measure built from coherent states.
An idealized joint measurement of position and momentum was described
by a coherent state quantum measure in \sca{Arthurs \& Kelly}
\cite{ArtK}, using infinitely many projectors
$|\alpha\rangle\langle\alpha|$ to all
possible coherent states $|\alpha\rangle$, where $\alpha$ is a complex
phase space variable. By discretizing this using a
\bfi{partition of unity} (i.e., a collection of finitely many smooth
nonnegative functions $e_k$ on phase space summing to 1),
these projectors can be grouped into finitely many positive operators
\[
P_k:=\pi^{-1}\int d\alpha e_k(\alpha)|\alpha\rangle\langle\alpha|
\]
corresponding to finite resolution measurements, making it look more
realistic. This would be suitable as a simple analytic example for
presentation in a course.
But to check how accurate an actual joint measurement of position and
momentum fits this construction for some particular partition of unity
would be a matter of quantum tomography!

\subsection{Realistic measurements of position}

In textbooks one often finds idealized hypothetical measurements of
operators with a continuous spectrum, for example position operators.
Realistic position measurements have limited accuracy and range only,
and no sharp boundaries between the position ranges where a particular
detection element responds. This can be modeled by quantum measures
based on a partition of unity on configuration space, analogous to the
above construction for coherent states. Details are given, e.g., in
\sca{Ali \& Emch} \cite{AliE.meas}.

\subsection{Measuring particle tracks}\label{ss.tracks}

\nopagebreak
\hfill\parbox[t]{10.8cm}{\footnotesize

{\em It is a little difficult to picture how it is that an outgoing 
spherical wave can produce a straight track [...]
The wave mechanics unaided ought to be able to predict the possible
results of any observation that we could make on a system, without 
invoking, until the moment at which the observation is made, the 
classical particle-like properties of the electrons or 
$\alpha$-particles forming that system.}

\hfill Nevill F. Mott, 1929 \cite{Mott}
}

\bigskip

\nopagebreak
\hfill\parbox[t]{10.8cm}{\footnotesize

{\em What is the quantum ensemble to which the quantum theory of
scattering applies?}

\hfill Brigitte Falkenburg, 2007 \cite[p.174]{Fal}
}

\bigskip

A simple conceptual analysis of how particle tracks are described by 
quantum mechanical correlations of atoms excited by a spherical wave
representing the particle state before detection was given by 
\sca{Mott} \cite{Mott}; for later refinements see, e.g., the historical
account in \sca{Figari \& Teta} \cite{FigT}. However, this standard 
treatment does not explicitly adreess the measurement aspects. 
In modern particle detectors, particle tracks are measured in 
a much more elaborate way. 

Indeed, in experimental practice in general, measurement is often a 
fairly complex procedure -- far more complex than the idealized 
statement of Born's rule or Mott's analysis would suggest. It involves 
not just reading a pointer or observing ionized atoms. One must also 
decide on the relevant aspects of the situation at hand and make a 
theoretical model of the facts assumed to be known, resulting in a
description of the quantities that count as measurement results.

Often it also involves nontrivial calculations from raw observations 
to arrive at the actual measurement results. This corresponds to the
basic description, which (according to the introductory quote by Peres) 
should be given in laboratory terms only.
As a consequence it may be difficult to describe the resulting 
measurement procedure explicitly in mathematical terms, but the 
quantum measure exists by the general analysis above. 

As a concrete complex example close to experimental practice, we 
consider a reasonably realistic version of a time projection 
chamber\footnote{
The description of the STAR time projection chamber in
\sca{Anderson} et al. \cite[ Section 5.2]{And} mentions only 2 layers,
so one has to use linear tracks. The LHC uses more layers and a
helical track finder, see \sca{Aggleton} et al. \cite[ Section 5]{Agg}.
} 
(TPC) for the measurement of properties of particle tracks. Here
obtaining the measurement results requires a significant amount of
nontrivial computation, not just pointer readings.

In a TPC, what emanates from the source measured passes an arrangement
of wires arranged in $L$ layers of $w$ wires each and generates electric
current signals, ideally exactly one signal per layer. From these
signals, time stamps and positions are computed by a least squares
process (via the Kalman filter), assuming that the track is a helix.
This is the case for a charged particle in a constant magnetic field,
experiencing energy loss in the chamber due to the induced ionizations.
From the classical tracks reconstructed by least squares, the momentum
is computed in a classical way.

The detector can be described by a quantum measure with an operator for
each of the $w^L$ possible signal patterns. 
\at{actually a contraction -- contributions with the same measurement 
results are summed} 
The value assignment is done by a
nontrivial computer program for the least squares analysis and initially
produces a whole particle track.  Part of the information gathered
is discarded; one typically records the computed energy, position and
velocity (or momentum if the mass is known). Momentum and energy are the
quantities of interest for scattering, but for secondary decays one
also needs the decay position. Thus one measures 7-dimensional phase
space vectors (including 3 position coordinates, 3 momentum coordinates,
and the energy) as a very complicated function of the signal patterns.

\newpage
\section{Quantum fields and macroscopic quantum systems}
\label{s.qfields}

Observations with highly localized detectors naturally lead to the 
notion of quantum fields whose quantum values encode the local 
properties of the universe. Quantum values appear to be the objective, 
reproducible elements of reality in the sense of 
\sca{Einstein, Podolski \& Rosen} \cite{EinPR}, while probabilities 
appear-- as in classical mechanics -- only in the context of statistical
measurements. 

Macroscopic systems are extended quantum systems whose observable 
features can be correctly described by local equilibrium thermodynamics,
as predicted by nonequilibrium statistical mechanics. Sources and 
detectors are macroscopic systems whose properties match the DRP
discussed in Section \ref{s.measP}. There is a fuzzy boundary between 
the macroscopic (directly observable) and microscopic (only inferable) 
part of the universe, characterized by the freedom in choosing a
Heisenberg cut somewhere within this fuzzy boundary.

\subsection{Extended sources}\label{s.extended}

\nopagebreak
\hfill\parbox[t]{10.8cm}{\footnotesize

{\em Correlations in space and time of the light intensity from 
extended sources [...] were analyzed [...] primarily in terms of field 
oscillations.}

\hfill U. Fano, 1961 \cite[p.539]{Fan}
}

\bigskip

An important idealization we have used so far was in the usage of beams.
Until now we have modeled our quantum systems as quantum beams, in 
analogy to classical optical beams. In many contexts, the latter are 
known to be very useful approximations, but under the condition that 
we may work with \bfi{geometric optics} rather than with the full 
electromagnetic field. Treating electromagnetic fields as 
light beams as in geometric optics is based on the paraxial, 
quasimonochromatic approximation, where light is essentially restricted 
to move along a single ray, and oscillate with a fixed classical 
frequency. 

But light is electromagnetic radiation, and has a wave-like character, 
hence cannot be confined to a ray but is spatially extended.
Most sources of light radiate in many directions unless the light is 
filtered through a tiny hole. Thus instead of thinking of a source as 
a tiny hole through which a beam leaves, we need to think it more 
realistically as an extended surface (\sca{Fano} \cite{Fan}) that emits 
at many points of the surface light in many directions. Whatever part 
of the emitted light reaches the detector has a cumulative additive 
effect on the detection rates.
 
The state of the beam will become correspondingly complex. Indeed, the
space representing classical light unrestricted in a region of free 
space is an infinite-dimensional state space, namely the space of 
all smooth solutions of the free Maxwell equations in this region. 
The quantities whose quantum value is observable are therefore linear 
operators on this space.  

The notion of a quantum beam must be generalized in the same way when
the conditions for the geometric optics approximation are not satisfied.
Again, the state space will become much bigger, typically 
infinite-dimensional.

The work for quantum tomography scales much stronger than linearly
with this dimension. Therefore the dimension of the state space of a 
quantum system strongly constrains how well we can determine a quantum 
state experimentally. In high (and even more in infinite) dimensions, 
quantum tomography becomes approximate even beyond the error introduced 
by the limited accuracy of statistical expectations, as it requires to 
truncate a large (or even infinite) basis of this space to a small, 
manageable dimension. This is done by limiting the energy range within 
which the quantum representation is accurately modeled, and living 
with the resulting modeling and prediction errors.

\subsection{The emergence of quantum fields}\label{s.fields}

\nopagebreak
\hfill\parbox[t]{10.8cm}{\footnotesize

{\em If, without in any way disturbing a system, we can predict with
certainty (i.e., with probability equal to unity) the value of a
physical quantity, then there exists an element of physical reality
corresponding to that physical quantity.}

\hfill Einstein, Podolsky, and Rosen, 1935 \cite[p.777]{EinPR}
}

\bigskip

In Subsection \ref{ss.compos}, we defined complex sources and detectors 
as combinations of a source or detector with one or more filters. With 
the description of a medium as a limit of many very narrow filters we 
may extend our view of sources and detectors. As in Subsection 
\ref{s.nonstat}, we consider a source as a surface in space where 
quantum beams exit, and can be in principle observed. 
Similarly, we consider a detector as a surface in space where 
quantum beams enter and can trigger one of the detection elements on
the other side. This is equivalent to wrapping the original source in
layers of filters formed by the medium around the source, and the
original detector in layers of filters formed by the medium around the
detector. This gives a new perspective on the interpretation of our
investigations. We can make the delineation of the source and the 
detector arbitrary depending on which part of the medium we count to 
where.

Now suppose that we have a detector that is sensitive only 
to quantum beams entering a tiny region in space, which we call the 
detector's \bfi{tip}. We assume that we can move the detector such that 
its tip is at an arbitrary point $x$ in the medium, and we consider a 
fixed source, extended by layers of the medium so that $x$ is at the
boundary of the extended source. The measurement performed in this 
constellation is a property of the source. The results clearly depend 
only on what happens at $x$, hence they may count as a measurement of 
a property of whatever occupies the space at $x$. Thus we are entitled 
to consider it as a \bfi{local property of the world} at $x$ at the 
time during which the measurement was performed. (Using coincidence 
measurements analogous to those used in quantum instrument tomography 
we could even measure bilocal properties of the world, given as the 
quantum values of operators depending on two positions.)

In general, the quantity whose quantum value is measured by the detector
depends on $x$, weakly or strongly depending on whether the quantum 
beams interact weakly or strongly with the medium. Therefore our set-up 
actually measures the quantum value $\<A(x)\>$ of a position-dependent 
quantity $A(x)$. We interpret $A(x)$ as an operator-valued 
\bfi{quantum field} whose quantum values describe a property of the 
world in all points $x$ accessible to the measurement device. If we 
idealize the detector as having an infinitely sharp tip consisting of 
a single point only, and being placed everywhere, we arrive at the view 
that all operationally verifiable local properties of the world are 
described by quantum values of suitable quantum fields. In particular,
beam properties become field properties at the points where the 
beam passes. Thus, from the new perspective, what we previously called 
quantum beams are just quantum fields concentrated along a narrow 
path. This is completely analogous to the way optical beams arise 
already in classical optics. 

Realistic measurements need to take account of the extension of the tip.
The result is therefore a weighted average over the region covered by 
the tip. Hence the detector measures the quantum value $\<\wt A(x)\>$ 
of a \bfi{smeared field}
\[
\wt A(x)=\int ds w(s)A(x+s)
\]
with a suitable weight function $w(s)$ depending on the tip.

Note that our reasoning is precisely the same as the reasoning used in 
classical physics to ascribe hydromechanic, electromagnetic, or 
gravitational properties to regions in space by placing detectors at
arbitrary points in these regions. The only difference in the quantum 
case is the quantum nature of the measurements, which lead to quantum 
values expressed by quantum expectations of fields rather than the 
fields themselves. 

By construction, the quantum values $\<A(x)\>$ of fields are 
operationally quantifiable properties in regions where they are 
sufficiently stationary, sufficiently slowly varying in space, and 
accessible by a detector whose tip is sensitive to these properties. 
They are \bfi{elements of reality} in the sense of 
\sca{Einstein, Podolski \& Rosen} \cite{EinPR}.

\subsection{Macroscopic quantum systems}\label{s.macro}

\nopagebreak
\hfill\parbox[t]{10.8cm}{\footnotesize

{\em Um zur Beobachtung zu gelangen, muss man also irgendwo ein 
Teilsystem aus der Welt ausschneiden und \"uber dieses Teilsystem eben 
'Aussagen' oder 'Beobachtungen' machen. Dadurch zerst\"ort man dort den 
feinen Zusammenhang der Erscheinungen und an der Stelle, wo wir den 
Schnitt zwischen dem zu beobachtenden System einerseits, dem Beobachter 
und seinen Apparaten andererseits machen, m\"ussen wir Schwierigkeiten 
f\"ur unsere Anschauung erwarten. [...] Jede Beobachtung teilt in 
gewisser Weise die Welt ein in bekannte und unbekannte oder besser: 
mehr oder weniger genau bekannte Gr\"ossen.\\
(To arrive at an observation one must somehow cut a 
subsystem out of the world and make 'statements' or 'observations'.
This destroys the fine connection of appearances, and we must expect 
difficulties for our intuition at the place where we make the cut 
between the system to be observed and the observer and his equipment. 
[...] Every observation splits in some sense the world into known and 
unknown, or better: more or less precisely known, quantities.)}

\hfill Werner Heisenberg, 1927 \cite[p.593f]{Hei.Como}
}

\bigskip

\nopagebreak
\hfill\parbox[t]{10.8cm}{\footnotesize

{\em
A satisfactory theory of the measuring process must start from a
characterization of the macroscopic properties of a large body. Such
properties must have an objective character.}

\hfill Daneri, Loinger and Prosperi, 1962 \cite[p.305]{DanLP}
}

\bigskip

\nopagebreak
\hfill\parbox[t]{10.8cm}{\footnotesize

{\em Quantum mechanics is concerned with macroscopic phenomena, which 
are not perturbed by observation.}

\hfill Nicolas van Kampen, 1988 \cite[Theorem II, p.98]{vKam1988}
}

\bigskip

\nopagebreak
\hfill\parbox[t]{10.8cm}{\footnotesize

{\em A quantum mechanical measuring apparatus consists of a 
macroscopic system prepared in a metastable state.}

\hfill Nicolas van Kampen, 1988 \cite[Theorem V, p.102]{vKam1988}
}

\bigskip

Our exposition clarified the meaning of the mathematical concepts
used in terms of measurement, hence defining their operational meaning. 
Everything is formulated in terms of objects and observations that we 
think we understand, though they themselves are in fact macroscopic 
quantum objects. Everything related to measurements is necessarily 
macroscopic. The microscopic information is only in the states, 
quantities, and quantum values used in the models -- these talk about 
particles or fields, depending on how one models the situation at hand.

The computational machinery of statistical mechanics provides a 
quantitatively correct deterministic dynamics for the quantum values of 
smeared fields describing matter in local equilibrium, 
\bfi{Local equilibrium} is the state of matter where the constituents 
have a mean free path small compared to the system size, hence no 
sensible asymptotic particle picture exists. This contrasts to the case 
where the system is very small (mesoscopic or microscopic) or the mean 
free path is so large that the system can be considered as a quantum 
beam. In the intermediate regime --  for very dilute bulk matter, or 
for studying details of collision experiments during the interaction 
of colliding quantum beams -- a microlocal (kinetic) picture is needed.

The book by \sca{Calzetta \& Hu} \cite{CalH.book} show how local 
equilibrium is derived from quantum field theory through a 
coarse-graining approach. The coarse-graining involved in the local 
equilibrium assumption leads naturally to dissipation and nonlinearity. 
These two features combine to yield very rich dynamical effects. These
include in particular the decay of near-equilibrium states to 
equilibrium and the phenomenon of metastability.

We call a quantum system \bfi{macroscopic} if its 
observable, coarse-grained aspects can be described by local equilibrium
thermodynamics. This additional assumption beyond our only postulate 
(DRP) is based on well-known results from statistical mechanics. The
\bfi{classical deterministic behavior} of macroscopic systems is an 
automatic consequence of local equilibrium. This is discussed in more 
detail in \sca{Neumaier \& Westra} \cite{NeuW} and \sca{Neumaier} 
\cite{Neu.CQP}.

The quantum state and the quantum measure, measurable through quantum 
tomography by observing approximations to quantum values, are 
collective properties of macroscopic objects, namely a macroscopic 
\bfi{source} and a macroscopic \bfi{detector}. The 
\bfi{detection elements} are given by the metastable regions of the 
detector with a macroscopically visible binary output, corresponding to 
the breaking of metastability in the associated nonlinear dynamics. 
Detectors and instruments therefore produce classical, irreversible 
results.

From this perspective, the \bfi{Heisenberg cut} can be seen as an 
imagined borderline in the fuzzy boundary between the macroscopic 
(directly observable) and the microscopic (only inferable) part of 
the universe. The output of a source (hence the state of a quantum beam)
and the input of a corresponding detector (hence the observed relation 
between input and detection events) depend on the details of this cut, 
but (as long as everything is linear, as we assumed) the results are 
consistent no matter how the cut is drawn.

\newpage
\section{On the foundations of quantum mechanics} \label{s.found}

\nopagebreak
\hfill\parbox[t]{10.8cm}{\footnotesize

{\em My 'orthodoxy' is not identical to that of Bohr, nor to that of 
Peierls, to mention two especially eminent examples. Hence I must
state my definition of 'orthodoxy'.
}

\hfill Kurt Gottfried, 1991 \cite[p.36]{Got}
}
\bigskip

\nopagebreak
\hfill\parbox[t]{10.8cm}{\footnotesize

{\em Orthodox QM, I am suggesting, consists of shifting between two 
different ways of understanding the quantum state according to context: 
interpreting quantum mechanics realistically in contexts where 
interference matters, and probabilistically in contexts where it does
not. Obviously this is conceptually unsatisfactory (at least on any 
remotely realist construal of QM) -- it is more a description of a 
practice than it is a stable interpretation. [...]
The ad hoc, opportunistic approach that physics takes to the 
interpretation of the quantum state, and the lack, in physical practice,
of a clear and unequivocal understanding of the state -- this is the 
quantum measurement problem.
}

\hfill David Wallace, 2016 \cite[p.22,p.24]{Wallace}
}
\bigskip

\nopagebreak
\hfill\parbox[t]{10.8cm}{\footnotesize

{\em More than a century after the birth of quantum mechanics, 
physicists and philosophers are still debating what a 'measurement' 
really means.}

\hfill DiVincenzo and Fuchs, 2019 \cite[p.2]{DiVF}
}

\bigskip

The traditional foundation of quantum mechanics depends heavily  -- far
too heavily -- on the concept of (hypothetical, idealized) measurements,
{\em exactly} satisfying Born's rule, a nontrivial technical rule far
from being intuitive. This -- almost generally assumed -- exact validity
without a precise definition of the meaning of the term measurement
is probably the main reason why, nearly 100 years after the discovery
of the basic formal setting for modern quantum mechanics, these
foundations are still unsettled. No other scientific theory has such
controversial foundations which have persisted for so long.

The source of this poor state of affairs is that Born's rule for
projective measurements, the starting point of the usual
interpretations, constitutes a severe idealization of measurement
processes in general. Except in a few very simple cases, it is too far
removed from experimental practice to tell much about real measurements,
and hence about how quantum physics is used in real applications. 
Foundations that provide a safe ground for interpreting reality cannot 
start with idealized concepts only. 

Thus it is worthwhile to reconsider the measurement problem in terms of 
the new foundations presented in this paper, which apply without any 
idealization. This is done in the present section. Apart from the 
quantum measurement problem, and what the present paper contributes 
towards its solution, this section also discusses philosophical 
considerations related to the development, namely

\pt
the nature of the basic substance of quantum matter -- particles or 
fields;

\pt 
the applicability of quantum mechanics to the whole universe; 

\pt
questions of reproducibility and objectivity;

\pt
the nature of probability and Born's rule within quantum mechanics;

\pt
the role of knowledge in the description of physical phenomena; and

\pt 
the thermal interpretation of quantum physics, which extends the 
discussion of observable sources and detectors in terms of quantum 
values represented by quantum expectations to arbitrary quantum systems 
even in their unobservable aspects.

\subsection{The quantum measurement problem}\label{ss.measP}

\nopagebreak
\hfill\parbox[t]{10.8cm}{\footnotesize

{\em The effect of a single scattering is calculated and shown to damp 
the off-diagonal elements in the reduced spatial density matrix by a 
factor that is simply the Fourier transform of the probability 
distribution for different momentum transfers. [...] Sec. 5 contains a 
brief discussion of what interpretational problems of quantum mechanics 
the decoherence approach does and does not solve.}

\hfill Max Tegmark, 1993 \cite[p.2]{Teg}
}

\bigskip

\nopagebreak
\hfill\parbox[t]{10.8cm}{\footnotesize

{\em Usually the system and its properties -- whatever can be said about
them -- are regarded as fundamental. Physical systems are considered
objectively, outside the context of measurements, real or potential.
Within this traditional viewpoint the question for physics is this:
What are the properties of physical systems, and what laws do they obey?
Notoriously this leads to the quantum measurement problem. Having
considered physical systems to obey quantum mechanics, what are we to
make of measurements? How does measurement fit into the framework of
quantum theory?}

\hfill Caves, Fuchs, Manne, and Renes, 2004 \cite[p.194]{CavFMR}
}

\bigskip

\nopagebreak
\hfill\parbox[t]{10.8cm}{\footnotesize

{\em There is no consistent quantum measurement theory, unless an
appropriate reinterpretation of what it means for an observable to have
a definite value can be found.}

\hfill Busch and Lahti, 2009 \cite[p.375]{BusL3}
}

\bigskip

\nopagebreak
\hfill\parbox[t]{10.8cm}{\footnotesize

{\em
The measurement problem is a central issue in quantum foundations, 
because textbook quantum mechanics uses the idea of a measurement to 
give a physical interpretation to probabilities generated from a 
quantum wave function, but never explains the measurement process 
itself in terms of more fundamental quantum principles.}

\hfill Robert Griffith, 2017 \cite[p.1]{Gri.m}
}

\bigskip

\nopagebreak
\hfill\parbox[t]{10.8cm}{\footnotesize

{\em Decoherence actually aggravates the measurement problem: where 
previously this problem was believed to be man-made and relevant only 
to rather unusual laboratory situations, it has now become clear that 
''measurement'' of a quantum system by the environment (instead of by 
an experimental physicist) happens everywhere and all the time: hence 
it remains even more miraculous than before that there is a single 
outcome after each such measurement.}

\hfill Klaas Landsman, 2017 \cite[p.443]{Lan2017} 
}

\bigskip

\nopagebreak
\hfill\parbox[t]{10.8cm}{\footnotesize

{\em We need to take the existence of measurement outcomes as a priori
given, or otherwise give an account outside of decoherence of how
measurement outcomes are produced, because the property of classicality
is ultimately a statement about measurement statistics.}

\hfill Maximilian Schlosshauer, 2019 \cite[p.72]{Schl2019}
}

\bigskip

In a \bfi{measurement}, we extract within some measurement tolerance
an unambiguous value from an \bfi{environment} and simultaneously make
a claim that this value reveals a property of the system at the
measurement time. Here \bfi{unambiguous} means that the uncertainty is
significantly smaller than the error tolerance claimed for the
measurement.

The \bfi{measurement problem} is the problem to show convincingly how 
Born's rule \gzit{e.BornEx} can be justified entirely in terms of the 
unitary dynamics of a larger quantum model containing a measured 
quantum system and a quantum detector measuring it. For such quantum 
models of measurements see \sca{Busch \& Lahti} \cite{BusL2}, several 
chapters in \sca{Busch} et al. \cite{BusGL, BusLPY}, and the detailed 
study by \sca{Allahverdyan} et al. \cite{AllBN0,AllBN1}.

Based on the present approach, we may give a reasonably precise 
formulation. A complete solution of the \bfi{measurement problem} would 
consist of three separate parts:

1. A derivation of the states $\rho$ and the operators $P_k$ from the
microscopic description of typical macroscopic quantum systems that 
serve as sources and detectors. This is the \bfi{classicality problem} 
of quantum measurement.

2. A description of a {\em single} measurement, deriving both the 
measurement result and its accuracy from the state and the dynamics of 
the composite quantum system formed by a measured system and a detector.
This is the \bfi{definite outcome problem} of quantum measurement.

3. To show that a single particle moving along a beam triggers at most 
one of an array of detection elements. This is the 
\bfi{unique outcome problem} of quantum measurement.

The  \bfi{classicality problem} is solved in principle by 
\bfi{decoherence} (\sca{Zurek} \cite{Zur}); very detailed expositions
can be found in \sca{Schlosshauer} \cite{Schl.book, Schl2019}. 
In the decoherence approach, a \bfi{quantum system} $S$ localized in 
some region of spacetime is regarded together with its 
\bfi{environment} $E$, essentially the complement of $S$ in the 
universe. But to make computations feasible, the environment is in 
actual models of decoherence heavily simplified -- usually to a free 
thermal field. This changes the quantitative predictions; however, 
the qualitative aspects are unaffected since the most strongly 
interacting part of the environment gives the dominant contributions 
to the decoherence effect, and the details of the interaction do not 
matter much.

$S$ and $E$ are considered to be subsystems of a larger system 
$S+E$ with a unitary time evolution. The \bfi{density operator}
of $S$ is defined at each time as the 
\bfi{reduced density operator}\index{density operator!reduced}
inducedon the state space of $S$ by the quantum expectation mapping of 
the large unitary system $S+E$ at the same time. 

This density operator encodes all dynamical information about the 
system $S$ that can be obtained from the unitary dynamics of $S+E$, 
hence is an exact, complete and irreducible description of the quantum 
system $S$ at all times. Moreover, in some (often but not always good) 
approximation, the dynamics is of the Lindblad form \gzit{e.Lindblad}, 
or its generalization with an integral over uncountably many $\ell$s
in place of the sum. However, unlike the statistical mixtures of pure
states traditionally employed to define density operators, the reduced 
density operator does {\em not} have an interpretation as a statistical 
mixture of pure states. While it can be decomposed mathematically into 
many physically inequivalent such mixtures, none of these has any 
physical relevance. (It is therefore sometimes said to be an 
\bfi{improper mixture}.)

For many specific cases of interest, it can be shown that the 
interaction of the system with the environment ensures that the 
reduced density operators of nonisolated quantum systems (hence all
quantum systems whose environment is nonempty) tend to become 
essentially diagonal in a preferred 
\bfi{environmentally induced basis}\index{basis!environmentally induced}
determined by the environment. 

More precisely, for a system with a purely discrete spectrum, the 
off-diagonal matrix elements in the environmentally induced basis
decay exponentially (in these models), with an extremely short 
half-life called the \bfi{decoherence time}, For systems with a
continuous spectrum, one may use an overcomplete basis of coherent 
states; then there is still exponential decay, but the decoherence time 
of an off-diagonal matrix element now depends on how far away the labels
of the corresponding coherent states are, and goes to zero in the 
limit where these labels coalesce.
Mathematically (\sca{Tegmark} \cite{Teg}), this is a consequence of the 
Riemann--Lebesgue Lemma, which asserts that time integrals over signals 
with a purely continuous spectrum vanish in the limit where the time 
span gets arbitrarily large. 

While decoherence provides a satisfactory solution of the classicality 
problem, \sca{Schlosshauer} \cite[p.50]{Schl.book} and 
\cite[Section 7.1]{Schl2019} states explicitly (and with good reasons) 
that the definite outcome problem and the unique outcome problem, 
combined by him to the \bfi{problem of outcomes}, are not solvable by 
decoherence.

According to our discussion, the \bfi{definite outcome problem} is 
solved in principle by the fact that outcomes of experiments are 
observations (or results computed from these) of macroscopic detectors. 
These are given by the 
extensive and intensive quantities of their description in terms of
local equilibrium thermodynamics. The extensive quantities are quantum 
values of appropriate quantum field operators, while the intensive
quantities are determined by the extensive quantities through the 
machinery of local equilibrium thermodynamics. In particular, the 
outcomes are automatically definite, determined by the state.
From a mathematical perspective (cf. \sca{Neumaier} 
\cite[Section 10,5]{Neu.CQP}), the story told by decoherence in terms 
of averages is refined in the present approach to a different, more 
detailed story for each single case.

The \bfi{unique outcome problem} is a problem {\em only if one assumes} 
that there is such a thing as a {\em single particle moving along a 
beam}. Although this is commonly tacitly assumed in discussions of the 
foundations of quantum mechanics, it is by no means a necessary 
assumption. Indeed, we have taken care in the whole paper to
ensure that everything (and hence all of standard quantum mechanics) 
is valid without making this assumption. Thus the third problem 
diappears when one only assumes the presence of fields and regards
the notion of a particle as an approximate concept valid only in a 
semiclassical description. The extent to which it is meaningful to do 
this is discussed in the next subsection.

\subsection{Particles and fields}\label{ss.partF}

\nopagebreak
\hfill\parbox[t]{10.8cm}{\footnotesize

{\em The use of observations concerning the behaviour of particles in 
the atom rests on the possibility of neglecting, during the process of 
observation, the interaction between the particles, thus regarding them 
as free. [...]\\ 
The wave mechanical solutions can be visualised only in so far as they 
can be described with the aid of the concept of free particles. [...]\\ 
Summarising, it might be said that the concepts of stationary states 
and individual transition processes within their proper field of 
application possess just as much or as little 'reality' as the
very idea of individual particles.}

\hfill Niels Bohr, 1927 \cite[pp.587--589]{Boh1927}
}

\bigskip

\nopagebreak
\hfill\parbox[t]{10.8cm}{\footnotesize

{\em It seems to me certain that we have to give up the notion of an 
absolute localization of the particles in a theoretical model. This 
seems to me to be the correct theoretical interpretation of Heisenberg's
indeterminacy relation. [...] 
Only if this sort of representation of the atomistic structure be 
obtained could I regard the quantum problem within the framework of 
a continuum theory as solved.}

\hfill Albert Einstein, 1934 \cite[p.169]{Einstein1934}
}

\bigskip

\nopagebreak
\hfill\parbox[t]{10.8cm}{\footnotesize

{\em For many purposes the quantization of the electromagnetic field is 
not necessary at all, and the response of the photodetector can be 
understood even if we continue to picture the field in terms of 
classical electromagnetic waves, provided the photoelectrons are 
treated by quantum mechanics. The field then simply behaves as an 
external potential that perturbs the bound electrons of the 
photocathode. [...] As we shall see, for those electromagnetic fields 
for which an adequate classical description exists, the semiclassical 
and the fully quantized treatments of the photodetection problem yield 
virtually identical answers.
}

\hfill Mandel and Wolf, 1995 \cite[p.439]{ManW}
}

\bigskip

\nopagebreak
\hfill\parbox[t]{10.8cm}{\footnotesize

{\em In its mature form, the idea of quantum field theory is that 
quantum fields are the basic ingredients of the universe, and particles 
are just bundles of energy and momentum of the fields.}

\hfill Steven Weinberg, 1999 \cite[p.242]{Wei99}
}

\bigskip

\nopagebreak
\hfill\parbox[t]{10.8cm}{\footnotesize

{\em If QFT is about fields, how can its restriction to nonrelativistic 
phenomena be about particles? 
}

\hfill Art Hobson, 2013 \cite[p.212]{Hob}
}

\bigskip

What happens in the medium between source and detector is unmodelled in 
the pure measurement context assumed in our approach. Without further 
theoretical assumptions, it remains a matter of subjective belief.

The present approach works independent of the nature or even the 
presence of a mediating substance: What is measured are properties of 
the source, and this has a well-defined macroscopic existence. 
We have never needed to make an assumption on the nature of the medium 
passed from the source to the detector. Thus the present approach is 
indifferent to the microscopic cause of detection events. 
It does not matter at all whether one regards such an event as caused 
by a quantum field or by the arrival of a particle. 
In particular, a microscopic interpretation of the single detection 
events as arrival of particles is not needed, not even an ontological 
statement about the nature of what arrives.
Nor would these serve a constructive purpose.  

In the present stochastic description of quantum measurement it is 
sufficient that {\em something} passes from a source through the medium 
to a detector. Nothing at all depends on the properties of {\em what} 
is flowing. In quantum field theory, what is flowing are excitations of 
quantum fields, while in particle physics and in the quantum mechanics 
of atoms and molecules, what is flowing is taken to be a stream of 
particles, though the latter have very counterintuitive nonclassical 
properties. 

A particle detector therefore detects the potentially discrete response 
of macroscopic matter to an incident quantum field, but probably 
nothing else. It is therefore questionable whether a single response of 
a detection element measures anything beyond the macroscopic quantity 
that defines the response. 

This is reinforced by the observation that the typical effects of 
photodetection already appear in models of photodetection that treat 
the electromagnetic field as a classical external field but the detector
as a quantum system. For example, in the particle-free model discussed 
by \sca{Mandel \& Wolf} \cite{ManW} in Sections 9.1--9.3, a 
Poisson-distributed stochastic response pattern is found that is 
quantitatively identical with the effect predicted by their full 
quantum field theoretic treatment of photodetection in Section 14.2
with light beams in a coherent state.

However, observable properties are associated with arbitrary 
sufficiently slowly changing macroscopic regions of spacetime. Thus 
unlike the particle view, the field view emerged naturally from our 
discussion.

The quantum field point of view is indeed quite coherent (\sca{Neumaier}
\cite{Neu.CQP}) and leaves very little room for mystery. On the other
hand, a particle description makes physical sense only if the 
microscopic constituents of matter in some region of space can be 
considered to be free for a sufficiently long time, so that one can use 
a semiclassical approximation -- corresponding to \bfi{geometric optics}
for light. This is discussed in much more detail in Chapter 12 of my 
book \cite{Neu.CQP}. 

As we have seen in Subsection \ref{s.fields}, quantum values $\<A(x)\>$
of fields are elements of reality in the sense of 
\sca{Einstein, Podolski \& Rosen} \cite{EinPR}. Since the density 
operator of a sufficiently stationary quantum source can be determined 
in principle by quantum tomography from such quantum values, the density
operator is itself an element of reality. The analysis of the Penning 
trap experiments in Section \ref{s.specHigh} demonstrates the same 
for the density operators of single particles confined to a trap.
On the other hand, as the arguments of Einstein, Podolski \& Rosen 
together with experiments violating the Bell inequalities 
(\sca{Bell} \cite{Bel}, \sca{Aspect} et al. \cite{Asp}) amply 
demonstrate, individual particles inside entangled beams cannot be such 
elements of reality -- only the nonlocal multiparticle density operator 
describing whole entangled systems qualify here.

Thus from the perspective of the present considerations, quantum 
particles appear to be ghosts in the beams. This explains their spooky 
properties in the quantum physics literature!

\subsection{The universe as a quantum system}\label{s.univ}

\nopagebreak
\hfill\parbox[t]{10.8cm}{\footnotesize

{\em Since the interactions between macroscopic systems are effective 
at astronomical distances, the only ''closed system'' is the universe 
as a whole.}

\hfill H. Dieter Zeh, 1970 \cite[p.73]{Zeh1970}
}

\bigskip

The current approach gives a natural ontology for individual quantum 
systems -- both for thermal systems describing materials, sources and 
detectors, and for microscopic systems, as required by the ion trap 
experiments discussed in Section \ref{s.specHigh}. It makes even sense 
for the whole universe, hence meets Einstein's intuition 
in his quote given in Subsection \ref{ss.concepts}.

There is no known limit of validity of the principles of quantum 
physics. Therefore, good foundations for quantum physics must allow 
a consistent quantum description of the universe from the smallest to 
the largest levels of modeling, without having to introduce any change 
in the formal apparatus of quantum physics.
The foundations for quantum physics should therefore be formulated in a
way that they apply not only to small systems, but to large systems 
such as our solar system, and even to the largest physical system, 
the whole universe. 

Here the \bfi{universe} is understood to be 
the smallest closed physical system containing us, hence strictly 
speaking the only closed system containing us, and hence the only 
system to which unitary quantum physics applies without approximation. 
In particular, this implies that the universe is unique.
A detailed discussion of the state of the universe should include 
gravitation, and hence would touch on the difficult, unsolved problem 
of quantum gravity. A simpler task is to find at least a consistent 
interpretational framework in which to discuss these questions.

In some sense, quantum mechanics was always a theory of the whole 
universe. For example, the probabilistic interpretation of the wave 
function $\psi$ of a single particle asserts that there is a typically 
positive (though almost everywhere exceedingly tiny) probability 
$p(\Omega):=\int_\Omega dx |\psi(x)|^2$ for finding the particle in an 
arbitrary open region $\Omega$ of the universe, no matter how far away 
from where the particle was prepared. 

Since every property of a subsystem is also a property of the 
whole system, the state of the universe must be compatible with 
everything we have ever empirically observed in the universe. 
This implies that the state of the universe is highly constrained.
Knowing this state amounts to having represented all physics 
accessible to us by the study of its subsystems. This constitutes a 
very stringent test of adequacy of a putative state of the universe.
Knowing all the detailed properties, or finding its exact state is 
already out of the question for a small macroscopic quantum system such 
as a drop of water. Thus, as for a drop of water, one must be content
with describing the state of the universe approximately. But, as in 
case of a drop of water, there is no physical reason to question the 
existence of the state of the whole universe, even though many of its 
details may remain unknown for ever.

\subsection{Reproducibility and objectivity}\label{ss.obj}

\nopagebreak
\hfill\parbox[t]{10.8cm}{\footnotesize

{\em An experiment may always be regarded as the determination of the 
transmission probabilities for a finite number of operations.}

\hfill Haag and Kastler, 1964 \cite[p.850]{HaaK}
}

\bigskip

\nopagebreak
\hfill\parbox[t]{10.8cm}{\footnotesize

{\em The purpose of measurements is the determination of properties of
the physical system under investigation.
}

\hfill Busch, Lahti and Mittelstaedt, 1996 (\cite[p.25]{BusLM})
}

\bigskip

\nopagebreak
\hfill\parbox[t]{10.8cm}{\footnotesize

{\em Measurements are regarded metrologically to be better the lower
their uncertainty is. However, measurements must be reproducible,
because otherwise they lose their objective character and therefore
become meaningless.}

\hfill Semyon Rabinovich, 2005 (\cite[p.22]{Rab})
}

\bigskip

\nopagebreak
\hfill\parbox[t]{10.8cm}{\footnotesize

{\em In a fundamentally statistical theory like quantum mechanics the
results of individual measurements tell us almost nothing: It is always
the probability distribution of outcomes for a fixed experimental
arrangement which can properly be called the result of an experiment.}

\hfill Busch, Lahti and Werner 2014 \cite[p.5]{BusLW}
}

\bigskip

Reproducibility and observer independence are the hallmark of
scientific investigations. Reproducibility ensures that the parameters 
are indeed properties of the system measured, rather than an artifact 
of the observational procedure. Observer independence ensures that 
these properties are \bfi{objective}. 

We have seen in Section \ref{s.specHigh} that the model parameters 
specifying a quantum system apart from its state are operationally 
quantifiable, independent of an observer. The use of established 
statistical techniques ensures that this can be done in a reproducible 
way. Thus, to the extent the model descries a quantum system correctly, 
these parameters are objective properties of the quantum system. 

In the case of quantum tomography, the model is -- according to our 
discussion so far -- a finite-dimensional Hilbert space together with 
the mathematical machinery deduced from the detector response principle.
The state, an element of the quantum phase space, is a complex positive 
semidefinite Hermitian matrix of parameters that are not determined by 
the model, but by experiments, using the traditional statistical 
techniques universally agreed upon. The time dependence of the state 
is governed by differential equations whose parameters are objective 
system properties that can be determined by appropriate system 
identification techniques, to in principle arbitrary accuracy.

When a source is stationary, it has a time independent state. In this 
case, response rates and probabilities can also 
be measured in principle with arbitrary accuracy. These probabilities, 
and hence everything computable from them -- quantum values and 
the density operator but not the individual detector events -- are 
operationally quantifiable, independent of an observer, in a 
reproducible way. Thus the density operator is an objective property
of a stationary quantum system, in the same sense as in classical 
mechanics, positions and momenta are objective properties.

Through quantum tomography, the quantum state of a sufficiently 
stationary source, the quantum measure of a measurement device, and 
the transmission operator of a sufficiently linear and stationary filter
can in principle be determined with observer-independent protocols. 
Thus they are objective properties of the source, the measurement 
device, or the filter, both before and after measurement. Measurements 
only serve to determine or to check their quantitative form, thereby 
revealing or confirming preexisting properties of the source, the 
detector, and the filter, or demonstrating deviations from their 
assumed behavior.

\subsection{Knowledge and probability}\label{ss.know}

\nopagebreak
\hfill\parbox[t]{10.8cm}{\footnotesize

{\em The Compton effect has as its consequence that the electron is 
caused to jump [...]
As our knowledge of the system does change discontinuously at each 
observation its mathematical representation must also change 
discontinuously; this is to be found in classical statistical theories 
as well as in the present theory.}

\hfill Werner Heisenberg, 1930 \cite[p.35f]{Hei30}
}

\bigskip

\nopagebreak
\hfill\parbox[t]{10.8cm}{\footnotesize

{\em In my view the most fundamental statement of quantum
mechanics is that the wavefunction, or more generally the density
matrix, represents our knowledge of the system we are trying to
describe. I shall return later to the question "whose knowledge?" [...]
\\
More precisely, [...] the initial values represent knowledge usually 
obtained from observations.
}

\hfill Rudolf Peierl, 1991 \cite[p.19]{Pei}
}

\bigskip

\nopagebreak
\hfill\parbox[t]{10.8cm}{\footnotesize

{\em This, I see as the line of attack we should pursue with relentless 
consistency: the quantum system represents something real and 
independent of us. [...]\\
What might have been [Einstein's] greatest achievement in building 
general relativity? [...] It was in identifying all the things that 
are 'numerically additional' to the observer-free situation -- i.e., 
those things that come about purely by bringing the observer 
(scientific agent, coordinate system, etc.) back into the picture. 
This was a breakthrough. [...] \\
The dream I see for quantum mechanics is just this. Weed out all the 
terms that have to do with gambling commitments, information, 
know\-ledge and belief, and what is left behind will play the the role 
of Einstein's manifold.}

\hfill Christopher Fuchs, 2003 \cite[p.989f]{Fuc2003}
}

\bigskip

\nopagebreak
\hfill\parbox[t]{10.8cm}{\footnotesize

{\em One can still come across articles negating quantum jumps -- and 
any other kind of discontinuities, for that matter -- or taking them 
as a sudden increase of our knowledge of the system rather than a 
physical phenomenon.}

\hfill  de la Pe\~na, Cetto and Vald\'es-Hern\'andez, 2020 
\cite{dlPenCV}
}

\bigskip

A physical model (phase space, dynamical law, form of the Hamiltonian)
determines the subject matter of a physical investigation and fixes the 
language for its discussion. In physics, it therefore plays the same 
role that axioms play in mathematics. Although different models can be 
chosen to discuss the same real situation, we usually do not perceive 
this choice as a restriction of objectivity in science since all that 
is to choose is made explicit. 

Subjectivity in the choice of the model is the kind of subjectivity one 
has everywhere in physics. It has nothing to do with subjective 
knowledge of details about a particular system under experimental 
investigation commonly discussed in the foundations of quantum physics 
and modelled by subjective probabilities. Indeed, it seems pointless 
to try to assign subjective probabilities for decisions to use a 
particular model as the subject of discourse.

In the quartic oscillator model discussed in Subsection 
\ref{s.systemId}, the only subjectivity is in the choice of the ansatz 
for the Hamiltonian and the use of classical mechanics or quantum 
mechanics. There is no subjectivity in the parameters: If the 
parameters do not agree with the true parameters within the limits of 
the model accuracy claimed, more and more independent measurements 
lead sooner or later to a statistically arbitrarily significant 
discrepancy with experiments. 

Precisely the same holds in quantum tomography of stationary quantum 
systems with a low-dimensional state space. Again, the only 
subjectivity is in the choice of the model -- in this case the Hilbert 
space representing the quantum system and the assumption inherent in 
the detector response principle that states are modelled by density 
operators. Everything else 
can be determined and checked completely independent of any subjective 
knowledge. Assuming that a quantum system is in a state different from 
the true state simply leads to wrong predictions that can be falsified 
by sufficiently long sequences of measurements. Nothing depends on the 
knowledge of an observer. The latter can be close to the objective 
truth or far away -- depending on how well informed the observer is.

The assignment of states to stationary sources is as objective as any 
assignment of properties to macroscopic objects. Thus the 
\bfi{knowledge} people talk about when referring to the meaning of a 
quantum state resides solely in what is encoded in (and hence 
''known to'') the model used to describe a quantum system -- the 
details of the dynamics and the initial conditions, and not to any 
subjective mind content of a knower! In particular, the situation is 
precisely the same as in classical mechanics, where the models also 
depend on this kind of knowledge (or prejudice, if the model is chosen 
based on insufficient knowledge). 

As quantum values of members of a quantum measure, all probabilities 
are objective \bfi{frequentist probabilities} in the sense employed 
everywhere in experimental physics -- classical and quantum. 
That the probabilities are only approximately measurable as relative 
frequencies simply says that -- like all measurements -- probability
measurements are of limited accuracy only.
Similarly, as a property of macroscopic instrument, the collapse and
its generalizations are objective phenomena -- not subjective 
updates of knowledge.

Thus by shifting the attention from the microscopic structure to the
experimentally accessible macroscopic equipment (sources, detectors,
filters, and instruments), we have got rid of all potentially 
subjective elements of quantum theory.

\subsection{The thermal interpretation of quantum physics}
\label{ss.thermal}

\nopagebreak
\hfill\parbox[t]{10.8cm}{\footnotesize

{\em We assume that (i) the density matrix is observable and
(ii) any observable is a function of the density components.}

\hfill Lajos Diosi, 1988 \cite[p.2887]{Dio}
}

\bigskip

\nopagebreak
\hfill\parbox[t]{10.8cm}{\footnotesize

{\em The idea of unsharp objectification arises if one intends to leave
quantum mechanics intact and still tries to maintain a notion of real
and objective properties.
}

\hfill Busch, Lahti and Mittelstaedt, 1996 (\cite[p.127]{BusLM})
}

\bigskip

\nopagebreak
\hfill\parbox[t]{10.8cm}{\footnotesize

{\em 
We shall study a simple measurement problem -- the measurement of the
diameter of a disk. [...] It may happen that the difference of the
measurements in different directions exceeds the permissible error of a
given measurement. In this situation, we must state that within the
required measurement accuracy, our disk does not have a unique diameter,
as does a circle. Therefore, no concrete number can be taken, with
prescribed accuracy, as an estimate of the true value of the measurable
quantity.}

\hfill Semyon Rabinovich, 2005 (\cite[p.11]{Rab})
}

\bigskip

The analysis given in the present paper strongly suggests 
that we should reject the traditional convention that declares the 
eigenvalues of operators to be the true values in a measurement. This 
is done explicitly in the \bfi{thermal interpretation of quantum 
physics} introduced in 
\sca{Neumaier} \cite{Neu.IIfound,Neu.IIIfound,Neu.IVfound}; a detailed,
definitive account is in my recent book {\it Coherent Quantum Physics}
(\sca{Neumaier} \cite{Neu.CQP}); see especially Section 9.2. 
In the preceding, we essentially derived the main features of 
the thermal interpretation, including everything that had only been 
postulated in my book, from the single and simple assumption (DRP), 
the detector response principle.

The thermal interpretation gives a new foundational perspective on 
quantum mechanics and suggests different questions and approaches than 
the traditional interpretations. It extends the discussion of 
observable sources and detectors in terms of quantum values represented 
by quantum expectations to arbitrary quantum systems even in their 
unobservable aspects. According to the thermal interpretation, the 
objective (reproducible) properties of any quantum system are the 
quantum values encoded mathematically in the quantum expectations, and 
what can be computed from these. In particular, the Born--von Neumann 
link of measurement results to eigenvalues (which did not show up 
anywhere in our treatment except for highly idealized projective 
measurements) is replaced by a link of reproducible measurement results 
to quantum values, computable as quantum expectations. This link is 
amply corroborated by the fact that the extensive quantities featured 
in equilibrium thermodynamics appear in statistical mechanics as the
quantum expectations of conserved quantities. As a consequence, there 
are no obstructions to treating large systems including observers, and 
even the whole universe, as quantum systems. 

The thermal interpretation generalizes the well-known fact that
in equilibrium statistical thermodynamics, all extensive quantities are
represented by quantum expectations. It proclaims
-- in direct opposition to the tradition created in 1927 by Jordan,
Dirac, and von Neumann -- the alternative convention that the true
values are the quantum expectations rather than the eigenvalues. 

In the thermal interpretation of quantum physics, given a particular 
instance of a quantum system described by a model at a given time, the 
state defines all its properties, and hence what \bfi{exists}\footnote{
This gives a clear formal meaning to the notion of existence. Whether
something that exists in this model sense also exists in Nature depends
on how faithful the model is to the corresponding aspect of Nature.
} 
in the system at that time. The objective \bfi{properties} of the system
are given by quantum values and what is computable from these. 
All properties of a quantum system at a fixed time depend on the state 
$\<\cdot\>$ of the system at this time and are expressed in terms of 
definite, but uncertain values of the quantities. As discussed in detail 
in \sca{Neumaier} \cite{Neu.CQP}, the identification of 
these formal properties with real life properties is done by means of

\bfi{(CC)} \bfi{Callen's criterion} (cf. \sca{Callen} \cite[p.15]{Cal}):
{\it Operationally, a system is in a given state if its properties are
consistently described by the theory for this state.
} 

This is a concise version of the principle of identification suggested
by \sca{Eddington} \cite[p.222]{Edd}; cf. the quote in Subsection 
\ref{ss.perspective}. Callen's criterion is enough to find out in each 
single case how to approximately determine the uncertain value of a 
quantity of interest. 

This alternative convention also matches the actual practice in quantum 
information theory, where states of individual quantum systems are 
manipulated, transmitted, and measured, hence their properties 
(i.e., whatever is computable from the state) are treated as objective 
properties. The density operator may be viewed simply as a 
calculational tool for obtaining these objective properties, and
in particular quantum values and their uncertainties.

In the thermal interpretation, every observable scalar or vector 
quantity $A$ has an associated intrinsic state-dependent uncertainty 
$\sigma_A$ within which it can be (in principle) determined. The idea 
is that the quantum value $\ol A=\<A\>$ itself has no direct operational
meaning; only the fuzzy region of $\xi\in\Cz^m$ with $|\xi-\ol A|$ 
bounded by the uncertainty $\sigma_A$ or another small multiple of 
$\sigma_A$ is meaningful. When measuring $A$ it is meaningless to ask 
for more accuracy than the uncertainty $\sigma_A$, just as meaningless 
as to ask for the position of a doughnut to millimeter accuracy. 

This is standard engineering practice when considering the diameter of a
disk that is not perfectly circular. The uncertainty is in the imprecise
definition, just as that in the position of an extended object such as 
a doughnut. In particular, the description of a quantum particle as 
having momentum $p$ and being at position $q$ is as unsharp as the 
description of a classical signal as having frequency $\nu$ at time $t$.
Even formally, the concepts are
analogous and share the uncertainty relation, known in signal analysis
as the Nyquist theorem, and discovered in the quantum context by 
Heisenberg. The analogy is especially clear in quantum field
theory, where on the one hand position and time and on the other hand 
momentum and energy, related to frequency by the Rydberg--Ritz formula 
\gzit{e.RR}, are described on an equal footing.

As we have seen, the simplest quantum system, a qubit, was already 
described by \sca{Stokes} \cite{Sto} in 1852, in terms essentially 
equivalent to the thermal interpretation -- except that the intrinsic 
uncertainty had not yet been an issue, being at that time far below the 
experimentally realizable accuracy. 

\bigskip 

In the thermal interpretation, measurement outcomes are defined 
as quantum values of macroscopic detector quantities strongly
correlated -- in the way phenomenologically discussed in Section 
\ref{s.measP} -- with the microscopic quantities to be measured.
The thermal interpretation regards the measurement result $a_k$ for
measuring $A$ as an approximation of the true value $\ol A$ of $A$, 
which by \gzit{e.Esigma} has a typical error $|a_k-\ol A|$ of at least 
$\sigma_A$. 

Statistics becomes important whenever a quantum value has too much 
quantum uncertainty, and only then. The uncertainty in single 
measurements can be reduced -- as within classical physics -- by 
calculating statistical means of measurement results from an ensemble 
of quantum systems prepared in the same state. 
This is done by independently replicating a quantity $A$ of interest a 
number of times as $A_1,\ldots,A_N$, and taking the mean $\ol a$ of the 
measurement results $a_k$ of the $k$th copy $A_k$ of $A$. From a 
statistical point of view, the $a_k$, and hence their mean $\ol a$ are
random variables, and we may consider the statistical expectation
$\E(\ol a)$ of the mean $\ol a$, assuming many repetitions of the 
sequence of experiments from which $\ol a$ is computed. By Born's rule 
(BR) in expectation form, 
\[
\E(\ol a)=\frac{1}{N}\sum_{k=1}^N\E(a_k)
=\frac{1}{N}\sum_{k=1}^N\<A_k\>=\<\wh A\>
\] 
is the quantum expectation of the mean quantity
\[
\wh A:=\frac{1}{N}\sum_{k=1}^N A_k.
\]
Since the $A_k$ are independent, it is easy to see that the uncertainty 
of $\wh A$ is $\sigma_{\wh A}=\sigma_A/\sqrt{N}$. Thus $\wh A$ has a 
quantum uncertainty smaller than $A$ by a factor $\sqrt{N}$, common to 
all statistical methods. Since all $A_k$ are replicates of the same
$A$ we have $\<A_k\>=\<A\>$, hence the quantum values of $\wh A$ 
and $A$ agree.

This is the reason why we can obtain in principle arbitrarily accurate 
approximations of quantum values by averaging measurements -- the true 
reason why quantum tomogrpahy works: In the limit $N\to\infty$ of 
arbitrarily many repetitions, the statistical mean value of the 
approximations approaches the quantum value $\ol A$, but their standard 
deviation approaches zero!

Using an ergodic argument as in Subsection \ref{s.imagined}, we may
apply this to accurately observe $\ol A$ for a single sufficiently 
stationary quantum system by regarding the $\ol A(t_k)$ at sufficiently
different times $t_k$ as independent replications of $A$. This gives
further support to the quantum tomography process for instationary 
systems discussed in Subsection \ref{s.nonstat}. 

Thus the foundations invoked by the thermal interpretation are 
essentially the foundations used everywhere for uncertainty 
quantification, just slightly extended to accommodate quantum effects 
by not requiring that quantities commute.

\newpage
\section{Conclusion}\label{s.conc}

\nopagebreak
\hfill\parbox[t]{10.8cm}{\footnotesize

{\em In conclusion it may be mentioned that the present theory suggests 
a point of view for regarding quantum phenomena rather different from 
the usual ones. One can suppose that the initial state of a system 
determines definitely the state of the system at any subsequent time. 
[...] One can obtain a good deal of information (of the nature of 
averages) about the values at the subsequent time considered as 
functions of the initial values. The notion of probabilities does not 
enter into the ultimate description of mechanical processes.}

\hfill Paul Dirac, 1927 \cite[p.641]{Dirac1927}
}

\bigskip

\nopagebreak
\hfill\parbox[t]{10.8cm}{\footnotesize

{\em It is perhaps remarkable that all the complexities of the
observational process can be subsumed under so simple a form.}

\hfill Carl Helstrom, 1976 \cite[p.59]{Hel}
}

\bigskip

In this paper, the standard kinematical, statistical, dynamical, and 
spectral rules of introductory quantum mechanics were derived with 
little effort from a single assumption -- the detector response 
principle (DRP) introduced in Subsection \ref{ss.meas}. At the same 
time, we found the conditions under which these standard rules are 
valid.

Though very elementary, our notion of a detector and the associated
formal notion of a discrete quantum measure is very flexible and
accounts for all basic aspects of practical measurement processes.

The new approach is indifferent to the microscopic cause of 
detection events. It cleanly separates the experimentally relevant 
formal aspects of the quantum measurement process from the controversial
issues of quantum foundations. In this way, a clear formal definition 
of measurement is possible without touching the quantum measurement 
problem related to the dynamical origins of the measurement results.

No idealization is involved. It is absolutely clear what a measurement 
amounts to, and how accurately it may represent theoretical quantum 
values in operational terms. It was pointed out that to fully solve the 
quantum measurement problem, more research is needed on the 
characterization of quantum systems that are nonstationary on 
experimentally directly accessible time scales.

Quantum theory is much more than a probability theory. It predicts
lots of nonprobabilistic stuff, such as spectra of all sorts,
mechanical, optical and electrical properties of materials, stability
of chemical compounds and chemical reaction rates. These
nonprobabilistic predictions of quantum theory are derived from the
deterministic evolution of the quantum state. The quantum evolution is
characterized in the conservative case as Hamiltonian dynamics on the 
Poisson manifold of normalized density operators. 

Quantum expectations
figuring in the derivations of these results have the meaning of 
objective quantum values, independent of a probability interpretation 
The latter is needed only when averaging actual repeated measurements of
individually unpredictable events. Thus the situation is completely 
analogous to that of classical mechanics, and a notion of intrinsic 
quantum probability is superfluous.

The link to experimental practice was established through careful 
definitions in terms of quantum tomography, replacing the traditional 
a priori link via idealized projective measurements.
As a result, there is no obstruction to treating large systems including
observers, and even the whole universe, as quantum systems. A special 
role for a notion of particles, a sharp classical/quantum interface, an 
experimenter/observer, or their knowledge, was nowhere needed.
In particular, the only reasons are gone why, in spite of their strange 
features, interpretations such as Bohmian mechanics or Many-Worlds have 
become respectable among some physicists.

The new approach described in this paper is closer to actual practice 
than the traditional foundations. Moreover it is more general, and 
therefore more powerful. Finally, it is simpler and less technical than 
the traditional approach, and the standard tools of quantum mechanics 
are not difficult to derive. This makes the new approach suitable for 
introductory courses on quantum mechanics. Thus there is no longer an 
incentive for teaching quantum mechanics starting with a special, 
highly idealized case in place of the real thing.

\bigskip


\addcontentsline{toc}{section}{References}

\begin{thebibliography}{99}

\bibitem{Agg} R. Aggleton et al.,
An FPGA based track finder for the L1 trigger of the CMS
experiment at the High Luminosity LHC,
J. Instrumentation 12 (2017), P12019.

\bibitem{AkhCDK} S. Akhmanov, A. Chirkin, K. Drabovich, A. Kovrigin, 
R. Khokhlov and A. Sukhorukov, 
Nonstationary nonlinear optical effects and ultrashort light pulse 
formation,
IEEE J. Quantum Electronics, 4 (1968), 598--605.

\bibitem{AliE.meas} S.T. Ali and G.G. Emch,
Fuzzy observables in quantum mechanics,
J. Math. Phys. 15 (1974), 176--182.

\bibitem{AliM} R. Alicki and J. Messer,
Nonlinear quantum dynamical semigroups for many-body open systems,
J. Stat. Phys. 32 (1983)  299--312.

\bibitem{AllBN0} A.E. Allahverdyan, R. Balian and T.M. Nieuwenhuizen,
Determining a quantum state by means of a single apparatus,
Phys. Rev. Lett. 92 (2004), 120402.

\bibitem{AllBN1} A.E. Allahverdyan, R. Balian and T.M. Nieuwenhuizen,
Understanding quantum measurement from the solution of dynamical models,
Physics Reports 525 (2013), 1--166.

\bibitem{AllBN2} A.E. Allahverdyan, R. Balian and T.M. Nieuwenhuizen,
A sub-ensemble theory of ideal quantum measurement processes,
Annals of Physics 376 (2017), 324--352.

\bibitem{And} M. Anderson et al.,
The STAR Time Projection Chamber: A Unique Tool
for Studying High Multiplicity Events at RHIC,
Nuclear Instruments and Methods in Physics Research Section A:
Accelerators, Spectrometers, Detectors and Associated Equipment
499 (2003), 659--678.

\bibitem{App1998} D.M. Appleby,
Concept of experimental accuracy and simultaneous measurements of
position and momentum, Int. J. Theor. Phys. 37 (1998), 1491--1509.

\bibitem{App2016} D.M. Appleby,
Quantum errors and disturbances: Response to Busch, Lahti and Werner,
Entropy, 18 (2016), 174.

\bibitem{ArtK} E. Arthurs and J.L. Kelly,
BSTJ briefs: On the simultaneous measurement of a pair of conjugate
observables,
The Bell System Technical Journal, 44 (1965), 725--729.

\bibitem{Asp} A. Aspect,
Proposed experiment to test the nonseparability of quantum mechanics,
Phys. Rev. D 14 (1976), 1944--1951.
(Reprinted in \cite{WheZ}.)

\bibitem{BarGT} A. Barchielli, M. Gregoratti and A. Toigo,
Measurement uncertainty relations for discrete observables: Relative
entropy formulation,
Comm. Math. Phys. 357 (2018), 1253--1304.

\bibitem{Bel} J.S. Bell,
On the Einstein Podolsky Rosen paradox,
Physics 1 (1964), 195--200.
(Reprinted in \cite{WheZ}.)

\bibitem{Bell.against} J. Bell,
Against measurement,
Physics World 3 (1990), 33--40.

\bibitem{BelE} G. Belot and J. Earman, 
Chaos out of order: Quantum mechanics, the correspondence principle and 
chaos,
Stud. Hist. Phil. Mod. Phys. 28 (1997), 147--182.

\bibitem{BerHIW} J.C. Bergquist, R.G. Hulet, W.M. Itano and D.J. 
Wineland,
Observation of quantum jumps in a single atom,
Phys. Rev. Lett. 57 (1986), 1699--1702.

\bibitem{BerE} A.N. Beris and B.J. Edwards,
Thermodynamics of flowing systems with internal microstructure,
Oxford 1994.

\bibitem{BerSF} M. Bertolotti, L. Sereda and A. Ferrari,
Application of the spectral representation of stochastic processes to 
the study of nonstationary light radiation: a tutorial. 
Pure and Applied Optics: J. Eur. Opt. Soc. A 6 (1997), 153--171.

\bibitem{Boh1927} N. Bohr,
The quantum postulate and the recent development of atomic theory,
Lecture delivered on Sept. 16, 1927 in Como.
Published in: Nature, April 14 (1928), 580--590.

\bibitem{Bohr} N. Bohr,
On the notions of causality and complementarity,
Dialectica 2 (1948), 312.
Reprinted in: Science, New Ser. 111 (1950), 51--54.

\bibitem{BonT} E. Bonet-Luz and C. Tronci, 
Hamiltonian approach to Ehrenfest expectation values and Gaussian 
quantum states,
Proc. Royal Society A: Math. Phys. Engin. 472 (2016), 20150777.

\bibitem{BonL} R. Bonifacio and L.A. Lugiato,
Optical bistability and cooperative effects in resonance fluorescence,
Phys. Rev. A 18 (1978), 1129--1144.

\bibitem{Bor1927} M. Born,
Das Adiabatenprinzip in der Quantenmechanik,
Z. Phys. 40 (1927), 167--192.

\bibitem{BornHeisenberg1928} M. Born and W. Heisenberg,
La m\'ecanique des quanta,
pp. 143--178 in: 
\'Electrons et photons. Rapports et discussions du cinqui\`eme Conseil 
de physique tenu \`a Bruxelles du 24 au 29 Octobre 1927 sous les 
auspices de l'Institut international de physique Solvay
(H. A. Lorentz, ed.), 
Gauthier-Villars, Paris 1928.


\bibitem{Bor.Nat} M. Born, 
Natural philosophy of cause and chance,
Clarendon Press, Oxford 1949.


\bibitem{Bra} H.E. Brandt,
Positive operator valued measure in quantum information processing,
Amer. J. Physics 67 (1999), 434--439.

\bibitem{Bras} G. Brassard,
Is information the key?
Nature Physics 1 (2005), 2--4.

\bibitem{BreP.QC} H. Breuer and F. Petruccione,
Quantum measurement and the transformation from quantum to classical
probabilities,
Phys. Rev. A 54 (1996), 1146--1153.

\bibitem{BreP} H.-P. Breuer and F. Petruccione,
The theory of open quantum systems.
Oxford Univ. Press, New York, 2002.


\bibitem{BroG} L.S. Brown and G. Gabrielse,
Geonium theory: Physics of a single electron or ion in a Penning trap
Rev. Mod. Phys. 58 (1986), 233--311.

\bibitem{SIunits} Bureau International des Poids et Mesure,
The international system of units (SI),  9th edition 2019.
\url{https://www.bipm.org/en/publications/si-brochure}

\bibitem{Bus} P. Busch,
Quantum states and generalized observables: a simple proof of
Gleason's theorem,
Phys. Rev. Lett. 91 (2003), 120403.

\bibitem{BusGL} P. Busch, M. Grabowski and P. Lahti,
Operational quantum physics,
Springer, Berlin 1995.

\bibitem{BusHL} P. Busch, T. Heinonen and P. Lahti,
Heisenberg's uncertainty principle,
Phys. Rep. 452 (2007), 155--176.

\bibitem{BusL} P. Busch and P. Lahti,
Individual aspects of quantum measurements,
J. Phys. A: Math. Gen. 29 (1996), 5899--5907.

\bibitem{BusL2} P. Busch and P. Lahti,
The standard model of quantum measurement theory: history and
applications,
Found. Phys. 26 (1996), 875--893.

\bibitem{BusL3} P. Busch and P. Lahti,
Measurement theory,
pp.374--379 in:
Compendium of Quantum Physics (D. Greenberger et al., eds.),
Springer, Dordrecht 2009.

\bibitem{BusLM} P. Busch, P. Lahti and P. Mittelstaedt,
The quantum theory of measurement, 2nd ed.,
Springer, Berlin 1996.

\bibitem{BusLPY} P. Busch, P. Lahti, J. Pellonp\"a\"a and K. Ylinen,
Quantum Measurement,
Springer, Berlin 2016.

\bibitem{BusLW} P. Busch, P. Lahti and R.F.Werner,
Measurement uncertainty relations,
J. Math. Phys. 55 (2014), 042111.

\bibitem{Cal} H.B. Callen.
Thermodynamics and an introduction to thermostatistics,
2nd. ed.
Wiley, New York, 1985.

\bibitem{CalH.book} E. Calzetta and B.L. Hu,
Nonequilibrium quantum field theory,
Cambridge Univ. Press, New York 2008.

\bibitem{CavFMR} C.M. Caves, C.A. Fuchs, K.K. Manne and J.M. Renes,
Gleason-type derivations of the quantum probability rule for
generalized measurements.
Foundations of Physics 34 (2004), 193--209.

\bibitem{CavFS} C.M. Caves, C.A. Fuchs and R. Schack,
Unknown quantum states: the quantum de Finetti representation.
J. Math. Phys. 43 (2002), 4537--4559.

\bibitem{Choi} M.-D. Choi,
Completely Positive Linear Maps on Complex Matrices,
Linear Algebra Appl. 10 (1975), 285--290.

\bibitem{ChuN7} I.L. Chuang and M.A. Nielsen,
Prescription for experimental determination of the dynamics of a
quantum black box,
J. Modern Optics. 44 (1997), 2455--2467

\bibitem{DanLP} A. Daneri, A. Loinger and G.M. Prosperi,
Quantum theory of measurement and ergodicity conditions,
Nucl. Phys. 33 (1962), 297--319.

\bibitem{DArMP} G.M. D'Ariano, L. Maccone and P.L. Presti,
Quantum Calibration of Measurement Instrumentation.
Phys. Rev. Lett. 93 (2004), 250407.

\bibitem{DArP} G.M. D'Ariano and P.L. Presti,
Quantum tomography for measuring experimentally the matrix elements of
an arbitrary quantum operation,
Phys. Rev. Lett. 86 (2001), 4195--4198.

\bibitem{Dav69} E.B. Davies,
Quantum stochastic processes,
Comm. Math. Phys. 15 (1969), 277--304.

\bibitem{DavL} E.B. Davies and J.T. Lewis,
An operational approach to quantu-m probability,
Comm. Math. Phys. 17 (1970), 239--260.

\bibitem{dlPenCV} L. de la Pe\~na, A.M. Cetto and 
A. Vald\'es-Hern\'andez,
How fast is a quantum jump?
Phys. Lett. A 384 (2020), 126880.

\bibitem{deMuyK} W.M. de Muynck and J.M.V.A. Koelman,
On the joint measurement of incompatible observables in quantum
mechanics,
Physics Letters A 98 (1983), 1--4.

\bibitem{deMuy2002} W.M. de Muynck,
Foundations of Quantum Mechanics, an Empiricist Approach,
Kluwer, Dordrecht 2002.

\bibitem{Dio} L. Diosi,
Quantum stochastic processes as models for state vector reduction,
J. Physics A: Math.Gen. 21 (1988), 2885--2898.

\bibitem{Dirac1925} P.A.M. Dirac,
The fundamental equations of quantum mechanics,
Proc. Roy. Soc. Lond. A 109 (1925), 642--653.

\bibitem{Dirac1927} P.A.M. Dirac, 
The physical interpretation of the quantum dynamics,
Proc. Roy. Soc. Lond. A 113 (1927), 621--641.

\bibitem{Dir1} P.A.M. Dirac,
The principles of quantum mechanics, 1st ed.,
Oxford Univ. Press, Oxford 1930.

\bibitem{Dir73} P.A.M. Dirac,
Development of the physicist's conception of Nature,
Section 1 (pp. 1--14) in:  
The physicist's conception of Nature (J. Mehra, ed.),
Reidel, Dordrecht 1973.

\bibitem{DiVF} D.P. DiVincenzo and C.A. Fuchs,
Quantum foundations,
Physics Today 72 (2019), 2--3.

\bibitem{Edd} A.S. Eddington,	
The mathematical theory of relativity, 2nd ed.,
Cambridge Univ. Press, Cambridge 1930.

\bibitem{Ehr} P. Ehrenfest,
Bemerkung \"uber die angen\"aherte G\"ultigkeit der klassischen
Mechanik innerhalb der Quantenmechanik,
Zeitschrift f\"ur Physik 45 (1927), 455--457.

\bibitem{Einstein1934} A. Einstein,
On the method of theoretical physics, 
Philosophy of Science 1 (1934), 163--169.

\bibitem{EinPR} A. Einstein, B. Podolsky, and N. Rosen,
Can quantum-mechanical description of physical reality be considered
complete?
Phys. Rev. 47 (1935), 777--781.

\bibitem{Ein53} A. Einstein,
Elementare \"Uberlegungen zur Interpretation der Grundlagen der 
Quanten-Mechanik,
pp.33--40 in: Scientific papers presented to Max Born,
Oliver and Boyd, New York 1953.

\bibitem{Eng} B.G. Englert,
On quantum theory.
The European Physical Journal D 67 (2013), 238.

\bibitem{Epi} Epicurus, Letter to Herodotus,
in: S. Naragon,  Letter to Herodotus, quoted October 11, 2021 from 
\url{https://en.wikipedia.org/wiki/Letter_to_Herodotus}

\bibitem{Fan} U. Fano, 
Quantum theory of interference effects in the mixing of light from 
phase-independent sources. 
Amer. J. Phys. 29(1961), 539--545.

\bibitem{FaiR} P. Faist and R. Renner, 
Practical and reliable error bars in quantum tomography,
Phys. Rev. Lett. 117 (2016), 010404.

\bibitem{Fal} B. Falkenburg,
Particle metaphysics,
Springer, Berlin 2007.

\bibitem{FeyLS} R.P. Feynman, R.B. Leighton and M. Sands,
The Feynman lectures of physics, Vol. 3, 
Addison--Wesley 1971.

\bibitem{Fey1982} R.P. Feynman, 
Simulating Physics with Computers, 
Int. J. Theor. Phys. 2l (1982), 467--488.

\bibitem{FigT} R. Figari and A. Teta,
Emergence of classical trajectories in quantum systems: the cloud
chamber problem in the analysis of Mott (1929),
Arch. Hist. Exact Sci. 67 (2013), 215--234.

\bibitem{FlaGLE} S.T. Flammia, D. Gross, Y.K. Liu and J. Eisert,
Quantum tomography via compressed sensing: error bounds, sample 
complexity and efficient estimators,
New J. Phys. 14 (2012), 095022.

\bibitem{Fro2021} J.  Fr\"ohlich,
Relativistic quantum theory,
pp. 237--257 in: 
Do Wave Functions Jump? (V. Allori et al., eds.),
Springer, Cham 2021.

\bibitem{Fro} J.  Fr\"ohlich,
A brief review of the ''ETH-approach to quantum mechanics'',
pp. 21--45 in: 
Frontiers in Analysis and Probability (N. Anantharaman et al., eds.)
Springer, Cham 2020.

\bibitem{Fuc2003} C.A. Fuchs,
Quantum mechanics as quantum information, mostly,
J. Modern Optics 50 (2003), 987--1023.


\bibitem{GevSCK} T.V. Gevorgyan, A.R. Shahinyan, L.Y. Chew and
G.Y. Kryuchkyan,
Bistability and chaos at low levels of quanta,
Phys. Rev. E 88 (2013), 022910.

\bibitem{Gib} J.W. Gibbs,
Elementary principles in statistical mechanics,
Yale Univ. Press, New Haven, 1902.
Reprinted, Dover 1960.

\bibitem{GolvL} G.H. Golub and C.F. van Loan,
Matrix Computations, 4th ed.,
Johns Hopkins Univ. Press, Baltimore 2013.

\bibitem{GomLL} I. Gomez, M. Losada and O. Lombardi, 
About the concept of quantum chaos,
Entropy 19 (2017), 205.

\bibitem{GorKS} V. Gorini, A. Kossakowski and E.C.G. Sudarshan,
Completely positive dynamical semigroups of $N$-level systems,
J. Math. Phys.17 (1976), 821--825.

\bibitem{Got} K. Gottfried,
Does quantum mechanics carry the seeds of its own destruction?
Physics World 4 (1991), 34--40.

\bibitem{Grabow} M. Grabowski,
What is an observable?
Found. Phys. 19 (1989), 923--930.

\bibitem{GraFF} C. Granade, C. Ferrie and S.T. Flammia,
Practical adaptive quantum tomography,
New J. Physics 19 (2017), 113017.

\bibitem{Gri.m} R.B. Griffiths,
What quantum measurements measure,
Phys. Rev. A 96 (2017), 032110.

\bibitem{GriS} D.J. Griffiths and D.F. Schroeter,
Introduction to quantum mechanics,
Cambridge University Press, Cambridge 2018.

\bibitem{Haa} R. Haag, 
Local quantum physics, 2nd. ed.,
Springer, Berlin 1992.

\bibitem{HaaK} R. Haag and D. Kastler,
An algebraic approach to quantum field theory,
J. Math. Phys. 5 (1964), 848--861.

\bibitem{Ham} J. Hamhalter,
Quantum measure theory,
Springer, New York 2013.

\bibitem{Hei1925} W. Heisenberg, 
\"Uber quantentheoretische Umdeutung kinematischer und 
mechanischer Beziehungen,
Zeitschr. Physik 33 (1925), 879--893.
(English transl. pp. 261--276 in \cite{Wae67}.)

\bibitem{Hei.Como} W. Heisenberg, in:
Discussione sulla communicazione Bohr, pp. 589--598 in:
Proc. Int. Conf. Physicists, Como 1927.

\bibitem{Hei30} W. Heisenberg, 
The physical principles of the quantum theory, 
Univ. of Chicago Press, Chicago 1930.

\bibitem{Hel74} C.W. Helstrom,
"Simultaneous measurement" from the standpoint of quantum estimation
theory,
Found. Phys. 4 (1974), 453--463.

\bibitem{HelWD} M. Hell, M.R. Wegewijs and D.P. DiVincenzo,
Qubit quantum-dot sensors: Noise cancellation by coherent backaction, 
initial slips, and elliptical precession,
Phys. Rev. B 93 (2016), 045418.

\bibitem{HelK} K.-E. Hellwig and K. Kraus,
Pure operations and measurements,
Comm. Math. Phys. 11 (1969), 214--220.

\bibitem{Hel} C.W. Helstrom,
Quantum detection and estimation theory,
Academic Press, New York 1976.

\bibitem{Her} P. Hertz,
\"Uber die mechanischen Grundlagen der Thermodynamik,
Ann. Physik IV. Folge (33) 1910, 225--274.

\bibitem{Hob} A. Hobson,
There are no particles, there are only fields,
Amer. J. Physics 81 (2013), 211--223.

\bibitem{Hol1973} A.S. Holevo,
Statistical decision theory for quantum systems,
J. Multivariate Analysis 3 (1973), 337--394.

\bibitem{Hol1982} A.S. Holevo,
Probabilistic and statistical aspects of quantum theory,
North-Holland, Amsterdam 1982.
(2nd ed., 2011. Russian original: Moscow 1980.)

\bibitem{Hol2001} A.S. Holevo,
Statistical structure of quantum theory,
Springer, Berlin 2001.

\bibitem{Hol2012} A.S. Holevo,
Quantum systems, channels, information,
de Gruyter, Berlin 2012.

\bibitem{HonZ} M.O. Hongler and W.M. Zheng,
Exact solution for the diffusion in bistable potentials,
J. Statist. Phys, 29 (1982), 317--327.

\bibitem{IngA} R.L. Ingraham and G.L. Acosta, 
On chaos in quantum mechanics: the two meanings of sensitive dependence,
Phys. Lett. A 181 (1993), 450--452.

\bibitem{JagP} V. Jagadish and F. Petruccione,
An invitation to quantum channels,
Quanta 7 (2018), 54--67.

\bibitem{JezFH} M. Je\v zek, J. Fiur\'a\v sek and Z. Hradil,
 Quantum inference of states and processes,
Phys. Rev. A 68 (2003), 012305.

\bibitem{Jon} R.C. Jones,
New calculus for the treatment of optical systems. I.
Description and discussion of the calculus,
J. Optical Soc. Amer. 31 (1941), 488--493.

\bibitem{KanNO} N.H. Kaneko, S. Nakamura and Y. Okazaki,
A review of the quantum current standard,
Measurement Science and Technology 27 (2016), 032001.

\bibitem{Kap} R. Kapral,
Quantum dynamics in open quantum-classical systems,
J. Phys. Condensed Matter 27 (2015), 073201.

\bibitem{KapC} R. Kapral and G. Ciccotti,
Mixed quantum-classical dynamics,
J. Chem. Phys. 110 (1999), 8919--8929.


\bibitem{Ken} E.H. Kennard,
Zur Quantenmechanik einfacher Bewegungstypen,
Zeitschrift f. Physik 44 (1927), 326--352.

\bibitem{Koo} B.O. Koopman,
Hamiltonian systems and transformations in Hilbert space,
Proc. Nat. Acad. Sci. 17 (1931), 315--318.

\bibitem{Kos} A. Kossakowski,
On quantum statistical mechanics of non-Hamiltonian systems,
Rep. Math. Phys. 3 (1972), 247--274.

\bibitem{Kra} K. Kraus, General state changes in quantum theory,
Ann. Phys. 64 (1971), 311--335.

\bibitem{KruM} P. Kruszy\'nski and W.M. De Muynck,
Compatibility of observables represented by positive operator-valued
measures,
J. Math. Phys. 28 (1987), 1761--1763.

\bibitem{Landau1927} L. Landau,
Das D\"ampfungsproblem in der Wellenmechanik,
Z. Phys. 45 (1927), 430--441.

\bibitem{LL.3} L.D. Landau and E.M. Lifshitz,
Course of Theoretical Physics,
Vol. 3: Quantum mechanics, 3rd ed.,
Pergamon Press, 1977.


\bibitem{Lan2017} N.P. Landsman,
Foundations of quantum theory: from classical concepts to operator 
algebras,
Springer, Cham 2017.

\bibitem{Lan2019} N.P. Landsman,
Randomness? What randomness?
Manuscript (2019).\\
arXiv:1908.07068.

\bibitem{Leo} U. Leonhardt,
Measuring the Quantum State of Light,
Cambridge, 1997.

\bibitem{Leo.i} U. Leonhardt, 
Quantum physics of simple optical instruments. 
Rep. Progr. Phys. 66 (2003), 1207--1248.

\bibitem{LeoN} U. Leonhardt and A. Neumaier,
Explicit effective Hamiltonians for linear quantum-optical networks,
J. Optics B: Quantum Semiclass. Opt. 6 (2004), L1--L4.

\bibitem{LepS} L. Lepp\"aj\"arvi and M. Sedl\'ak,
Postprocessing of quantum instruments,
Phys. Rev. A 103 (2021), 022615.

\bibitem{Lin} G. Lindblad,
On the generators of quantum dynamical semigroups,
Comm. Math. Phys. 48 (1976), 119--130.

\bibitem{Nobel1989s} I. Lindgren, 
The Nobel Prize in Physics 1989. Award ceremony speech.
\url{https://www.nobelprize.org/prizes/physics/1989//ceremony-speech/}

\bibitem{LunFCP} J.S. Lundeen, A. Feito, H. Coldenstrodt-Ronge, 
K.L. Pregnell, C. Silberhorn, T.C. Ralph, J. Eisert, M.B. Plenio and 
I.A. Walmsley,   
Tomography of quantum detectors,
Nature Physics 5 (2009), 27--30.

\bibitem{LuiS} A. Luis and L.L. S\'anchez-Soto,
Complete characterization of arbitrary quantum measurement processes,
Phys. Rev. Lett. 83 (1999), 3573--3576.

\bibitem{Mal} E.-L. Malus,
Sur une propri\'et\'e de la lumi\`ere r\'efl\'echie,
M\'em. Phys. Chim. Soc. D'Arcueil 2 (1809), 143--158.
Reprinted and partially translated into English in \cite{Swi}.

\bibitem{ManW} L. Mandel and E. Wolf,
Optical coherence and quantum optics,
Cambridge Univ. Press, Cambridge 1995.

\bibitem{ManMSZ} V.I. Man'ko, G. Marmo, E.C.G. Sudarshan and F. 
Zaccaria, 
Wigner's problem and alternative commutation relations for quantum 
mechanics,
Int. J. Modern Physics B 11 (1997), 1281--1296.

\bibitem{MarR} J.E. Marsden and T.S. Ratiu,
Introduction to mechanics and symmetry,
Springer, New York 1994.

\bibitem{MarH} D. Marx and J. Hutter,
Ab initio molecular dynamics: Theory and implementation,
pp. 301--449 in:
Modern methods and algorithms of quantum chemistry 
(J. Grotendorst, ed.),
John von Neumann Institute for Computing, Research Center J\"ulich, 
J\"ulich 2000.

\bibitem{Mer} N.D. Mermin,
Making better sense of quantum mechanics, 
Rep. Prog. Phys. 82 (2018), 012002.

\bibitem{Mil} R.A. Millikan,
Recent Developments in Spectroscopy, 
Proc. Amer. Phil. Soc. 66 (1927), 211--230.

\bibitem{Mog} D. Mogilevtsev,
Calibration of single-photon detectors using quantum statistics,
Phys. Rev. A 82 (2010), 021807.

\bibitem{MohRL} M. Mohseni, A.T. Rezakhani and D.A. Lidar,
Quantum-process tomography: Resource analysis of different strategies,
Phys. Rev. A 77 (2008) 032322.

\bibitem{Mott} N.F. Mott,
The wave mechanics of $\alpha$-ray tracks,
Proc. Royal Soc. London A 126 (1929), 79--84.

\bibitem{NagSD} W. Nagourney, J. Sandberg and H. Dehmelt,
Shelved optical electron amplifier: Observation of quantum jumps,
Phys. Rev. Lett. 56 (1986), 2797--2799.

\bibitem{Nai} M.A. Naimark,
On a Representation of Additive Operator Set Functions (in Russian),
Dokl. Akad. Nauk SSSR. 41 (1943), 373--375.
English translation: M.A. Neumark,
C.R. (Doklady) Akad. Sci. URSS. (N.S.) 41 (1943) 359--361.

\bibitem{Neu.Ifound} A. Neumaier,
Foundations of quantum mechanics I. A critique of the tradition,
Manuscript (2019).
\url{https://arxiv.org/abs/1902.10778}


\bibitem{Neu.IIfound} A. Neumaier,
Foundations of quantum physics II. The thermal interpretation,
Manuscript (2019).
\url{https://arxiv.org/abs/1902.10779}

\bibitem{Neu.IIIfound} A. Neumaier,
Foundations of quantum mechanics III. Measurement,
Manuscript (2019).
\url{https://arxiv.org/abs/1902.10782}

\bibitem{Neu.IVfound} A. Neumaier,
Foundations of quantum physics IV. More on the thermal interpretation,
Manuscript (2019).
\url{https://arxiv.org/abs/1904.12721}

\bibitem{Neu.CQP} A. Neumaier,
Coherent Quantum Physics. A Reinterpretation of the Tradition,
de Gruyter, Berlin 2019.

\bibitem{Neu.qubit} A. Neumaier,
A classical view of the qubit,
Physics Forums Insights (March 2019). \\
{\footnotesize\url{https://www.physicsforums.com/insights/a-classical-view-of-the-qubit}}


\bibitem{NeuW} A. Neumaier and D. Westra,
Classical and quantum mechanics via Lie algebras.
Online book, 1st ed. 2008, 2nd ed. 2011.
\url{http://arxiv.org/abs/0810.1019}

\bibitem{vNeu1927} J. von Neumann,
Mathematische Begr\"undung der Quantenmechanik,
Nachr. Ges. Wiss. G\"ottingen, Math.-Phys. Klasse 1927 (1928), 1--57.

\bibitem{NieC} M.A. Nielsen and I.L, Chuang,
Quantum computation and quantum information: 10th Anniversary Edition,
Cambridge Univ. Press, Cambridge 2011.


\bibitem{Nok} J. Nokkala, 
Online quantum time series processing with random oscillator networks,
Manuscript (2021). 
arXiv:2108.00698

\bibitem{Oet} H.C. Oettinger,
Beyond Equilibrium Thermodynamics,
Wiley-Interscience 2005.  


\bibitem{Oza2021}
M. Ozawa, 
Quantum Measurement Theory for Systems with Finite Dimensional State 
Spaces,
Manuscript (2021).
\url{https://arxiv.org/abs/2110.03219}

\bibitem{Pei} R. Peierls, 
In defence of 'Measurement', 
Physics World 4 (1991), 19--20.

\bibitem{Peres} A. Peres,
Quantum theory: Concepts and methods,
Kluwer, Dordrecht 2002.

\bibitem{Per2003} A. Peres,
What's Wrong with these Observables?
Found. Phys. 33 (2003), 1543--1547.

\bibitem{PerO}
J.J.G. Perez and R. Ossikovski,
Polarized light and the Mueller matrix approach,
CRC Press, Boca Raton 2017.

\bibitem{PoiScH} H. Poincar\'e,
La science et l'hypoth\`ese,
Flammarion, 1902.
English translation:
Science and hypothesis,
Walter Scott, London 1905.

\bibitem{PoyCZ} J.F.Poyatos, J.I. Cirac and P. Zoller,
Complete characterization of a quantum process: the two-bit quantum
gate,
Phys. Rev. Lett. 78 (1997), 390--393.

\bibitem{Pru} E. Prugove\v{c}ki,
Information-theoretical aspects of quantum measurements,
Int. J. Theor. Phys. 16 (1977), 321--331.

\bibitem{QueBJ} N. Quesada, A.M. Bra\'nczyk and D.F. James,
Self-calibrating tomography for multidimensional systems,
Phys. Rev. A 87 (2013), 062118.

\bibitem{Rab} S.G. Rabinovich,
Measurement errors and uncertainties: Theory and practice.
Springer, New York 2005.

\bibitem{Req} R. Requist,
Hamiltonian formulation of nonequilibrium quantum dynamics: Geometric 
structure of the Bogoliubov-Born-Green-Kirkwood-Yvon hierarchy,
Phys. Rev. A 86 (2012), 022117.

\bibitem{RigAM} M. Rigo, G. Alber, F. Mota-Furtado and P.F. O'Mahony, 
Measurement-induced quantum fluctuations and bistability of a 
relativistic electron in a Penning trap,
Phys. Rev. A 58 (1998), 478--487.

\bibitem{Rob29} H.P. Robertson,
The uncertainty principle,
Phys. Rev. 34 (1929), 163--164.

\bibitem{Rov.rQM} C. Rovelli, 
Relational quantum mechanics.,
Int. J. Theor. Phys. 35 (1996), 1637--1678.

\bibitem{Schl.book} M. Schlosshauer,
Decoherence and the quantum-to-classical transition,
Springer, New York 2007.

\bibitem{Schl2019} M. Schlosshauer,
Quantum decoherence,
Physics Reports (2019).

\bibitem{Schr1985} F.E. Schroeck,
Compatible stochastic observables that do not commute,
Found. Phys. 15 (1985), 677--681.

\bibitem{Schr1989} F.E. Schroeck,
Coexistence of observables,
Int. J. Theor. Phys. 28 (1989), 247--262.

\bibitem{Schroedinger1958} E. Schr\"odinger,
Might perhaps energy be a merely statistical concept?
Il Nuovo Cimento (1955-1965) 9 (1958), 162--170.

\bibitem{SidOCK} J.S. Sidhu, Y. Ouyang, E.T. Campbell and P. Kok, 
Tight bounds on the simultaneous estimation of incompatible parameters,
Phys. Rev. X 11 (2021), 011028.

\bibitem{SimSNE} J.Y. Sim, J Shang, H.K. Ng and B.G. Englert,
Proper error bars for self-calibrating quantum tomography,
Phys. Rev. A 100 (2019), 022333.

\bibitem{Stil} W.S. Stiles,
Current problems of visual research, 
Nature 154 (1944), 290--293.

\bibitem{Sti} W.F. Stinespring,
Positive functions on $C^*$-algebras,
Proc. Amer. Math. Soc. 6 (1955), 211--216.

\bibitem{Sto} G.G. Stokes,
On the composition and resolution of streams of polarized light
from different sources,
Trans. Cambridge Phil. Soc. 9 (1852), 399--416.
Reprinted in \cite{Swi}.

\bibitem{Stro} F. Strocchi,
Complex coordinates and quantum mechanics,
Rev.f Mod. Phys. 38 (1966), 36--40.

\bibitem{SudMR} E.C.G. Sudarshan, P.M. Mathews and J. Rau,
Stochastic dynamics of quantum-mechanical systems,
Phys. Rev. 121 (1961), 920--924.

\bibitem{Sug} T. Sugiyama,
Finite Sample Analysis in Quantum Estimation,
Springer, Tokyo 2014.

\bibitem{Swi} W. Swindell,
Polarized light,
Dowden 1975.

\bibitem{TajO} D. Taj and H.C. \"Ottinger,
Natural approach to quantum dissipation,
Phys. Rev. A 92 (2015), 062128.

\bibitem{Teg} M. Tegmark, 
Apparent wave function collapse caused by scattering,
Found. Phys. Lett. 6 (1993), 571--590.


\bibitem{tHoo2021} Gerard 't Hooft,
Ontology in quantum mechanics
arXiv preprint arXiv:2107.14191, 2021

\bibitem{TraN} Q.H. Tran and K. Nakajima,
Learning Temporal Quantum Tomography, Manuscript (2021). 
arXiv:2103.13973

\bibitem{Uff1994} J. Uffink,
The joint measurement problem,
Int. J. Theor. Phys. 33 (1994), 199--212.

\bibitem{vDyckSD} R.S. Van Dyck Jr.,  P.B. Schwinberg and H.G. Dehmelt, 
Precise measurements of axial, magnetron, cyclotron, and 
spin-cyclotron-beat frequencies on an isolated 1-MeV electron,
Phys. Rev. Lett. 38 (1977), 310--314. 

\bibitem{vKam1988} N.G. van Kampen,
Ten theorems about quantum mechanical measurements,
Physica A: Stat. Mech. Appl. 153 (1988), 97--113.

\bibitem{Wae67} B.L. van der Waerden, 
Sources of quantum mechanics, 
North-Holland, Amsterdam 1967.

\bibitem{Var} Y.P. Varshni, 
Comparative study of potential energy functions for diatomic molecules,
Rev. Mod. Phys. 29 (1957), 664--682.

\bibitem{Wal12} D. Wallace,
Decoherence and its role in the modern measurement problem. 
Phil. Trans. Roy. Soc. A: 
Math. Phys. Eng. Sci. 370 (2012), 4576--4593.

\bibitem{Wallace} D. Wallace,
What is orthodox quantum mechanics?
Unpublished Manuscript (2016). 
\url{https://arxiv.org/abs/1604.05973}

\bibitem{WeiI} S. Weinberg,
The Quantum Theory of Fields,
Volume 1, Foundations.
Cambridge Univ. Press, Cambridge 1995.

\bibitem{Wei99} S. Weinberg,
What is quantum field theory, and what did we think it was?
pp. 241--251 in: 
Conceptual foundations of quantum field theory
(Tian Yu Cao, ed.)
Cambridge Univ. Press, Cambridge 1999.

\bibitem{Wei2017} S. Weinberg, 
The trouble with quantum mechanics,
The New York Review of Books 19 (2017), 1--7.

\bibitem{Wer} R. Werner,
Screen observables in relativistic and nonrelativistic quantum
mechanics,
J. Mat. Phys. 27 (1986), 793--803.

\bibitem{Wet} C. Wetterich,
Non-equilibrium time evolution in quantum field theory,
Phys. Rev. E 56 (1997), 2687--2690.

\bibitem{WheZ} J.A. Wheeler and W.H. Zurek (eds.),
Quantum theory and measurement.
Princeton Univ. Press, Princeton 1983.

\bibitem{Whi} P. Whittle,
Probability via expectation, 3rd ed.,
Springer, New York 1992.

\bibitem{Wik.Born} Wikipedia,
Born rule,
Web document.\\
\url{https://en.wikipedia.org/wiki/Born_rule}


\bibitem{Woi} P. Woit,
Quantum Theory, Groups and Representations: An Introduction,
Springer, New York 2017.

\bibitem{ZavB} A. Zavatta and M. Bellini,
The quantum picture of a detector,
Nature Photonics 6((2012), 350--351.

\bibitem{Zeh1970} H.D. Zeh,
On the interpretation of measurement in quantum theory, 
Foundations of Physics 1 (1970), 69--76.

\bibitem{ZhaF} W.M. Zhang and D.H. Feng,
Quantum nonintegrability in finite systems,
Phys. Reports 252 (1995), 1--100.

\bibitem{Zur} W.H. Zurek,
Decoherence, einselection, and the quantum origins of the classical,
Rev. Mod. Phys. 75 (2003), 715--775.

\end{thebibliography}
\end{document}